\documentclass[pre,superscriptaddress]{revtex4}

\usepackage{amsmath,amsfonts,graphicx,hyperref}
\usepackage{subfigure}
\usepackage{amssymb}
\usepackage{amsfonts}
\usepackage{color}
\usepackage[normalem]{ulem}

\DeclareMathOperator{\sech}{sech}

\begin{document}
	\newtheorem{corollary}{Corollary}[section]
	\newtheorem{remark}{Remark}[section]
	\newtheorem{definition}{Definition}[section]
	\newtheorem{theorem}{Theorem}[section]
	\newtheorem{proposition}{Proposition}[section]
	\newtheorem{lemma}{Lemma}[section]
	\newtheorem{help1}{Example}[section]

\def\re{\text{Re}}
\def\im{\text{Im}}
\def\labelitemi{$\blacktriangleright$}

\title{Extreme wave events for a nonlinear Schr\"odinger equation  
with linear damping and Gaussian driving}

\author{G. Fotopoulos}
\affiliation{Department of Electrical Engineering  and Computer Science,\\ Khalifa University P.O. Box 127788,  Abu Dhabi, United Arab Emirates}
\affiliation{Department of Mathematics, Statistics and Physics, College of Arts and Sciences, Qatar University, P.\,O. Box 2713, Doha, Qatar}
\author{D.\,J. Frantzeskakis}
\affiliation{Department of Physics, National and Kapodistrian University of Athens, Panepistimiopolis, Zografos, Athens 15784, Greece}
\author{N.\,I. Karachalios}
\affiliation{Department of Mathematics,\\ Laboratory of Applied Mathematics and Mathematical Modelling,\\ University of the Aegean, Karlovassi, 83200 Samos, Greece}
\author{P.\,G. Kevrekidis}
\affiliation{Department of Mathematics and Statistics, University of Massachusetts, Amherst MA 01003-4515, USA}
\author{V. Koukouloyannis}
\affiliation{Department of Mathematics,\\ Laboratory of Applied Mathematics and Mathematical Modelling,\\ University of the Aegean, Karlovassi, 83200 Samos, Greece}
\affiliation{Department of Mathematics, Statistics and Physics, College of Arts and Sciences, Qatar University, P.\,O. Box 2713, Doha, Qatar}
\author{K. Vetas}
\affiliation{Department of Mathematics, Statistics and Physics, College of Arts and Sciences, Qatar University, P.\,O. Box 2713, Doha, Qatar}

\begin{abstract}
We perform a numerical study of the initial-boundary value problem, with vanishing boundary 
conditions, of a driven nonlinear Schr\"odinger equation (NLS) with linear damping and a 
Gaussian driver. We identify Peregrine-like rogue waveforms, excited by two different types of vanishing initial data decaying at an algebraic or exponential rate. The observed extreme events emerge on top of a decaying support. 
Depending on the spatial/temporal scales of the driver, the transient dynamics -- 
prior to the eventual decay of the solutions -- may resemble the one in the semiclassical 
limit of the integrable NLS, or may, e.g., lead to large-amplitude breather-like patterns.
The effects of the damping strength and driving amplitude in suppressing 
or enhancing respectively the relevant features, as well as of the phase of the driver in the construction of a diverse array of spatiotemporal
patterns, are numerically analyzed. 
\end{abstract}

\maketitle

\section{Introduction}
The nonlinear Schr\"odinger (NLS) equation  
\begin{equation}
\label{eq0}
\mathrm{i}{{u}_{t}}+\frac{1}{2} {{u}_{xx}}+|u|^2u =0,
\end{equation}
is one of the universal
evolution equations  with a multitude of applications relevant to nonlinear wave propagation (from nonlinear optics and Bose-Einstein condensates, to plasma physics  and deep water waves) \cite{Ablo,ablo2}. 
Its counterparts which may incorporate damping, external driving or nonlinear gain /loss effects, define a fundamental class of infinite dimensional dynamical systems which may exhibit complex spatiotemporal behaviour.  In particular, the linearly damped and driven 
NLS equation, expressed in dimensionless form:
\begin{equation}
\label{eq1}
\mathrm{i}{{u}_{t}}+\frac{1}{2} {{u}_{xx}}+|u|^2u =f-\mathrm{i}\gamma u,
\end{equation}
is one of the prototypical partial differential equations which may demonstrate rich and complex dynamics \cite{NB86,CLM}. Here, $f=f(x,t)$ stands for the driving (or forcing) of the system, 
while $-\mathrm{i}\gamma u$ accounts for linear damping of strength $\gamma>0$; thus, 
Eq.~(\ref{eq1}) defines a non-autonomous perturbation of the integrable NLS (\ref{eq0}) ($\gamma=0$, $f=0$), 
\cite{nobe1,nobe2,Li,Wig1,kai,eli}. Generically, the complex dynamics of the system is 
captured by its global attractor, \cite{Ghid88,XW95,Goubet1,Goubet2,Goubet3,Goubet2a,Lauren95}.

In the present study on 
Eq.~(\ref{eq1}), we continue our explorations \cite{All2,ZNA2019} on the robustness of extreme wave events and rogue waves \cite{k2a,k2b,k2c,k2d}, in the presence of non-integrable perturbations of the integrable NLS (\ref{eq0}). 
We recall that in this limit, the manifestation of such events is
described by the class of rational solutions such
as the Peregrine rogue wave (PRW) and the
Akhmediev and Kuznetsov--Ma (KMb) breathers 
\cite{H_Peregrine,kuz,ma,akh,dt}.  The study of this class of
solutions, and accordingly, the question of their persistence under
the influence of additional physically realistic mechanisms is
of substantial interest, as such extreme waveforms have been
identified experimentally in numerous physical contexts
\cite{hydro,opt2,laser,He,plasma}, beyond their (often disastrous)
appearance in the oceans. These effects may be described by correction
terms
ranging from those incorporating higher-order derivatives, to
time-dependent coefficients associated to dispersion and nonlinearity
management. The same persistence question has been investigated also
for coupled systems. For results in the above directions, we refer to
\cite{devine,NR4Anki,NR5Wang,NRbor6,NRbor5a,calinibook,BorPT,BorCD,BorMAN,BorDPJ}
and references therein.  However, to the best of our knowledge, very little seems to be known for the damped and forced model \eqref{eq1}, particularly in the case of a spatiotemporally localized external forcing that we will consider herein. 

%
We will touch  upon this significant problem  by numerically simulating the initial-boundary value 
problem of Eq.~(\ref{eq1}), for vanishing boundary conditions:
$\lim_{|x|\rightarrow \infty} u(x,t)=0$, $t\ge 0$.
More precisely, we consider two types of vanishing initial data: The first are decaying algebraically (at a quadratic decay rate), and the second exponentially (possessing  a $\mathrm{sech}$-profile). The aim 
is to examine the effect of the rate of decay of the initial condition in the dynamics. 
Regarding the form of the driving, we will consider the case of a 
Gaussian profile. This is a second, physically relevant example of a driver we consider in our studies on Eq. \eqref{eq1}, since, in our recent work \cite{ZNA2019}, we examined
the case of time-periodic forcing. Concerning the magnitudes of the damping and forcing, 
we will use $\gamma\sim O(10^{-2})$, and forcing amplitude $\Gamma\sim O(1)$; 
such a choice may describe a simplified effect of wave amplification against small fluid viscosity \cite{Kharif1,Kharif2}. 

Our main findings can be summarized as follows. First, we find that PRW-solitonic structures, with profiles remarkably close to the analytical PRW of the integrable limit, emerge on top of a spatially extended 
and decaying support, formed at the initial stages of the evolution.
Second, we find that the subsequent dynamics remains proximal to that of the semi-classical limit of the NLS 
\begin{equation}
	\label{eq1sl}
	\mathrm{i}\epsilon{{u}_{t}}+\frac{\epsilon^2}{2} {{u}_{xx}}+|u|^2u =0,
\end{equation}
for $\epsilon\rightarrow 0$ \cite{BM1,BM2}, when supplemented with 
vanishing boundary conditions.
Spatiotemporal regions of distinct asymptotic regimes, separated by curves in the $x-t$-plane (known as nonlinear caustics), are identified; these curves bound the region where the space-time oscillations occur.  Remarkably, this semi-classical type dynamics is found to be persistent.
This is so for certain spatiotemporal scales where the spatial width of the driver dominates over its temporal width -- even far from the integrable limit.
This phenomenology differs nontrivially from the one observed in \cite{ZNA2019} in the presence of a time-periodic forcing. Therein, PRW-type waveforms emerge on the top of a finite background as explained in terms of the modulation instability of continuous wave (cw) solutions of the model. Nevertheless, in both cases of driving, it is justified  that the emergence of PRWs, excited by decaying initial conditions \cite{BM1,BM2,Rev1_b,BS}, is a universal
outcome, being robust in the dynamics of the 1D-damped and forced NLS. Third, we examine the dependencies of various characteristics of the emerging PRW-type extreme events, on the type of decay of the initial condition, its width, as well as, the damping strength and forcing amplitude. One of the most interesting findings concerns the effect of the phase of the driver on the formation of the emergent spatiotemporal patterns. 
Fourth, as the temporal localization width of the forcing is increasing, the transient dynamics prior to decay, is manifested by the 
formation of almost time-periodic breathing modes of (extremely, in some cases) 
large amplitude. We also establish a strong connection between the spatial
width of the driving and form/width of the emerging spatiotemporal patterns.
\begin{figure}[tbp!]
	\begin{center}
		\includegraphics[scale=0.11]{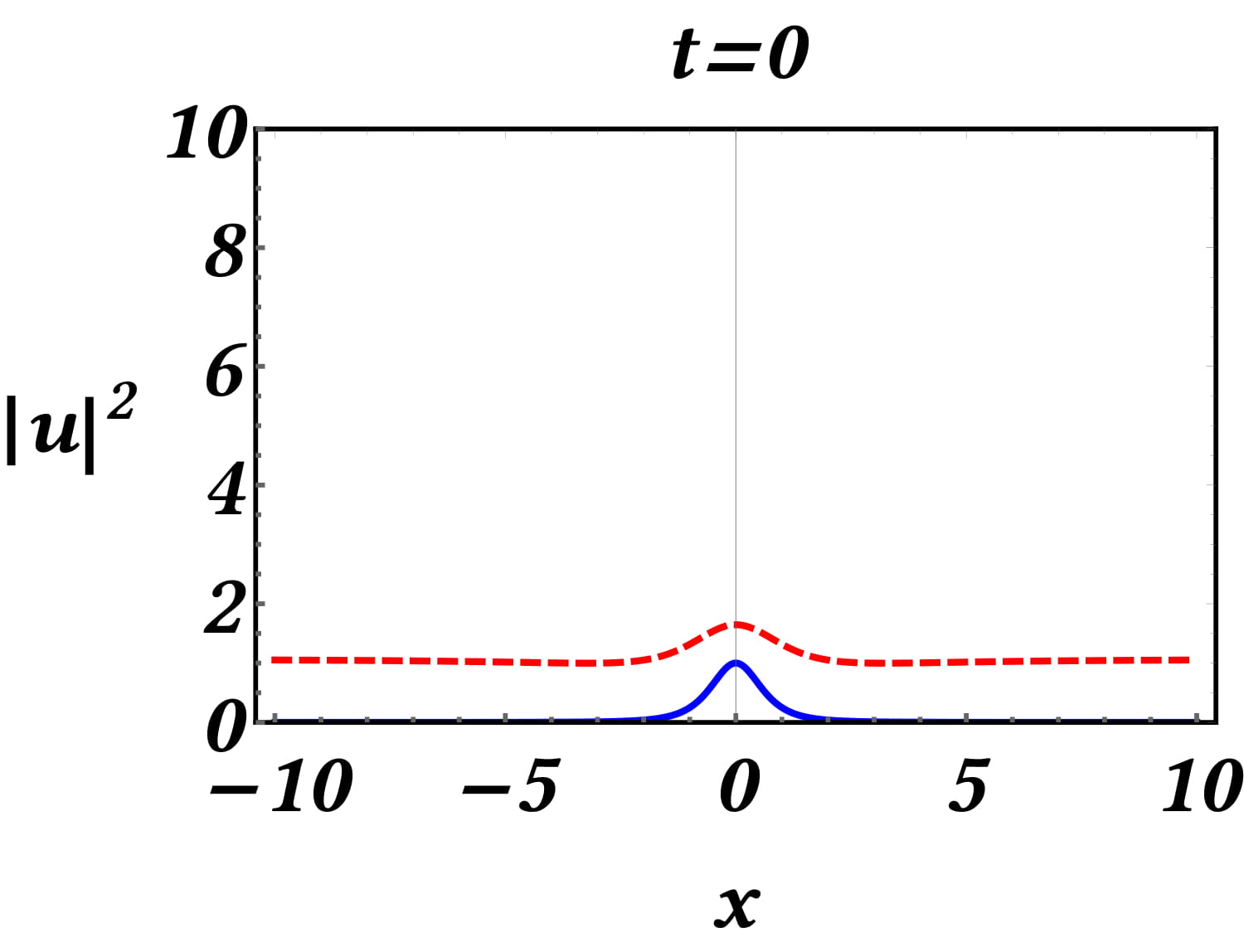}
		\includegraphics[scale=0.11]{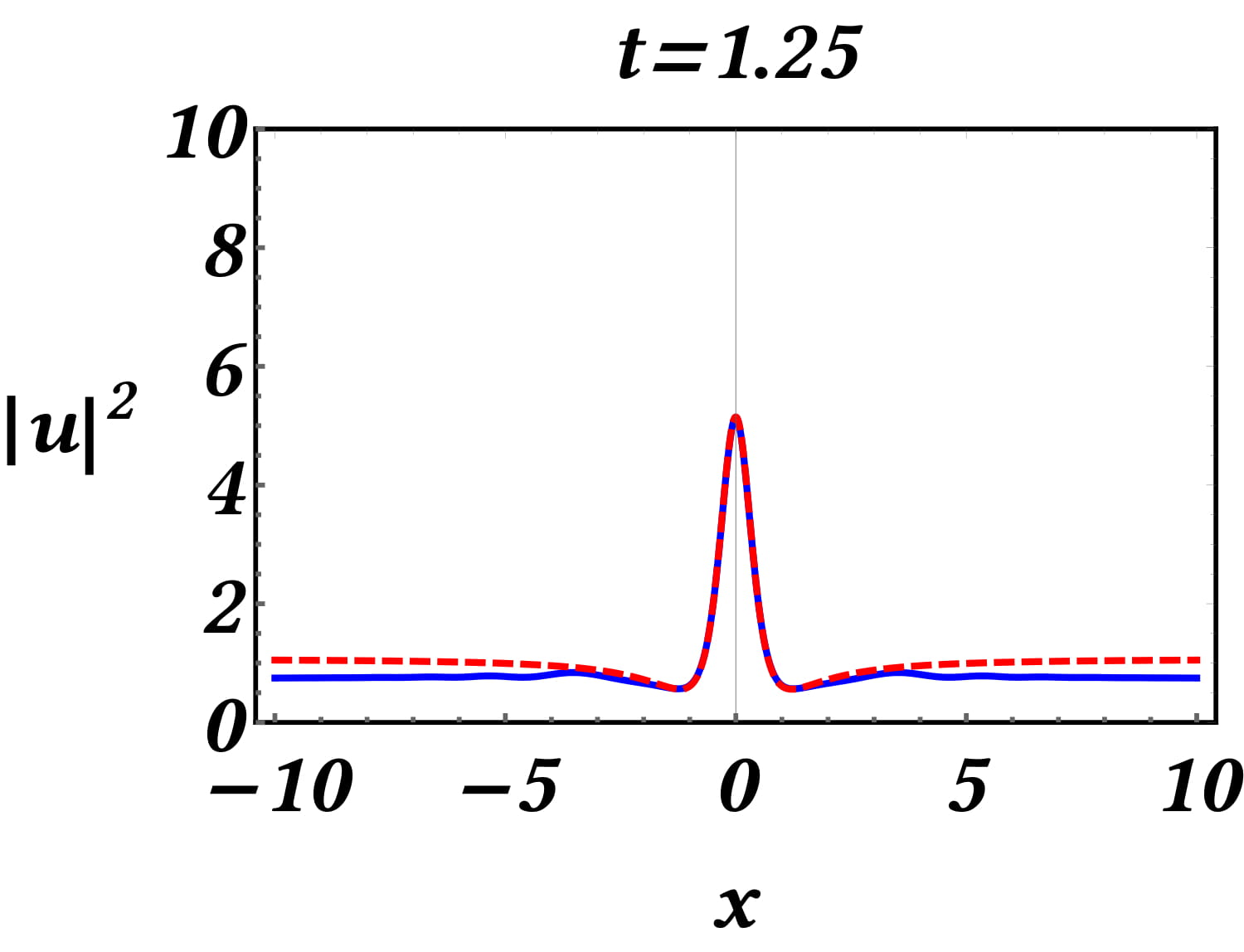}
		\includegraphics[scale=0.11]{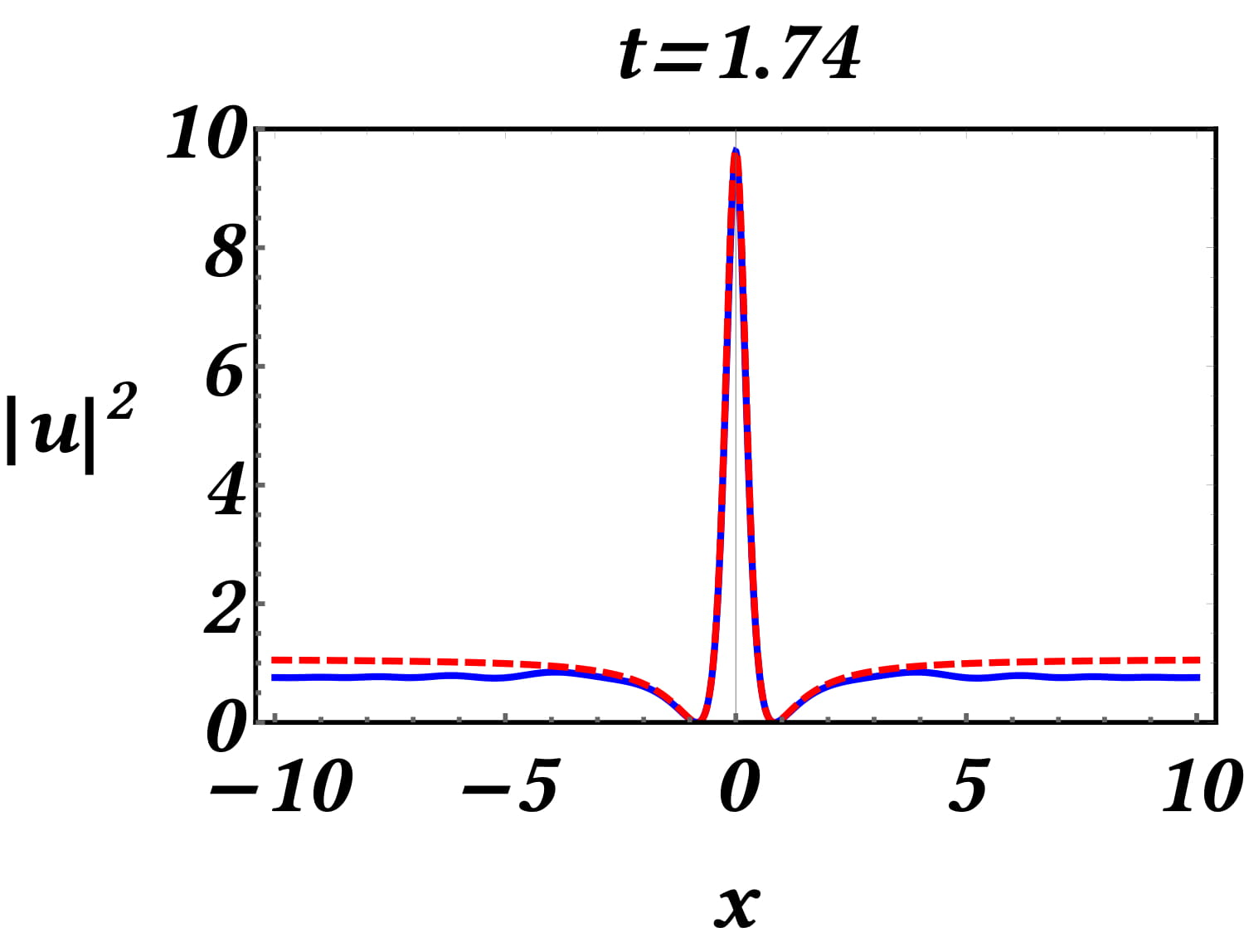}\\
		\includegraphics[scale=0.11]{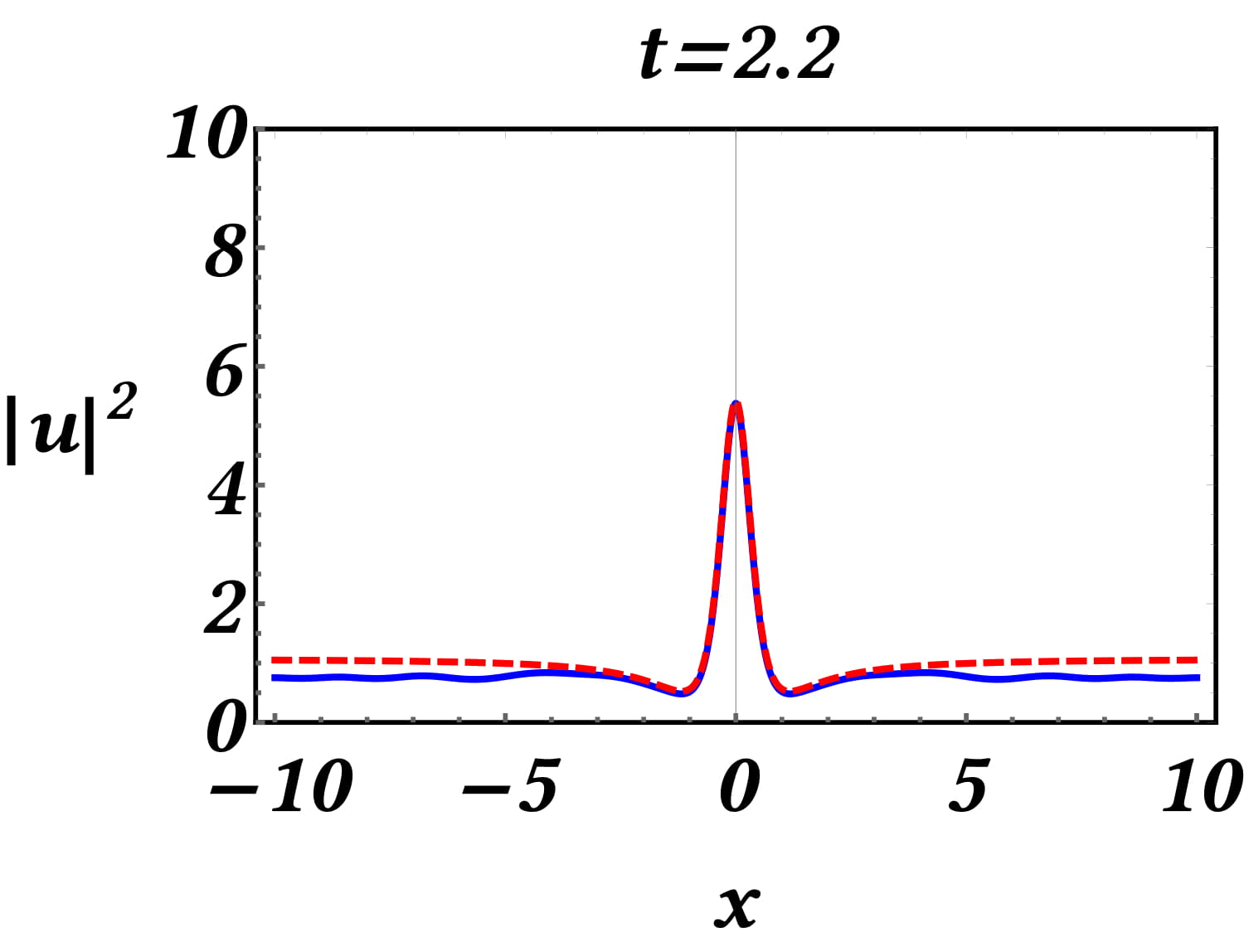}
		\includegraphics[scale=0.11]{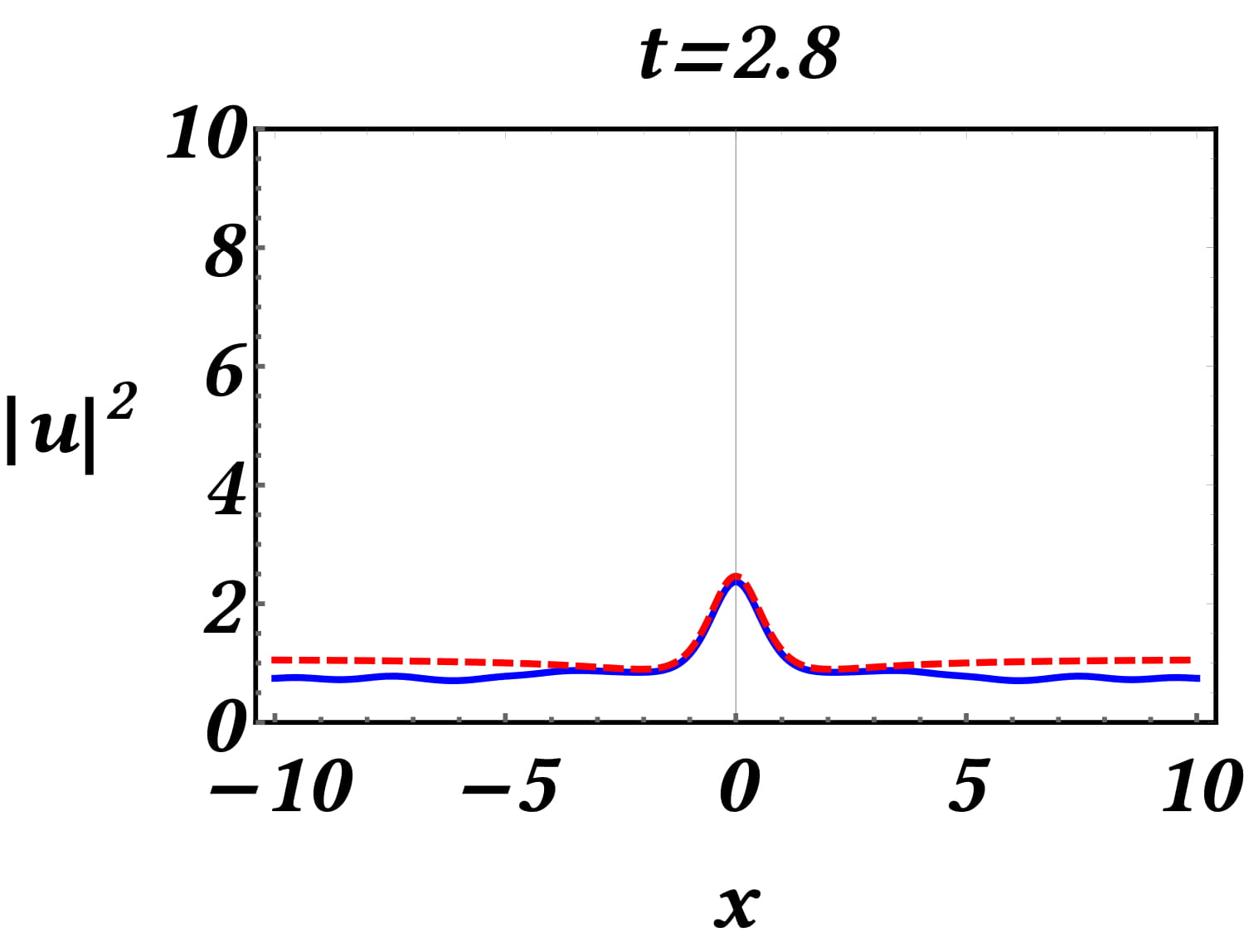}
		\includegraphics[scale=0.11]{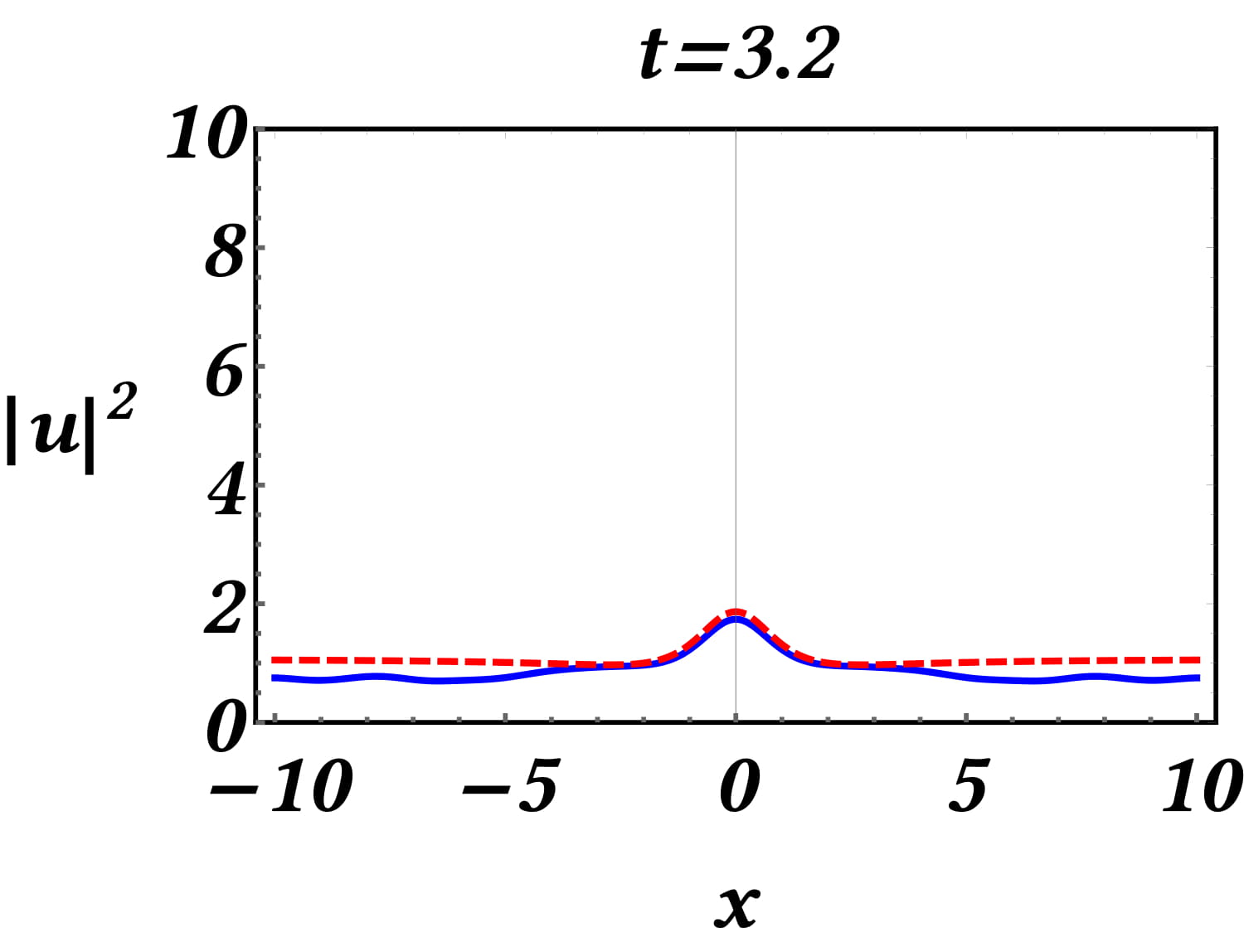}	\vspace{-0.5cm}
	\end{center}
	\caption{(Color Online) Snapshots of the evolution of the density $|u(x,t)|^2$ [solid (blue) curves], for the initial condition (\ref{eq4}). Parameters: $\gamma=0.01$, $L=250$,  Gaussian driving (\ref{wlf}), with  $\Gamma=1$, $\sigma_x=100$ and $\sigma_t=0.5$. The density of the numerical solution is compared against the density of 
		the PRW (\ref{PRWP}), $u_{\mbox{\tiny PS}}(x,t;1.74;1.07)$ [dashed (red) curves].}
	\label{figure1}
\end{figure}
The paper is structured as follows. 
In Section~\ref{numerical} we report the results of our numerical simulations. First we provide some information about the initial conditions and the type of forcing we are going to consider in this investigation. After that, in Section~\ref{sec:2a} we present our results on the emergence of extreme events, while in Section~\ref{sec:2b} we examine the spatiotemporal patterns that appear in the dynamics of the system. Finally, in Section~\ref{conclusions}, we summarize and discuss the implications of our results with an eye
towards future work currently in progress.

\begin{figure}[tbp!]
	\centering
	\begin{center}
		\includegraphics[scale=0.12]{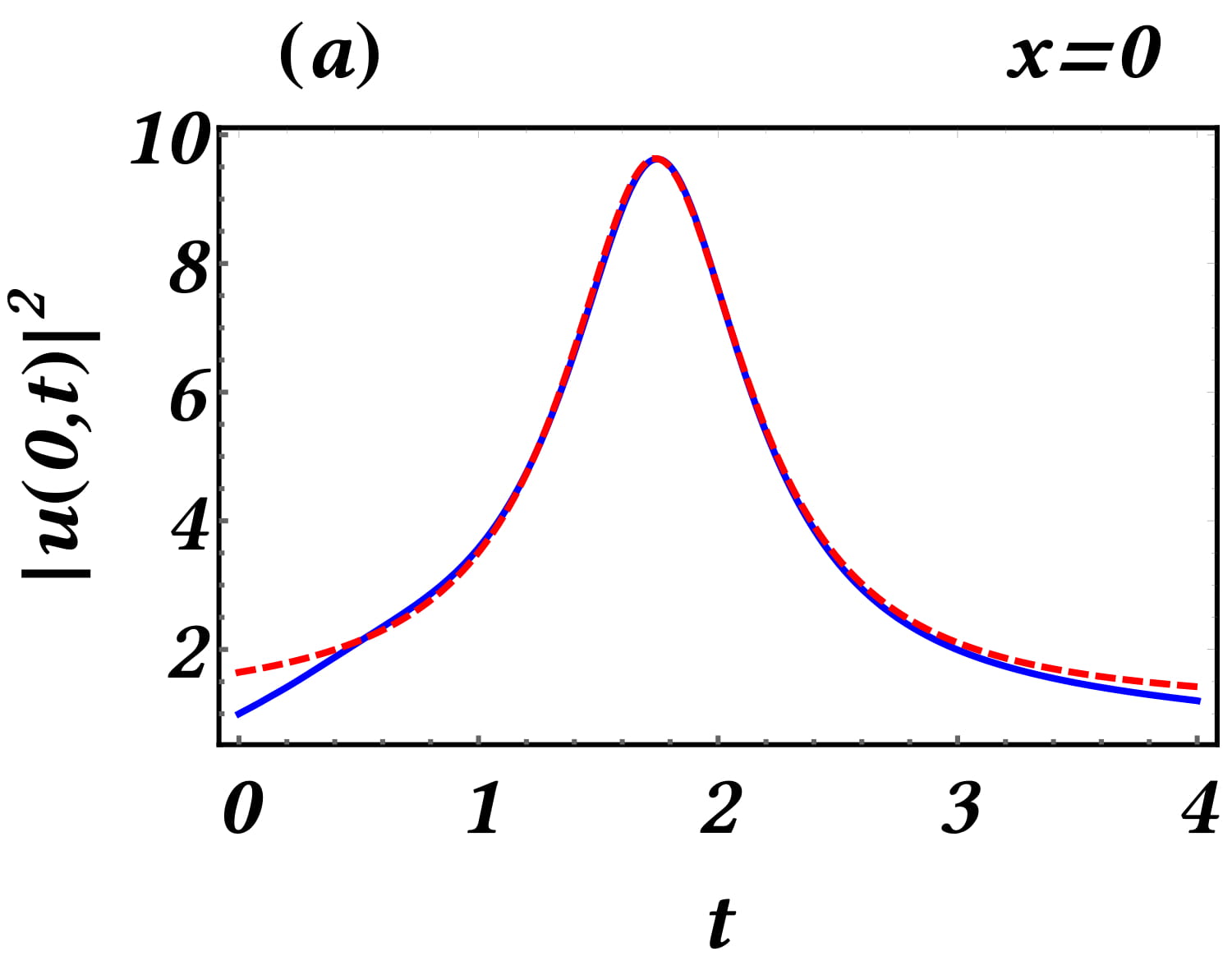}
		\includegraphics[scale=0.122]{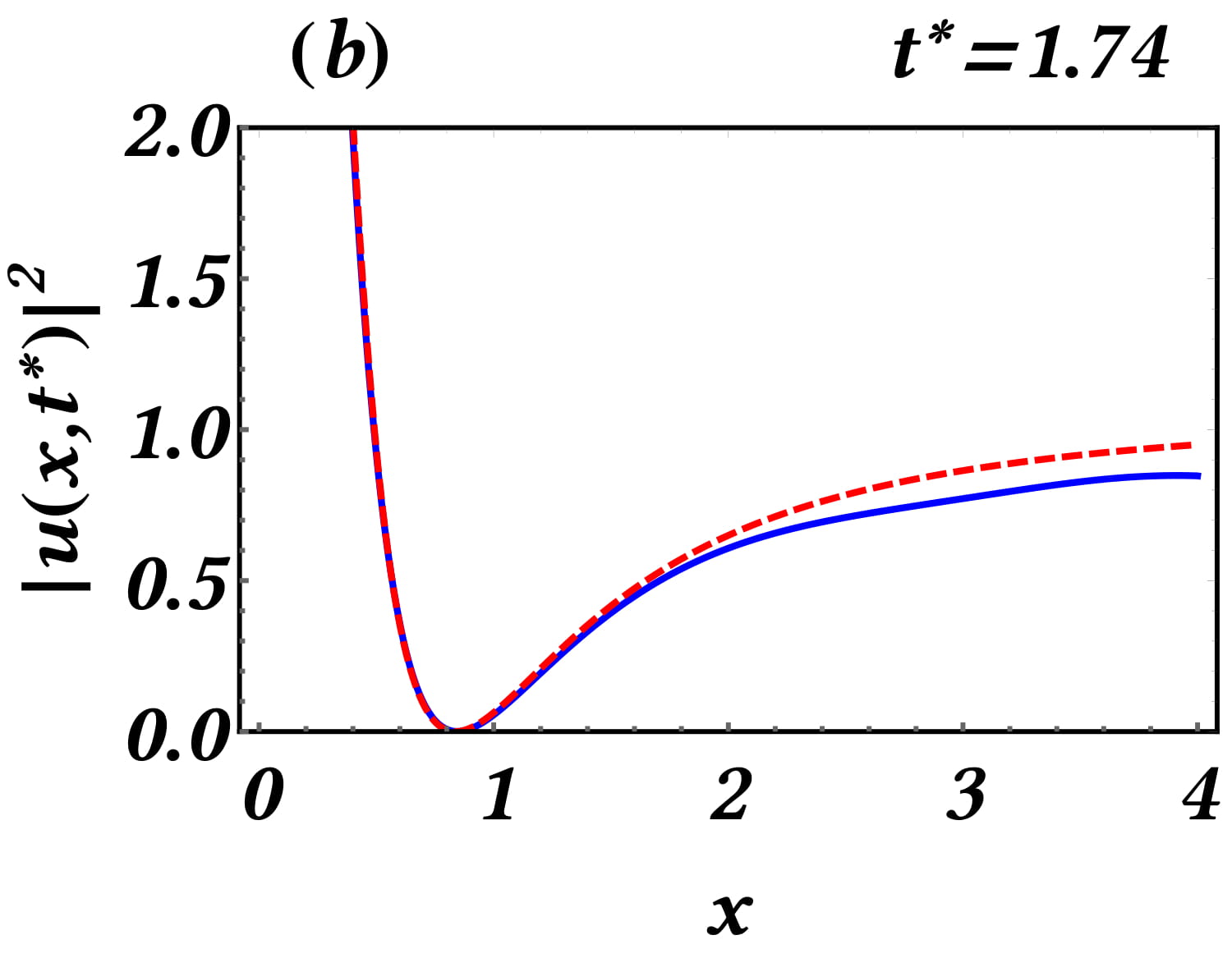}\\	
		\includegraphics[scale=0.122]{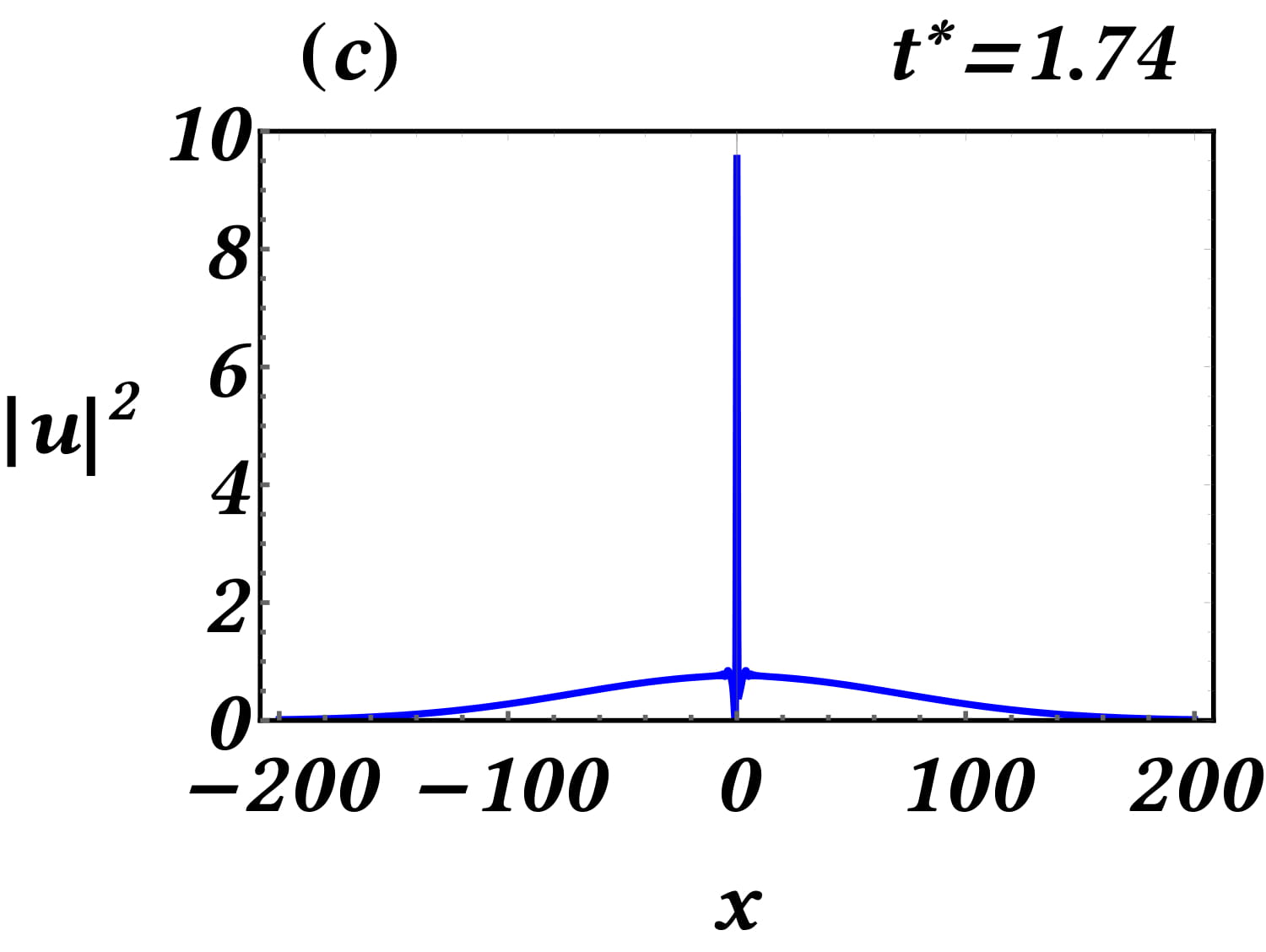}
		\includegraphics[scale=0.137]{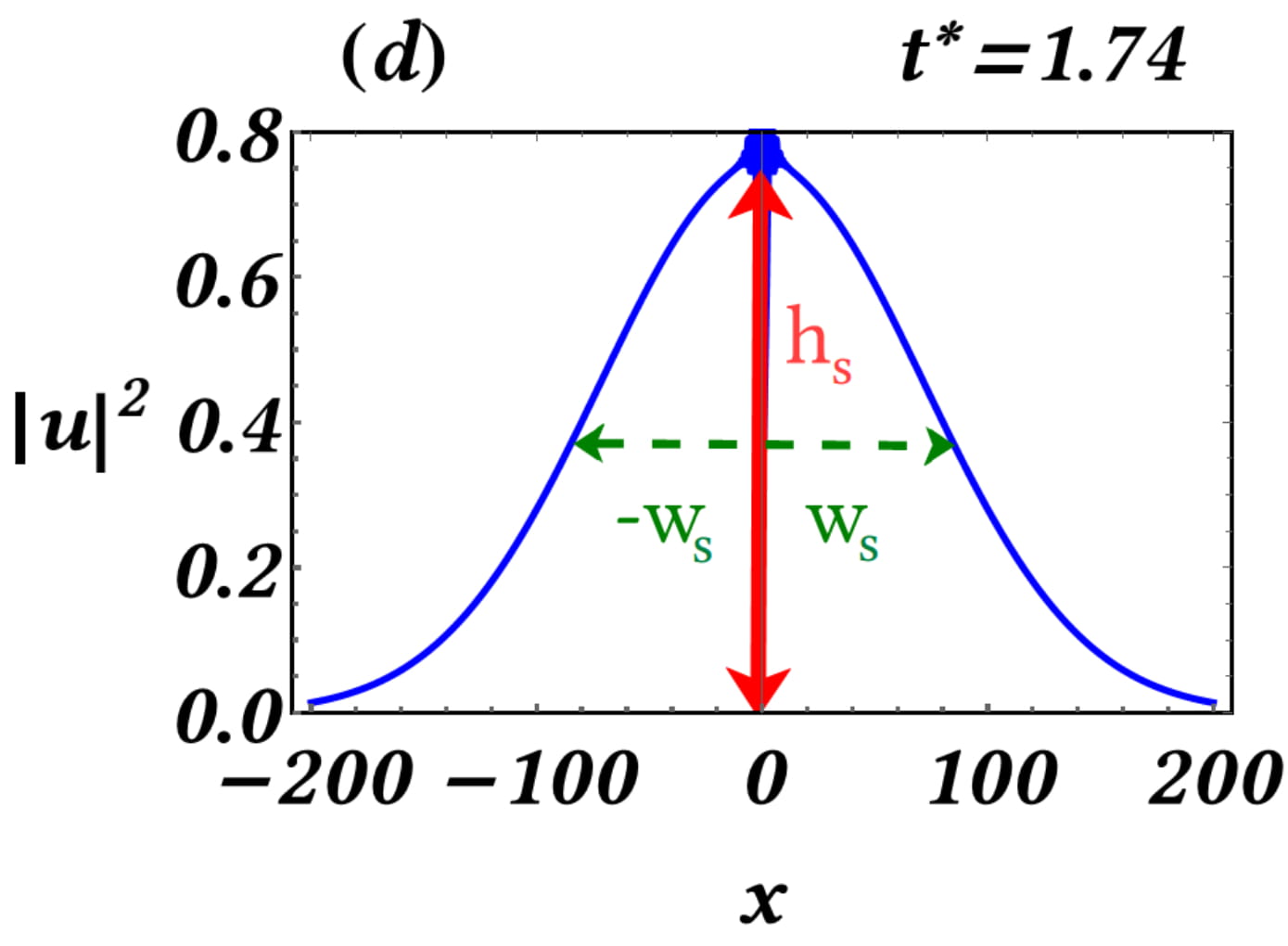}\vspace{-0.5cm}
	\end{center}
	\caption{(Color Online) Panel (a): evolution of the density of the center,
		$|u(0,t)|^2$, for the initial condition (\ref{eq4}) [solid (blue) curves], against the evolution of the density of the center of the PRW (\ref{PRWP}), $u_{\mbox{\tiny PS}}(0,t;1.74;1.07)$ [dashed (red) curves]. Panel (b): a detail of the spatial profile of the maximum event at $t^*=1.74$, close to the right of the two symmetric minima of the exact PRW $u_{\mbox{\tiny PS}}(x,t;1.74;1.07)$. Panel (c): an expanded view of the numerical density at time $t^*=1.74$, where the extreme event attains its maximum amplitude, for $x\in[-150,150]$. The PRW-like structure is formed on top of a decaying support. Panel (d): a magnification of the support of peak amplitude $h_s$ [continuous (red) vertical line] and half-width $w_s$.}
	\label{figure2}
\end{figure}
\section{Numerical investigations}\label{numerical}
In this section, we present the results of our direct numerical simulations for Eq.~(\ref{eq1}), 
with the vanishing boundary conditions discussed above, and initial data distinguished by the rate of their decay. For instance, we shall consider a quadratically decaying initial condition
\begin{equation}
u_0(x)=\frac{1}{1+x^2},
\label{eq4}
\end{equation}
and an exponentially decaying initial condition of the form
\begin{equation}
u_0(x)=\sech(\alpha x).
\label{eq4a}
\end{equation}
The parameter $ \alpha>0$, configures the width of the above ``bright-soliton" initial profile. For the driving force, we assume the form of a Gaussian function, 
centered at $(x=0,t=0)$, having spreads $\sigma_x,\;\sigma_t>0$, 
with respect to the space and time variable, respectively:
\begin{equation}
	\label{wlf}
	\begin{array}{c}
	 f(x,t)=g(x,t)\exp(\mathrm{i}\theta),\\[5pt]
\hspace{-1.5cm}\hbox{where}\qquad g(x,t)=\sqrt{2}\,\Gamma\exp\left(-\frac{x^2}{2\sigma_x^2}-\frac{t^2}{2\sigma_t^2}\right).
	 \end{array}
\end{equation}
The function $f$, serves as a simple phenomenological example, of a  spatiotemporal,  exponentially localized driver, of amplitude  $\Gamma>0$. The role of the phase parameter $\theta$ will be further elucidated below. 

The smallest value of the half-length of the spatial computational interval $[-L,L]$ 
considered in the simulations is $L=250$, so that effects that may be produced at the 
boundaries are negligible.
The system is integrated by using both Runge--Kutta and pseudo-spectral schemes (see also \cite{ZNA2019}), 
for both Dirichlet and  periodic boundary conditions on $[-L,L]$. 
Note that, there were no noticeable differences between the dynamics produced by the two numerical methods.   
%
\begin{figure}[tbp!]
	\hspace{-0.5cm}
	\begin{center}
		\includegraphics[scale=0.13]{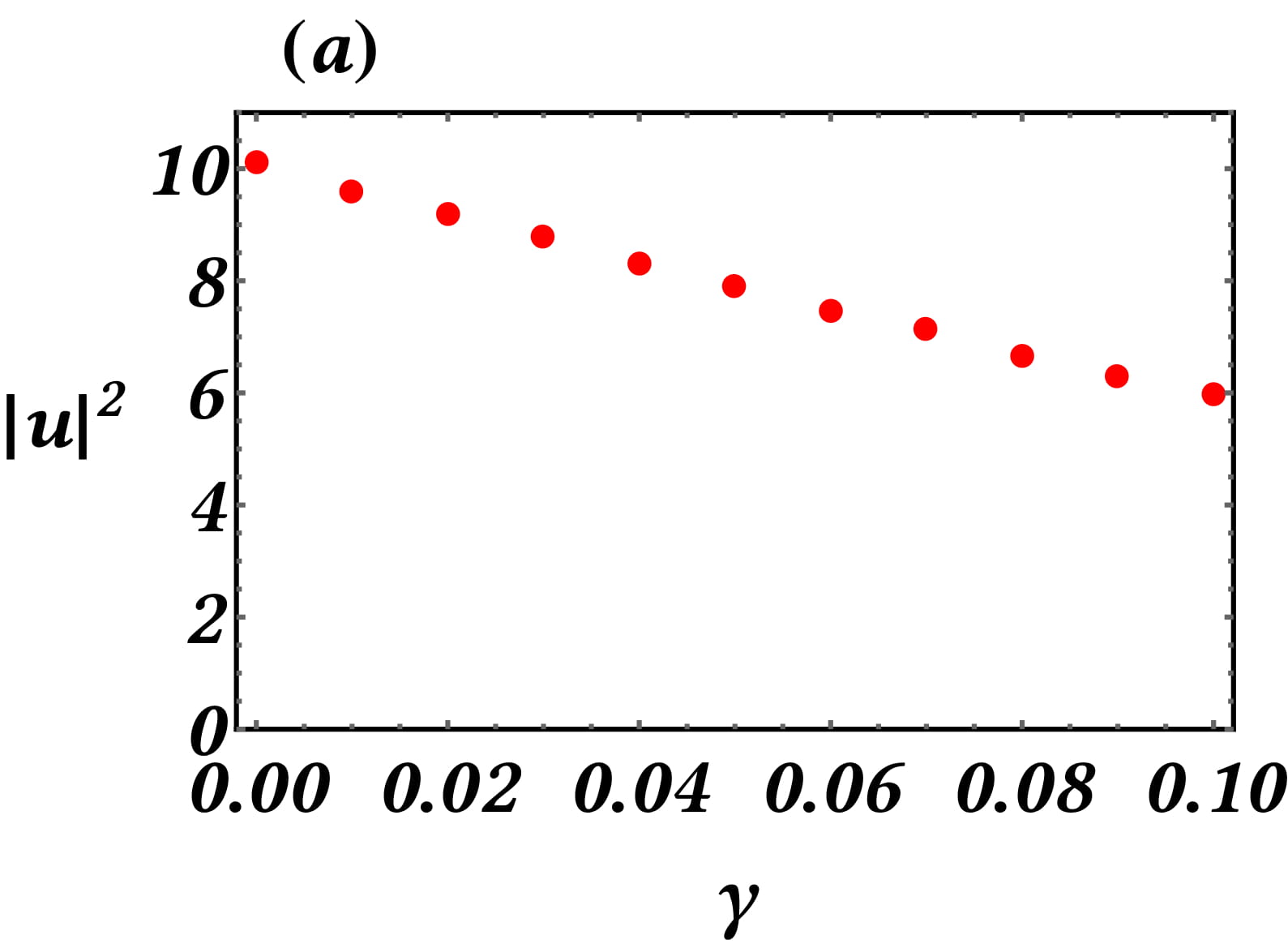}
		\includegraphics[scale=0.128]{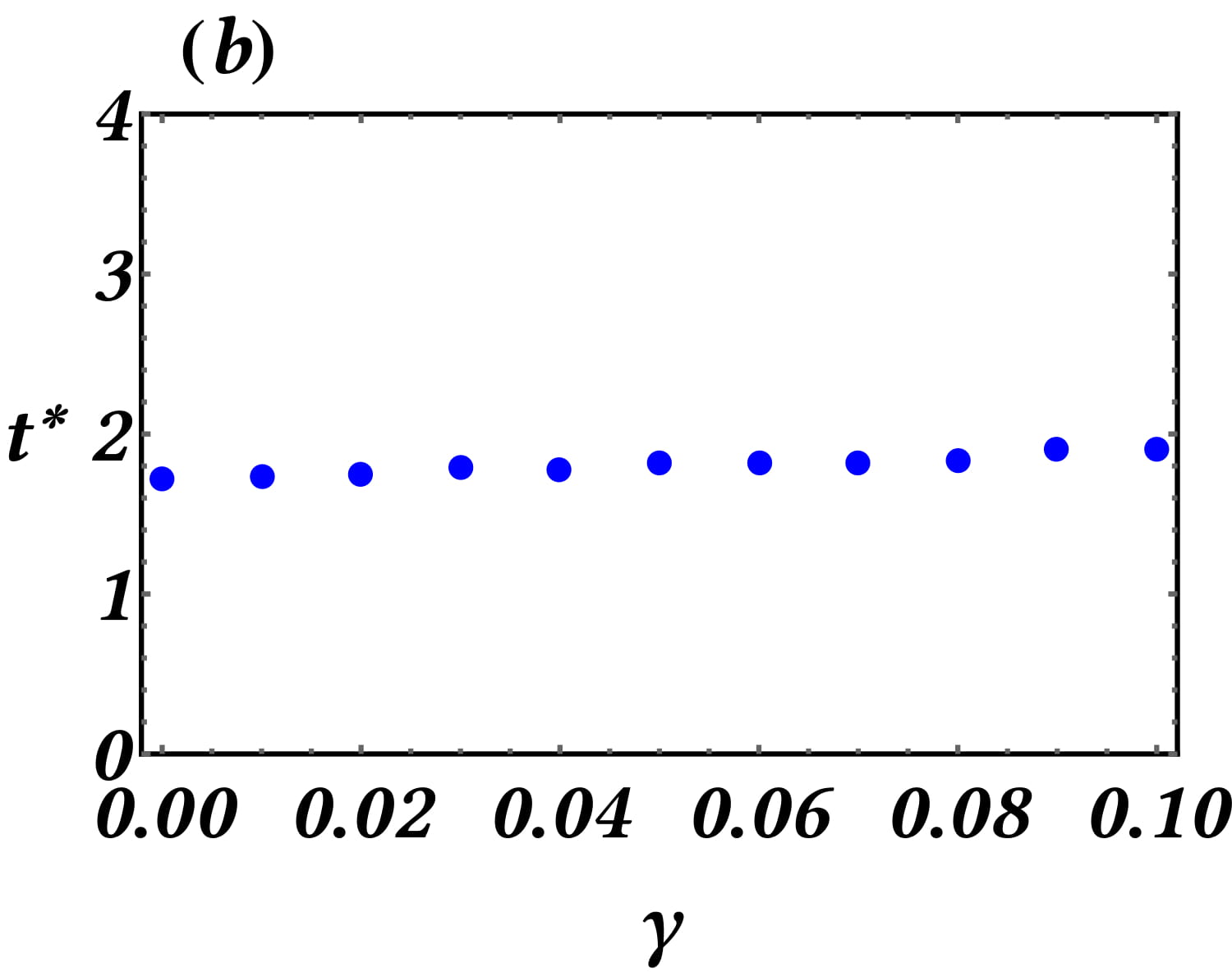}\\[5pt]
		\includegraphics[scale=0.135]{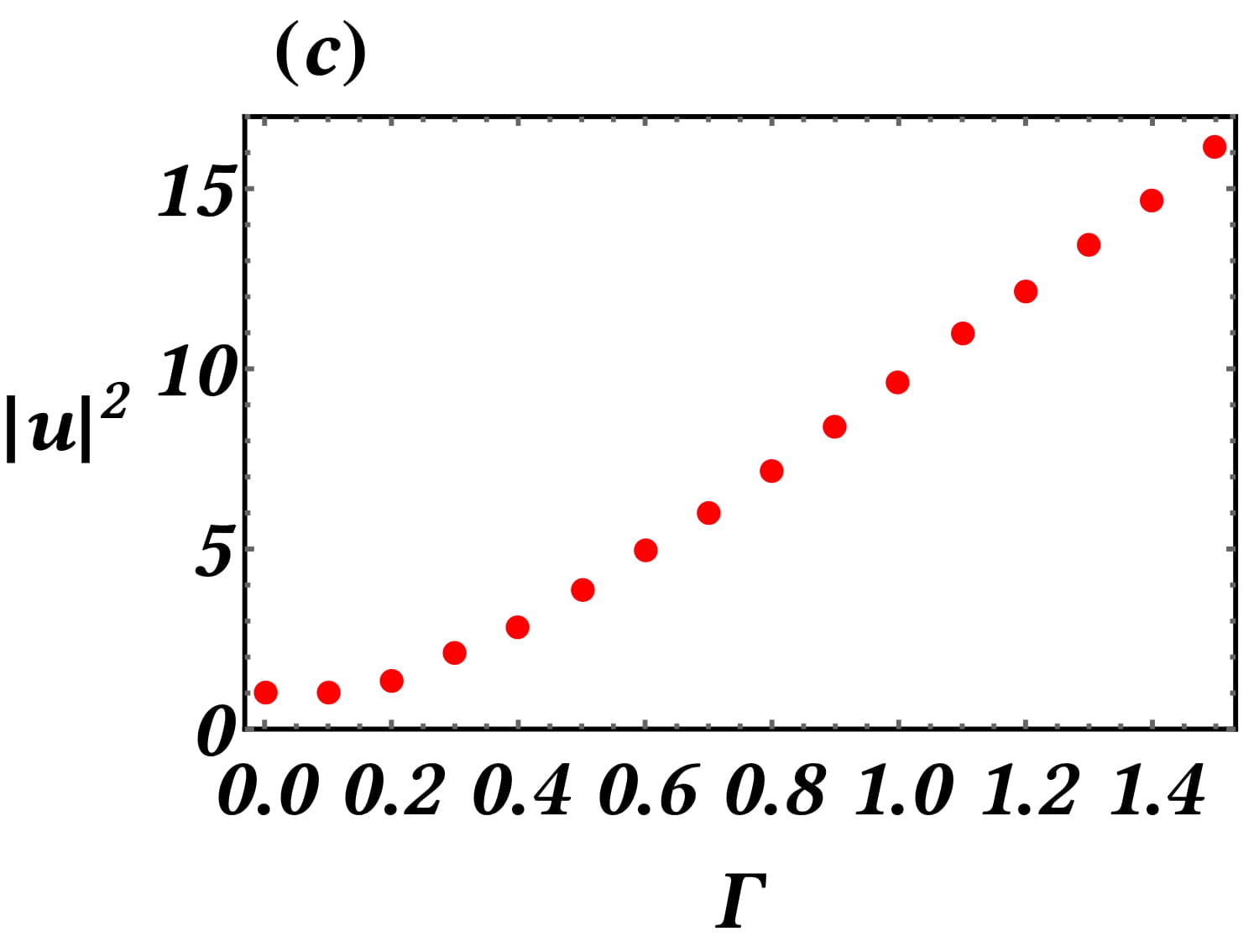}
		\includegraphics[scale=0.125]{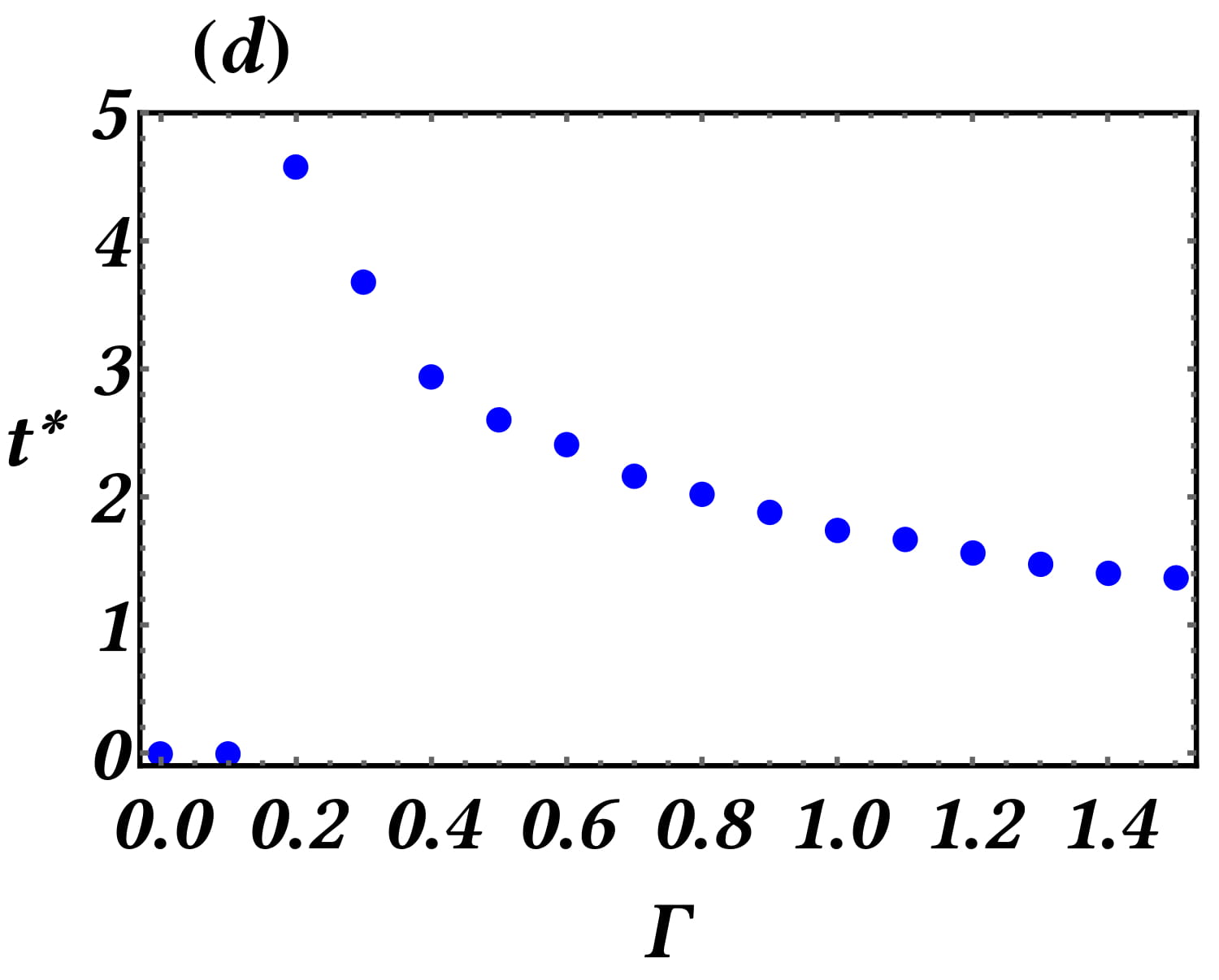}
	\end{center}
	\caption{Panels (a) and (b): Maximum amplitude of the extreme wave event and its occurrence time, as a function of $\gamma$, for fixed $\Gamma =1$. Panels (c) and (d): Maximum amplitude of the extreme event and its occurrence time, as a function of $\Gamma$, for fixed $\gamma =0.01$. Other parameters: $L=500, \sigma_t=0.5$ and $\sigma_x=100$.}
	\label{figure3}
\end{figure}
\subsection{Emergence of extreme wave events} \label{sec:2a}
\paragraph{Quadratically decaying initial condition.} We begin our numerical investigations by considering first the quadratically decaying initial condition \eqref{eq4}. In order to characterize the observed spatiotemporally localized waveforms as extreme, we compare in certain time intervals, the numerical solution against
the rescaled analytical PRW solution of the NLS (\ref{eq0}) \cite{BS}: 
\begin{eqnarray}
\label{PRWP}
u_{\mbox{\tiny	PS}}(x,t;t_0;P_0)=
 \sqrt{P_0}\left\{1-\frac{4\left[1+2\mathrm{i}P_0(t-t_0)\right]}{1+4P_0x^2+4P_0^2(t-t_0)^2}\right\}\mathrm{e}^{\mathrm{i}P_0(t-t_0)},
\end{eqnarray}
where 
$P_0$ stands for the power of the continuous wave background.

Figure~\ref{figure1}, shows snapshots of the evolution of the density $|u(x,t)|^2$
for  damping strength $\gamma=0.01$, forcing amplitude $\Gamma=1$ and spreads $\sigma_x=100$, $\sigma_t=0.5$.  We observe the emergence of a PRW solitonic waveform. The numerical solution is plotted by the continuous (blue) curve, against the dashed (red) curve depicting the evolution of the PRW-profile (\ref{PRWP}), $u_{\mbox{\tiny PS}}(x,t;1.74;1.07)$.  The maximum amplitude of the event is attained at $t=t^*=1.74$.
The snapshots, illustrate the time-growth and time decay of the solution and justify that its evolving profile is remarkably close to the analytical PRW of the integrable NLS (except for the region far from the core).

In particular, as shown in panel (a) of Fig.~\ref{figure2}, for $t\in[1,3]$,
the peak amplitude evolution
is nearly indistinguishable from that of the corresponding member
of the PRW family.
Panel (b) of Fig.~\ref{figure2}, shows a detail of the spatial profile of the maximum event at $t=1.74$, close to the right of the two symmetric minima of the exact PRW $u_{\mbox{\tiny PS}}(x,t;1.74;1.07)$. This detail illustrates that the emerged extreme event possesses an algebraic spatial decay rate close to that of the analytical PRW, at least near the minima.

Panel (c) of Fig.~\ref{figure2} offers another view of the extreme event 
occurring at $t=1.74$, 
highlighting a novel feature which totally differs from what  was observed before: the extreme event occurs on the top of an  {\em emergent decaying support} of the solution, as opposed to the
uniform background of the exact PRW waveforms, or even of the PRW-waveforms observed in the presence of gain/loss, time-periodic driver, or higher-order effects \cite{All2,ZNA2019,Yang1,Yang2}.  This support appears due to the existence of the specific type of driving -- see discussion below. Panel (d) of Fig.~\ref{figure2} depicts a magnification of this support.  For demonstration purposes of the results that follow, we denote by $h_s$ the maximum
(squared amplitude) of the support, marked by the vertical (red) line. By $w_s$, we denote the half-width of the support measured at $h_s/2$. The localization interval of the support is marked in the Figure by the horizontal (green, dashed-dotted) line. 
Our next investigations concern the dependence of various characteristics of the emergent extreme event on the system's parameters. Regarding the variation of the loss strength $\gamma$ (for fixed driving amplitude 
$\Gamma$) or $\Gamma$ (for fixed $\gamma$), we find the following. 
First, the increase of $\gamma$ results -- as may be expected -- in the decrease 
of the maximum amplitude of the first event $|u(0,t^*)|^2$, attained at $t=t^*$, 
and vice versa, see panel (a) of Fig. \ref{figure3}. On the other hand, in panel (b) of Fig. \ref{figure3}, we notice that  the increase of $\Gamma$ (for fixed $\gamma$) 
results in the increase of $|u(0,t^*)|^2$, and vice versa.
For the occurrence time of the extreme events, we see in panel (c) of
Fig. \ref{figure3}, that
although the increase of $\gamma$ affects significantly the maximum of the events, it has a small influence on the time of their occurrence. On the other hand, the increase of $\Gamma$ leads to a significant decrease in the occurrence time, as we can see in panel (d) of Fig. \ref{figure3}. 
\begin{figure}[tbp!]
	\centering
	\begin{center}
	\includegraphics[scale=0.13]{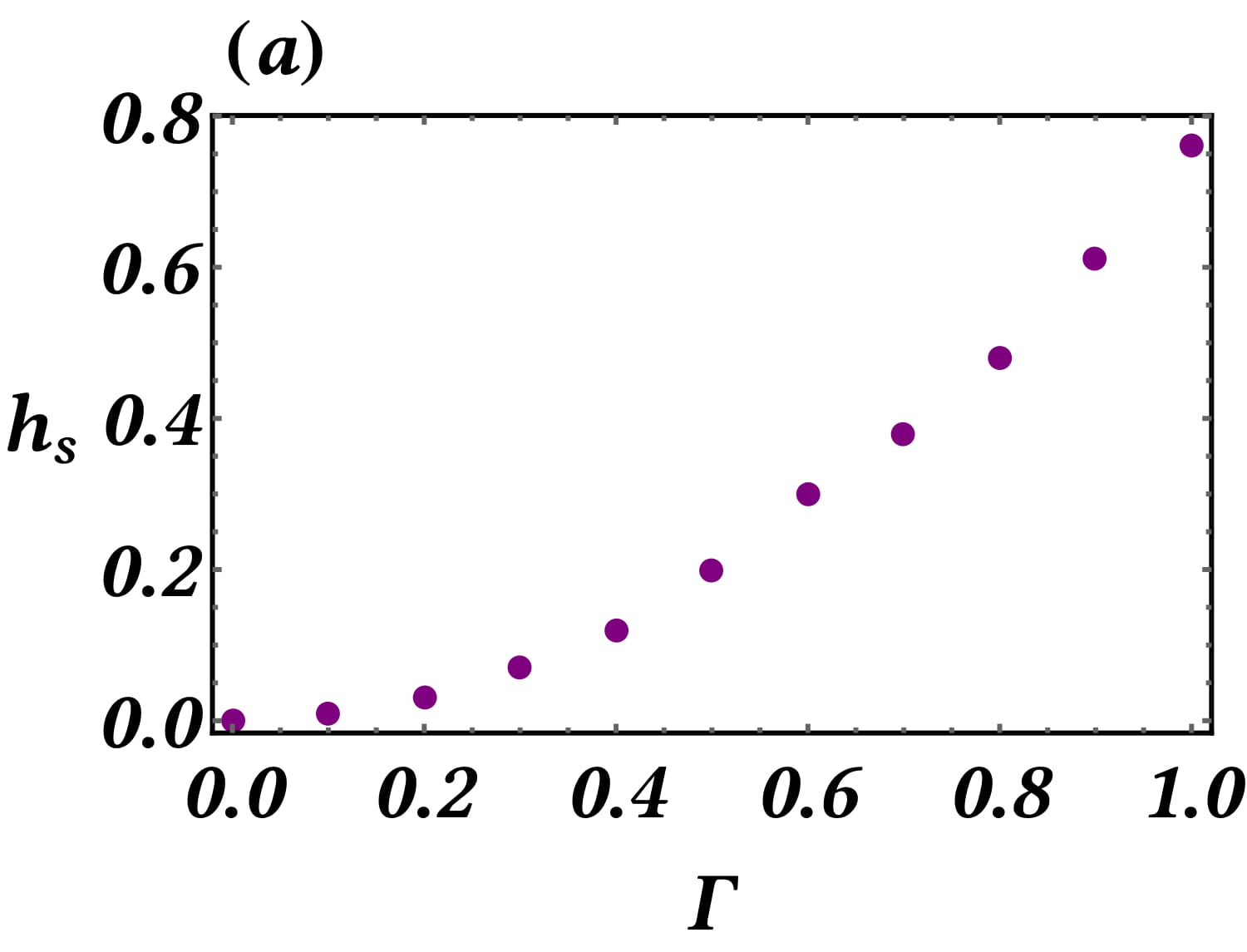}
	\hspace{0.1cm}\includegraphics[scale=0.13]{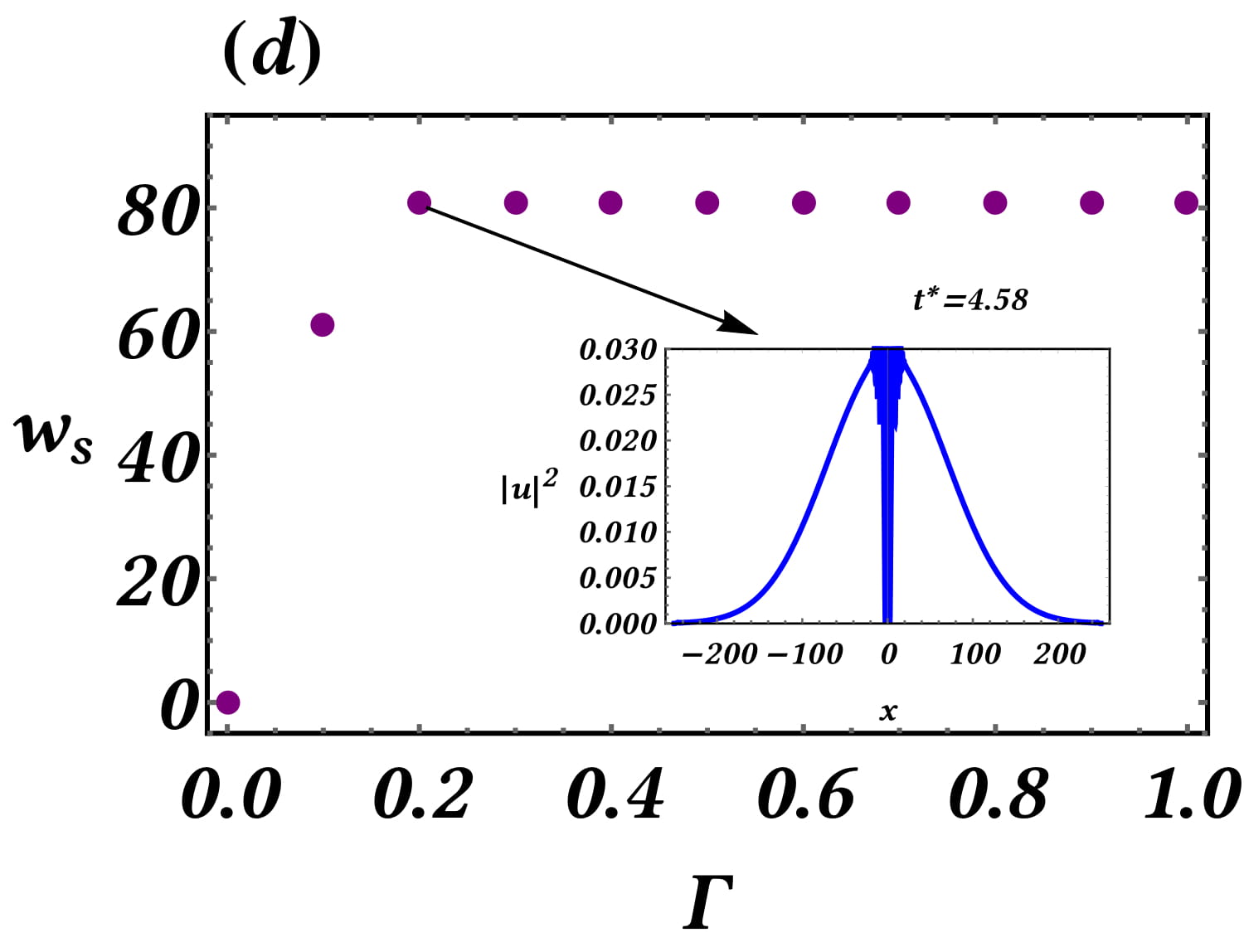}
	\end{center}
	\caption{(Color Online) 
Panel (a): The height of the support $h_s$ (see also  panels (c) and (d)  of Fig. \ref{figure2}), as a function of $\Gamma$, for fixed $\gamma=0.01$. Panel (b): The width of the support $w_s$ as a function of $\Gamma$, for fixed $\gamma=0.01$. The inset depicts the support of the solution at time $t^*=4.58$ which corresponds to the value $\Gamma=0.2$ of the forcing amplitude. }
	\label{figure4}
\end{figure}
\begin{figure}[tbh]
	\hspace{-0.5cm}
	\begin{center}
		\includegraphics[scale=0.13]{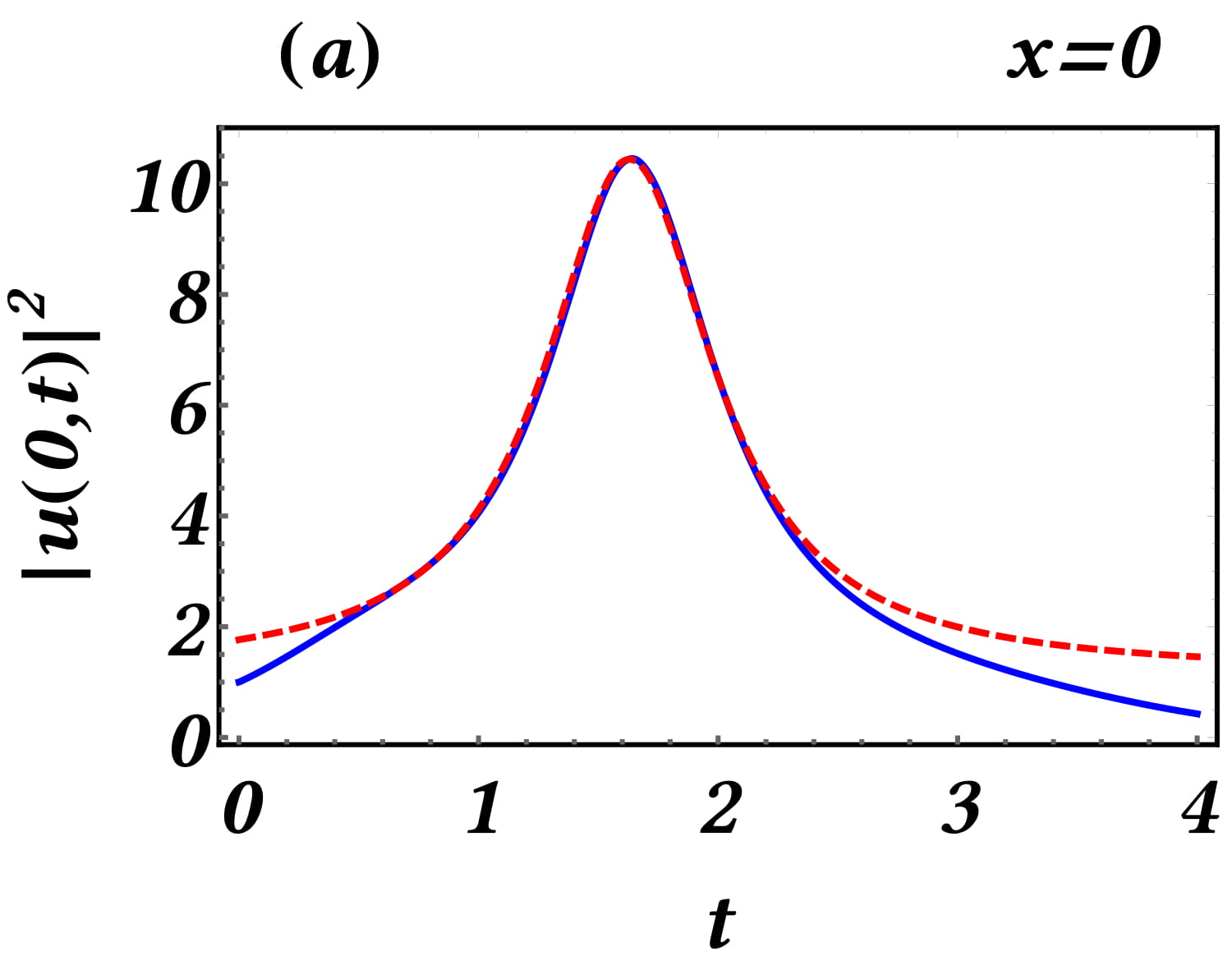}
		\includegraphics[scale=0.13]{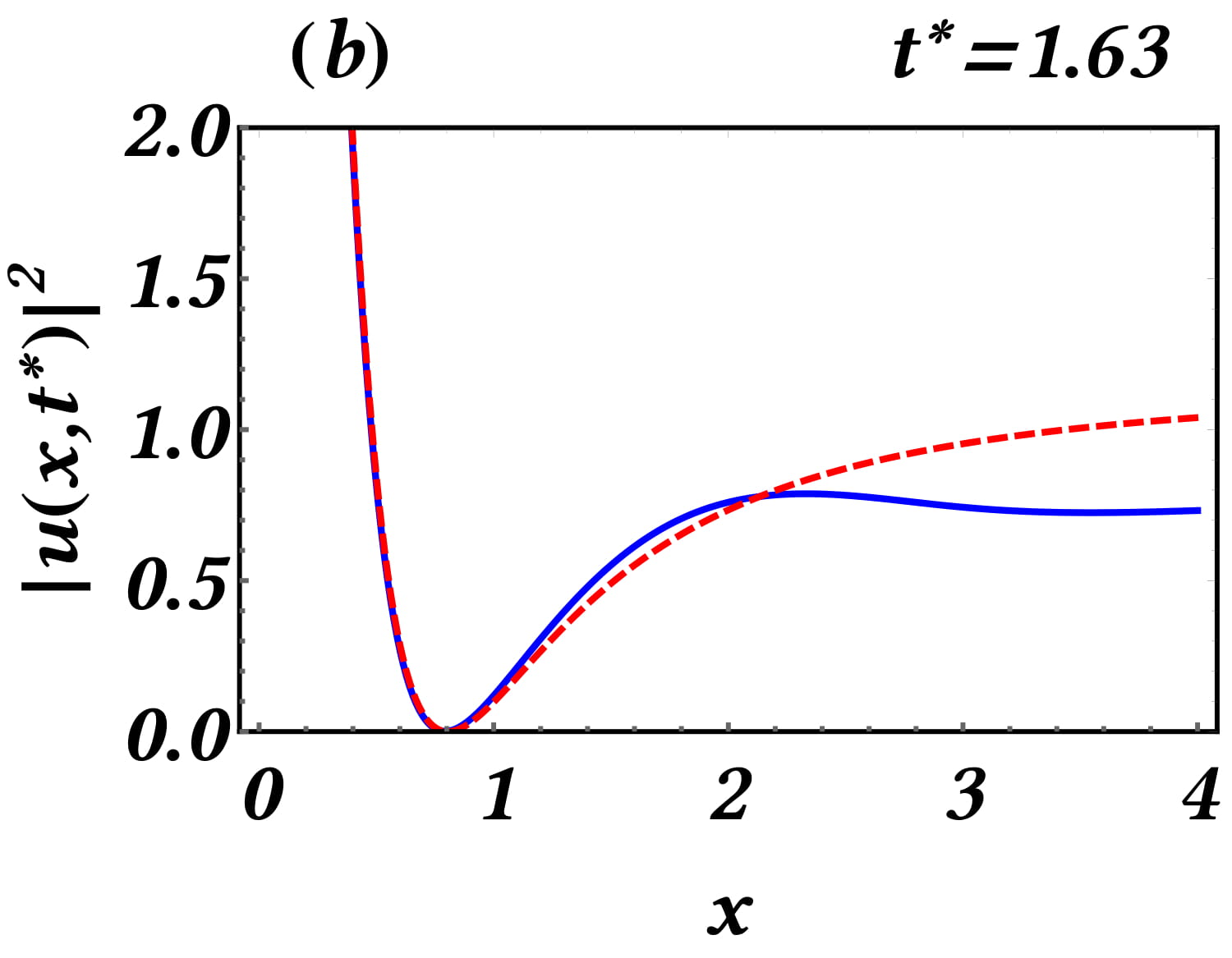}
	\end{center}
	\caption{(Color online). Extreme event emerging from the
          exponentially decaying initial condition \eqref{eq4a},
          against the PRW $u_{\mbox{\tiny PS}}(x,t;1.63;1.16)$ of the
          integrable limit. Panel (a): Evolution of the  density of
          the centers $|u(x,0)|^2$ and $u_{\mbox{\tiny
              PS}}(x,0;1.63;1.16)$. Panel (b): Comparison of the
          profiles close to the right one among  the symmetric minima of the extreme event at $t^*=1.63$. Parameters: $\gamma = 0.01, L=500$, $\Gamma=1$, $\sigma_x=100$ and $\sigma_t=0.5$.}
	\label{figure5}
\end{figure}	
We continue our study in figure \ref{figure4}, which illustrates the dependence of the height and of the width of the emerging support (shown in panels (c) and (d) of Fig. \ref{figure2}), on the amplitude of the driving, for fixed damping strength $\gamma=0.01$.   Panel (a) of figure~\ref{figure4} shows the  dependence of $h_s$ with respect to $\Gamma$.  The computations reveal the existence of a threshold value $\Gamma^*=0.2$, beyond which, the emergence of the decaying support becomes noticeable (i.e., its amplitude is $\gtrsim 10^{-3}$). Secondly, $h_s$ is an increasing function of $\Gamma$. 
Panel (b) of the same Figure shows the dependence of the  half-width  $w_s$ on $\Gamma$, which also becomes comparable to the half-width of the initial condition, when $\Gamma\geq \Gamma^*$. 
Another interesting observation is that $w_s$ 
approaches a constant value $w_s^*\sim 86$ 
for $\Gamma\geq 0.3$. The inset portrays a magnification of the support at $t=t^*=1.74$, for $\Gamma=1$. We observe that although we have considered a much larger spatial interval $[-L,L$] (as we have increased its half length to $L=500$),  the width of the support doesn't change significantly compared to the one of Fig.~\ref{figure2}(d).
%
%
\begin{figure}[tbp!]
	\hspace{-0.5cm}
	\begin{tabular}{cc}
		\includegraphics[scale=0.125]{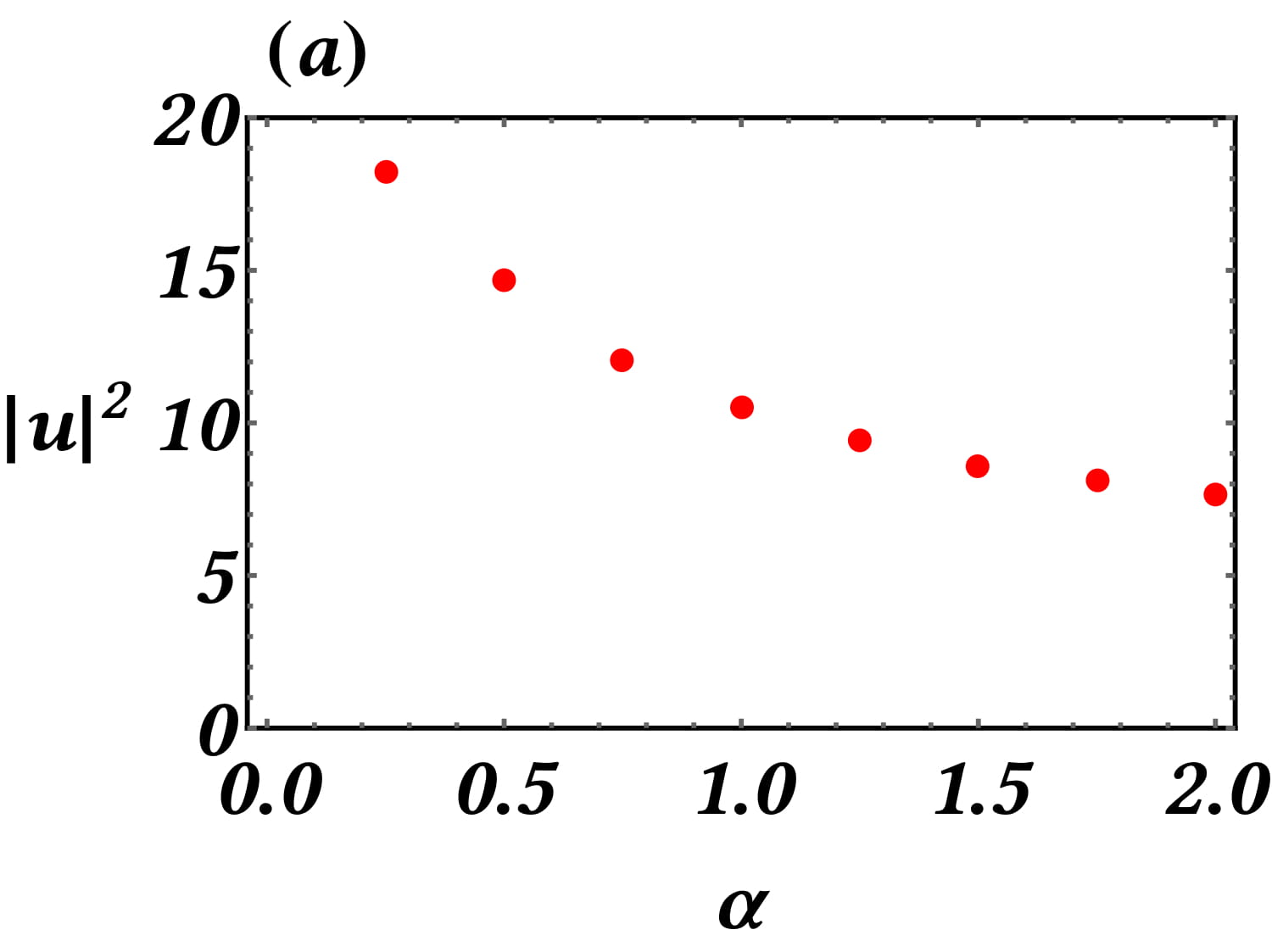}&
		\includegraphics[scale=0.12]{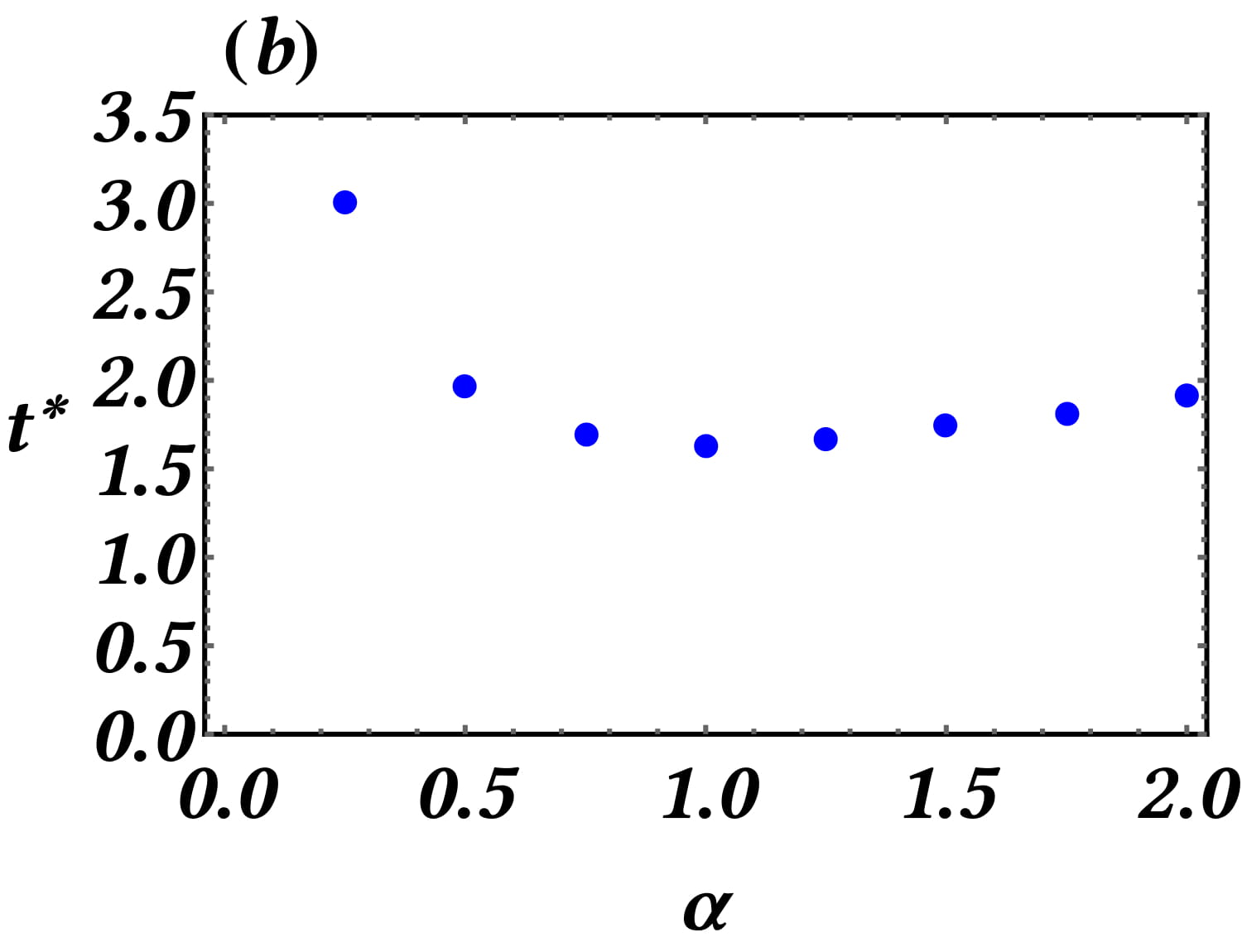}
	\end{tabular}
	\caption{Panel (a): The maximum  density of the extreme event (at the moment of its occurrence) as a function of the width $\alpha$ of the initial condition \eqref{eq4a}. Panel (b): The time of occurrence of the maximum density of the extreme event as a function of $\alpha$. Parameters: $\gamma = 0.01, L=500$, $\Gamma=1$, $\sigma_x=100$ and $\sigma_t=0.5$. }
	\label{figure6NN}
\end{figure}	
\begin{figure}[tbh!]
	\begin{center}
		\includegraphics[scale=0.12]{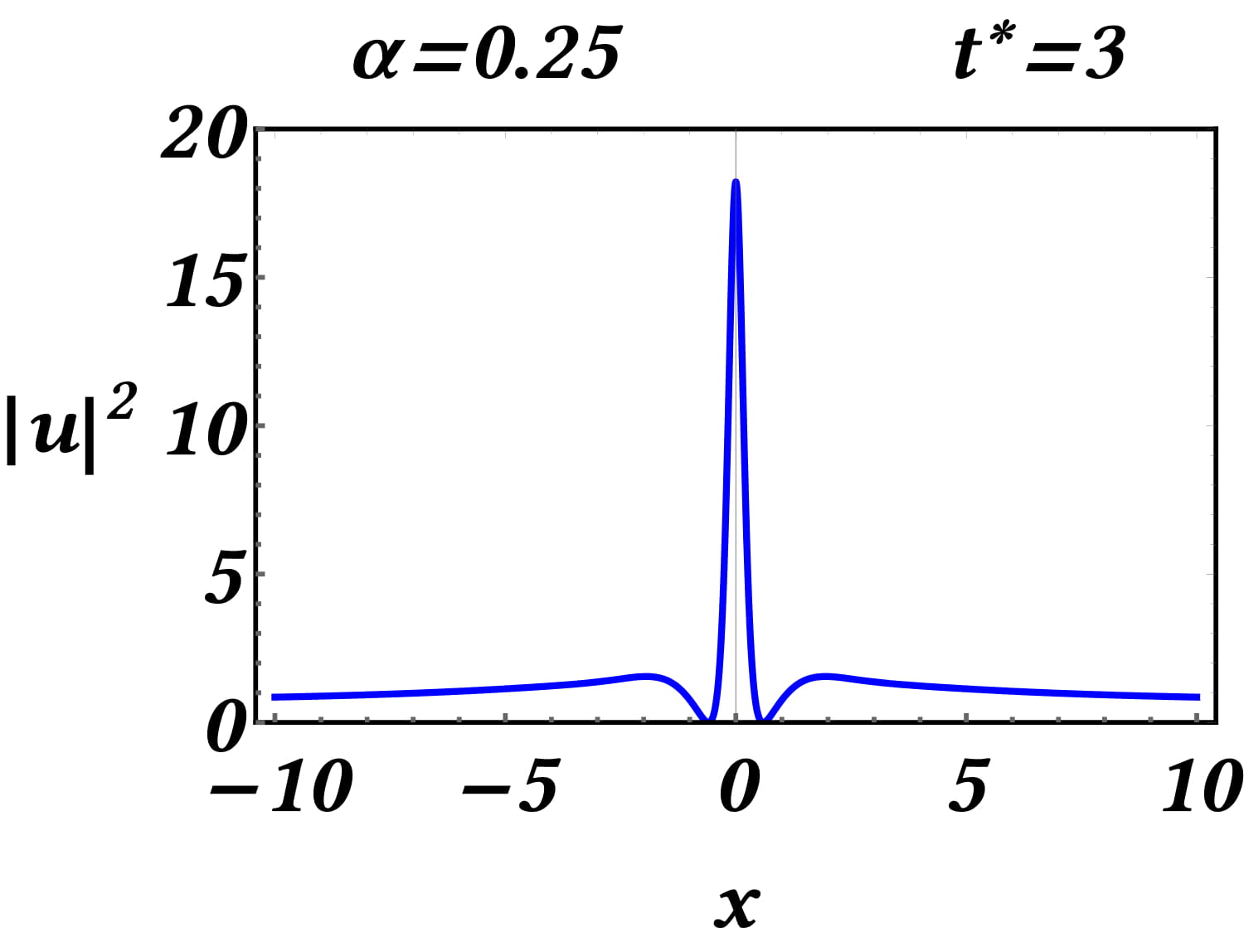}
		\includegraphics[scale=0.12]{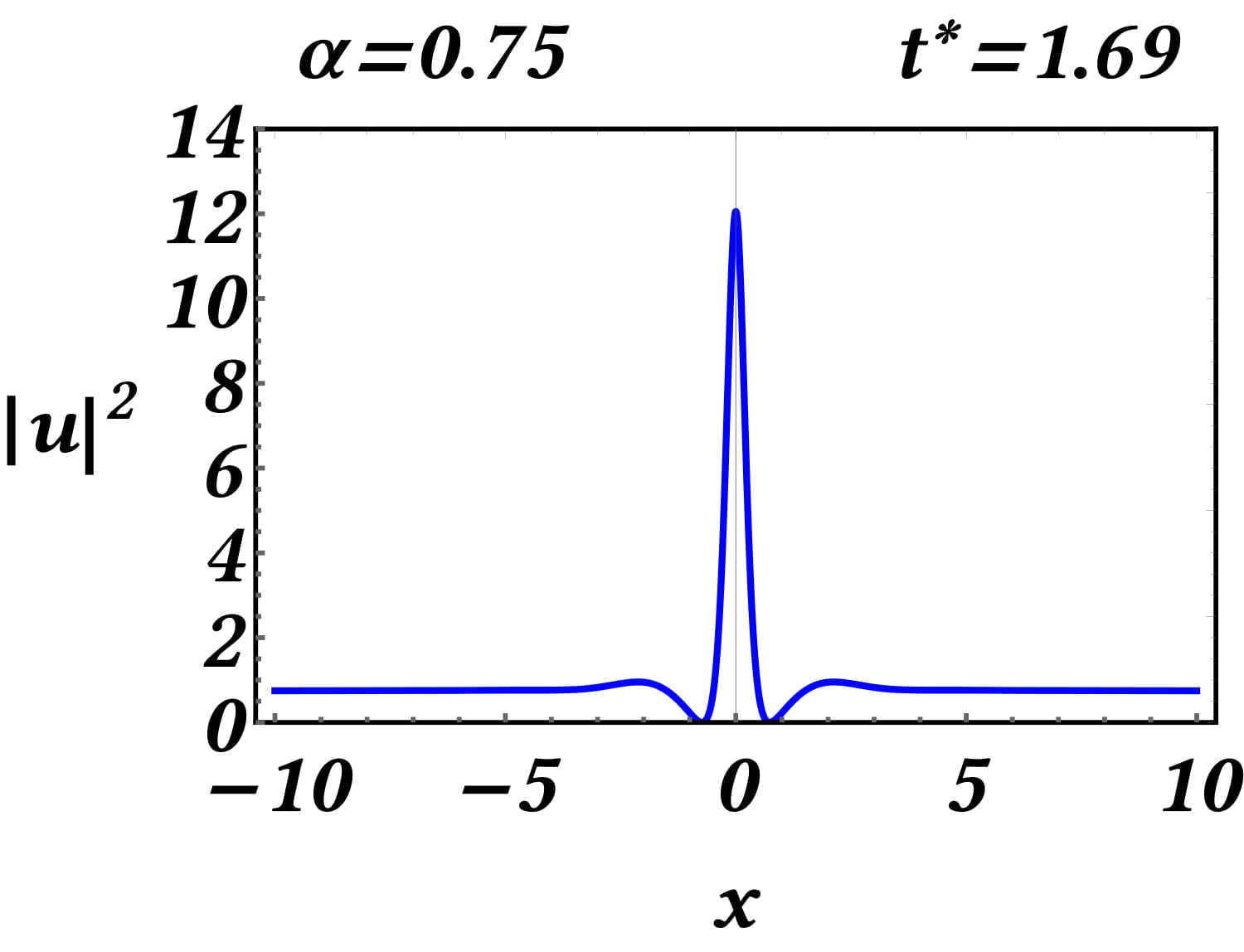}\\
		\includegraphics[scale=0.12]{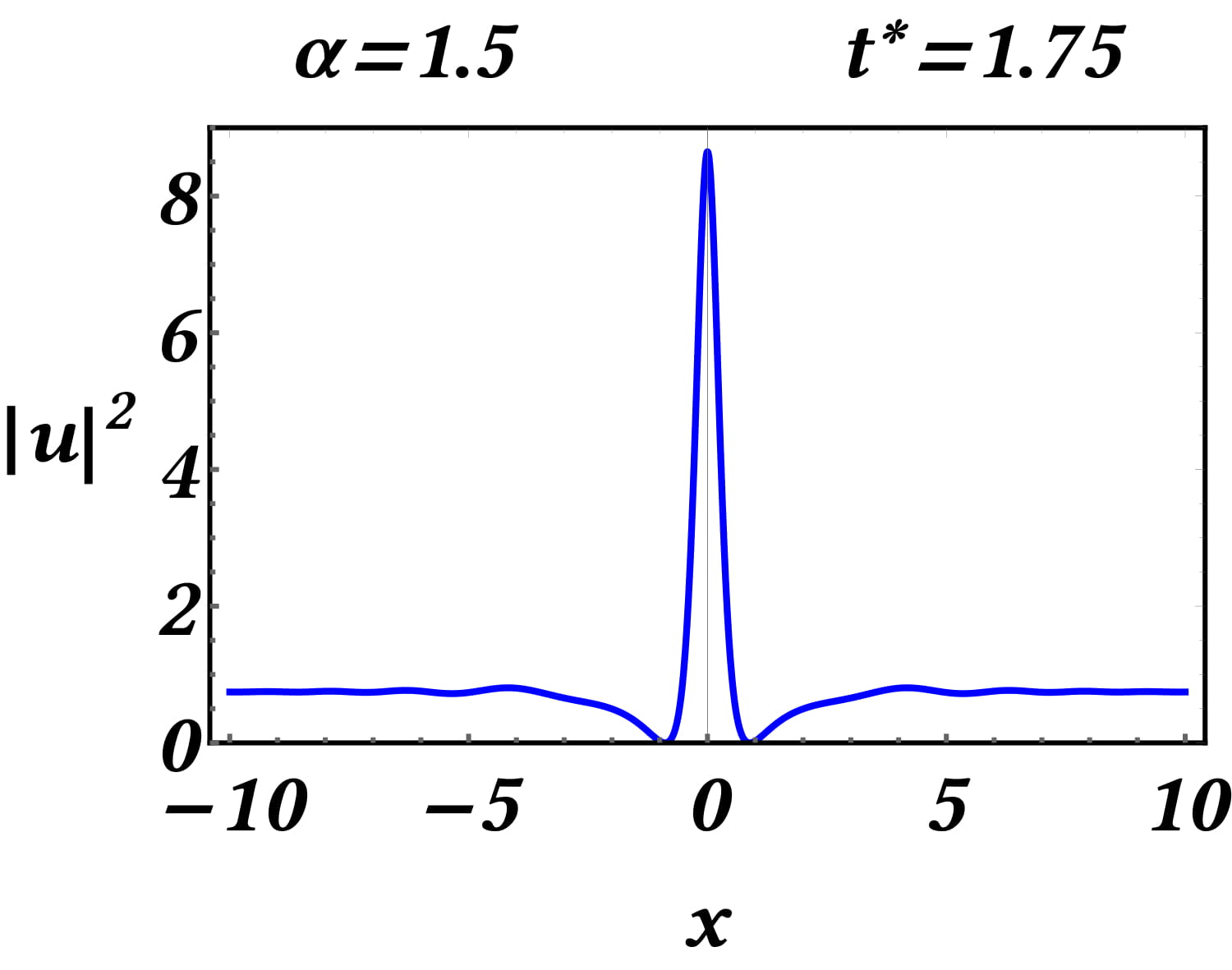}
		\includegraphics[scale=0.12]{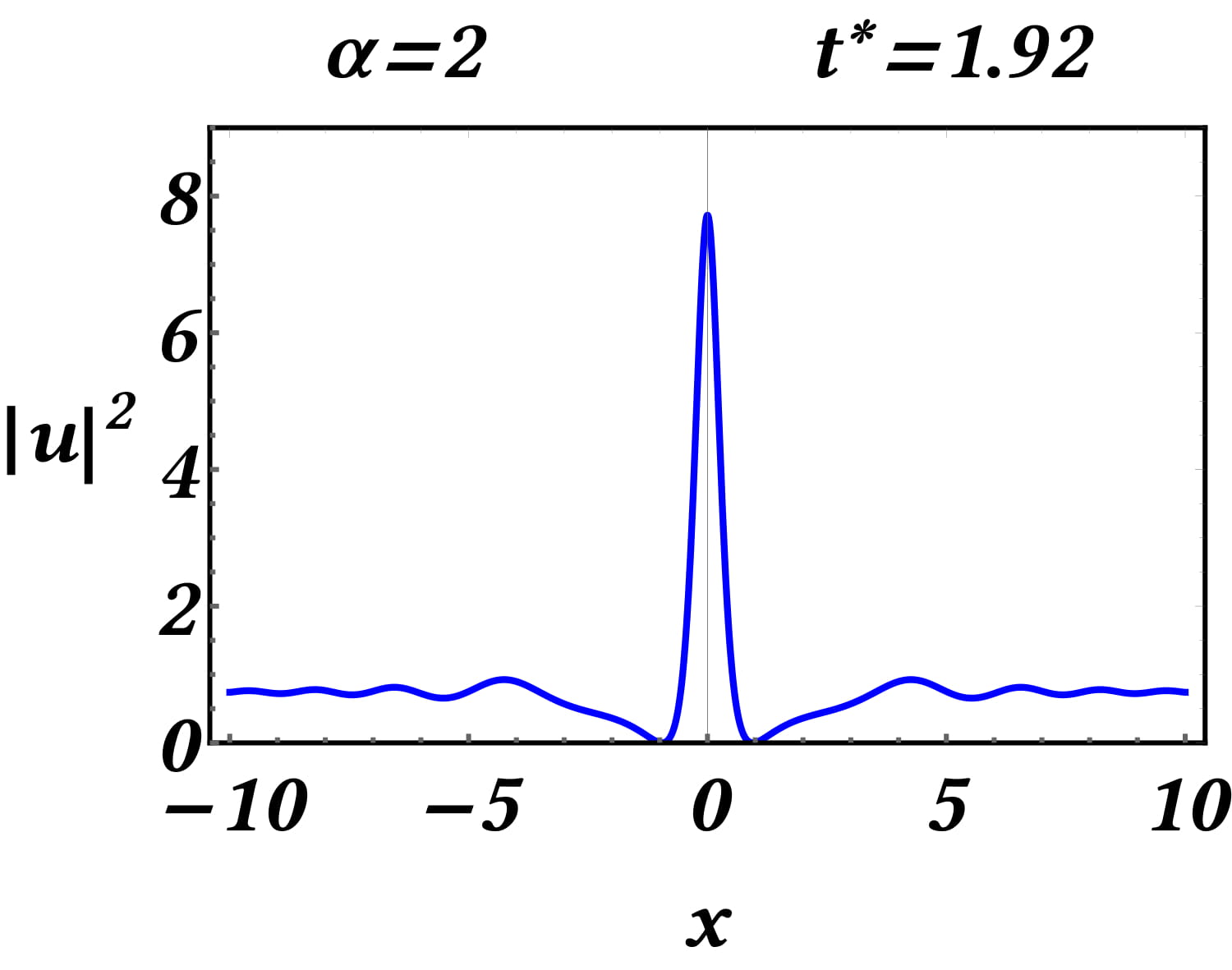}
	\end{center}
	\caption{Profiles of the extreme events for the $\sech$-initial data \eqref{eq4a} for increasing vales of $\alpha$ and their times of occurrence.  Parameters: $\gamma = 0.01, L=500$, $\Gamma=1$, $\sigma_x=100$ and $\sigma_t=0.5$.}
	\label{figure7NN}
\end{figure}
\paragraph{Exponentially decaying initial condition.} We conclude our study on the emergence of extreme events, with a short presentation of the relevant dynamics emerging from an exponentially decaying initial condition \eqref{eq4a}. 

Figure~\ref{figure5} illustrates the results of a similar study to the
one presented in panels (a)-(b) of Fig.~\ref{figure2}. Panel (a)
depicts the evolution of the density of the center $|u(x,0)|^2$ of the
solution, during the emergence of the first extreme event, as compared
to the evolution of the density of the center of the PRW \eqref{PRWP}
$u_{\mbox{\tiny PS}}(x,t;1.63;1.16)$.  We see that although for $t\in
[0,2.5]$ the agreement of the two curves is almost exact,  afterwards,
they diverge with a faster time-decay rate than in the case of the
algebraic initial condition \eqref{eq4}. In addition, in panel (b), we
compare the profiles of the two solutions, close to the right one
among
the symmetric minima. We observe again that at the core of the event
the two curves coincide and that a greater divergence between them is
exhibited at the beginning  of their tails.  Summarizing the above
results, we find that the PRW-solitonic structures emerging from the
algebraic initial data \eqref{eq4} are more proximal to the PRW
rational solution, than those emerged from the exponentially decaying
initial condition \eqref{eq4a}. The greater proximity in the first
case  can be explained by the fact that the algebraic initial
condition is functionally closer to the analytical PRW at $t=0$, as it
can be measured by a suitable-weighted norm \cite{All2}. Henceforth,
the emerged PRW-solitonic profile is expected accordingly, to be
closer to the analytical PRW, as a
result of the continuous dependence of solutions from the initial data.

For the $\sech$-profiled initial condition \eqref{eq4a}, we find similar dependencies for the maximum amplitude and the moment of  occurrence of the extreme event as functions of $\gamma$ and $\Gamma$ to those presented in Fig. \ref{figure3}, and are not shown here.  Instead, we present in Figure \ref{figure6NN} the results of a numerical study varying the width  $\alpha$ of the initial condition \eqref{eq4a}. Panel (a) shows that the maximum amplitude of the extreme event is a decreasing function of $\alpha$. This fact is connected with the value  
of the initial power (squared $L^2$ norm) of the initial condition $P[u_0(x)]=\int_{-\infty}^{+\infty}\sech^2(\alpha x)dx=2/\alpha>0$ which also is a decreasing function of $\alpha$. 
As a consequence, the localization mechanism acting on larger initial powers (associated to smaller values of $\alpha$) 
results in higher amplitudes of the extreme events.  Panel (b) shows
that the time of occurrence is initially decreasing as a function of
$\alpha$. For instance, more time is needed for events of increasing
amplitude to be formed in the case of smaller values of $\alpha$,
which are also associated to larger values of the initial power. Then,
the relevant time attains a minimum at $\alpha=1$, and afterwards, is slightly increasing. To analyse this interesting phenomenon further, we depict in Figure \ref{figure7NN}, profiles of the PRW events for selected values of $\alpha$ below and above $\alpha=1$; recall  that the maximum event for $\alpha=1$ is analysed in Fig. \ref{figure5}. For $\alpha<1$ and $\alpha>1$ it appears that the profile of the PRW-event gradually deviates
from the one of the PRW of the integrable limit \eqref{PRWP}. It becomes less proximal for increasing values of $\alpha>1$ than for decreasing
values $\alpha<1$.
It appears that $\alpha=1$ is the optimal width to define a most proximal
(to the original PRW \eqref{PRWP}) PRW-type event, thus needing the minimum time to be formed.
%
%
\begin{figure}[tbp!]
	\centering
	\includegraphics[scale=0.12]{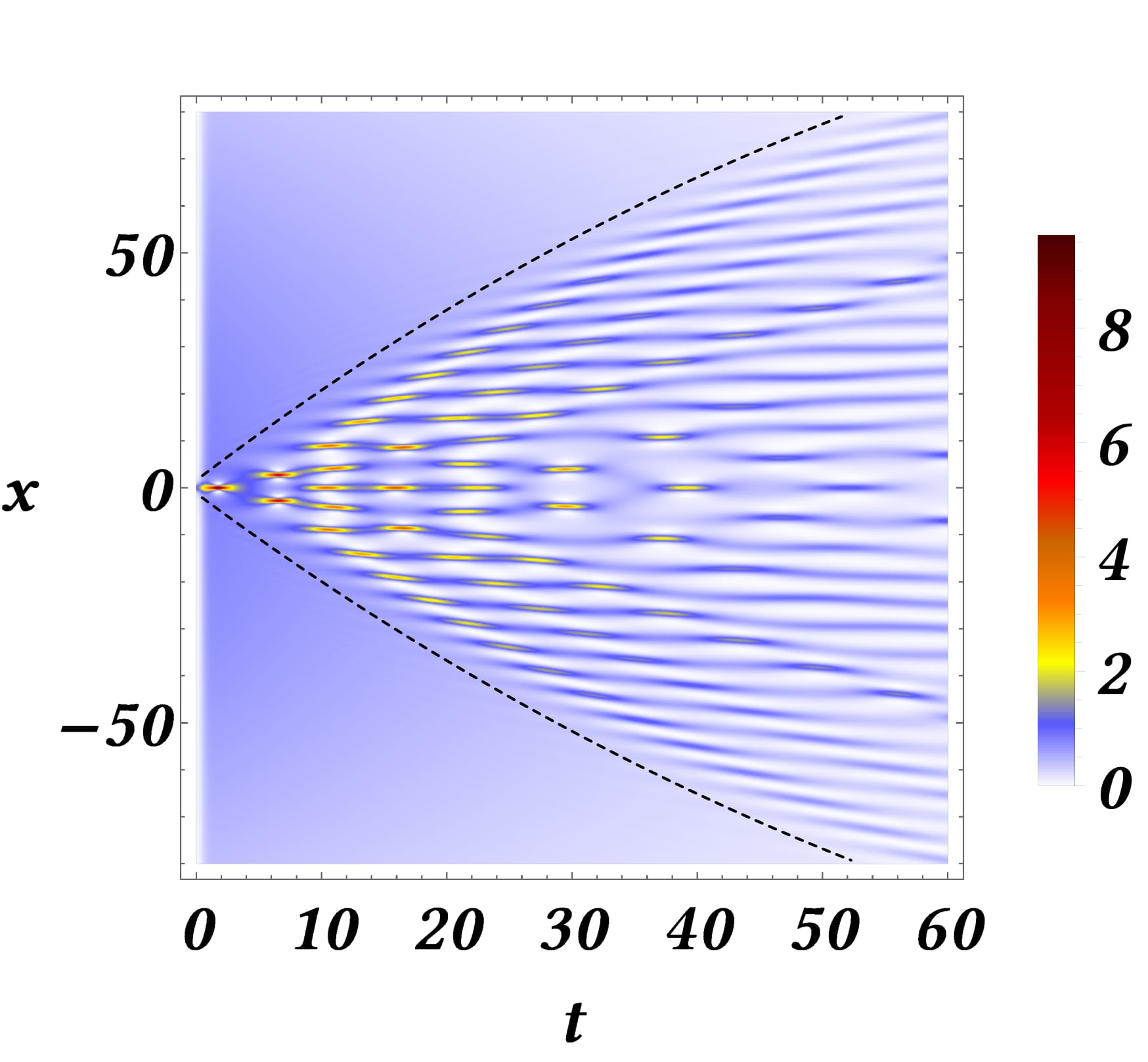}
	\includegraphics[scale=0.11]{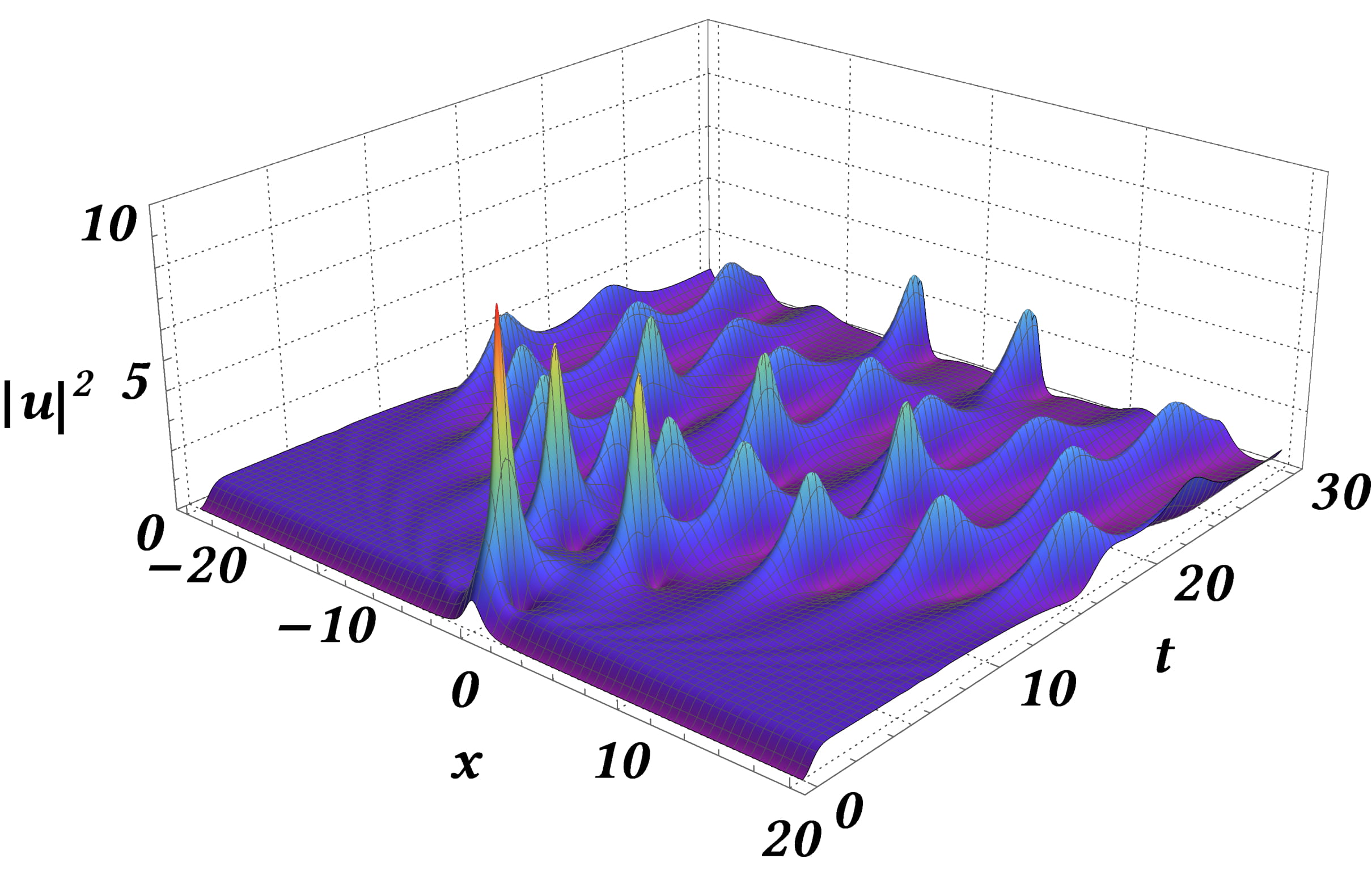}
	\caption{(Color Online) 
		Left panel: Contour plot of the spatiotemporal evolution of the density of the initial condition (\ref{eq4}), for the same set of parameters as in Figs.~\ref{figure1} and \ref{figure2}. The dotted (black) curves highlight nonlinear caustics. Right panel: A 3D-plot of the spatiotemporal evolution of the density of the right panel.}
	\label{figure8NN}
\end{figure}
\subsection{Emergence of Spatiotemporal Patterns} \label{sec:2b}
Beyond the formation of the PRW-extreme event, further interesting dynamical features arise at later times. This can be seen in the left panel of Fig.~\ref{figure8NN}, depicting  a contour plot of the spatiotemporal evolution of the density, for $x\in [-80,80]$, $t\in [0,60]$ and for the same parameters as in Figs.~\ref{figure1} and \ref{figure2}; a 3D-plot of this evolution is illustrated in the right panel of Fig.~\ref{figure8NN}.

Here, it should be pointed out that the observed dynamics, prior to the ultimate decay 
of the solution, is found to be partially reminiscent of that 
of the semi-classical NLS~(\ref{eq1sl}): this is due to the fact that 
the change of dimensionless space and time variables, respectively,
$x \rightarrow x/\sigma_x$ and 
$t \rightarrow t/\sigma_x$
(for $\sigma_x =100$, i.e., $\sigma_x^{-1} \sim \epsilon \ll1$) 
in Eq.~(\ref{eq1})  renders 
its left-hand side identical to that of Eq.~(\ref{eq1sl}). 
Indeed, as shown in Fig.~\ref{figure8NN}, and similarly to Refs.~\cite{BM1,BM2} where the 
semi-classical NLS was studied, distinct spatiotemporal regions are formed; these are 
separated by nonlinear caustics (dashed curves), 
which bound the pattern of the transient decaying spatiotemporal oscillations.  
%
In particular, 
in these works, a ``lattice'' of extreme events -- in the form of PRW structures -- was
found to occupy the region in between the caustics, at points corresponding to the poles of the 
tritronqu{\'e}e solution of the Painlev{\'e}-I equation.
In our case, 
in the region bounded by the caustics we observe the formation 
of PRW events as discussed above.
Yet, the 
pattern formed
appears to be 
closer to the PRW lattice of~\cite{BM1,BM2} in the early stage of the evolution, while, at its final stages, the dynamics is weakened due to the effect of the damping.
\begin{figure}[tbp!]
	\centering
	\begin{tabular}{cc}
		\includegraphics[scale=0.12]{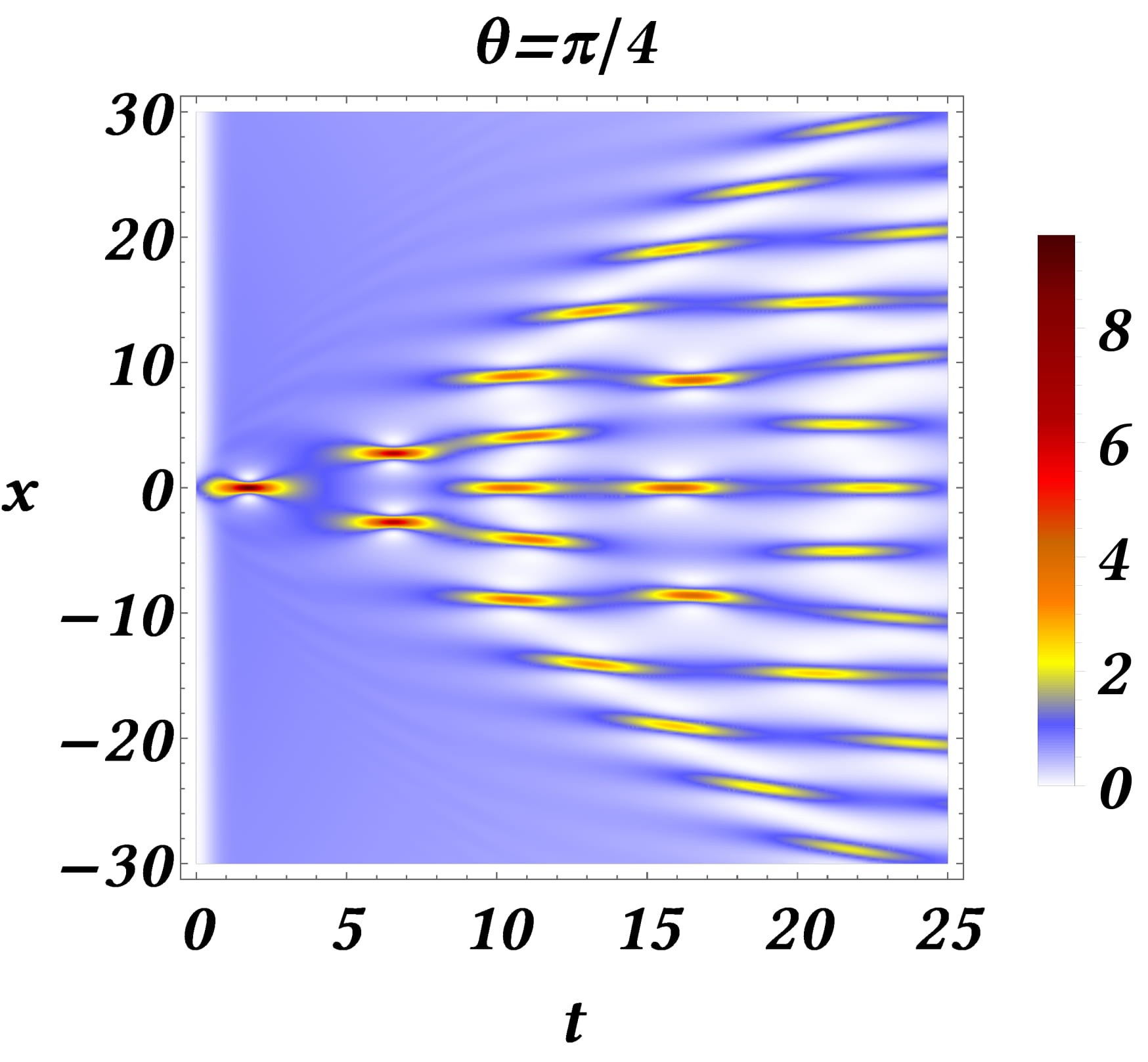}&\hspace{0.2cm}\includegraphics[scale=0.12]{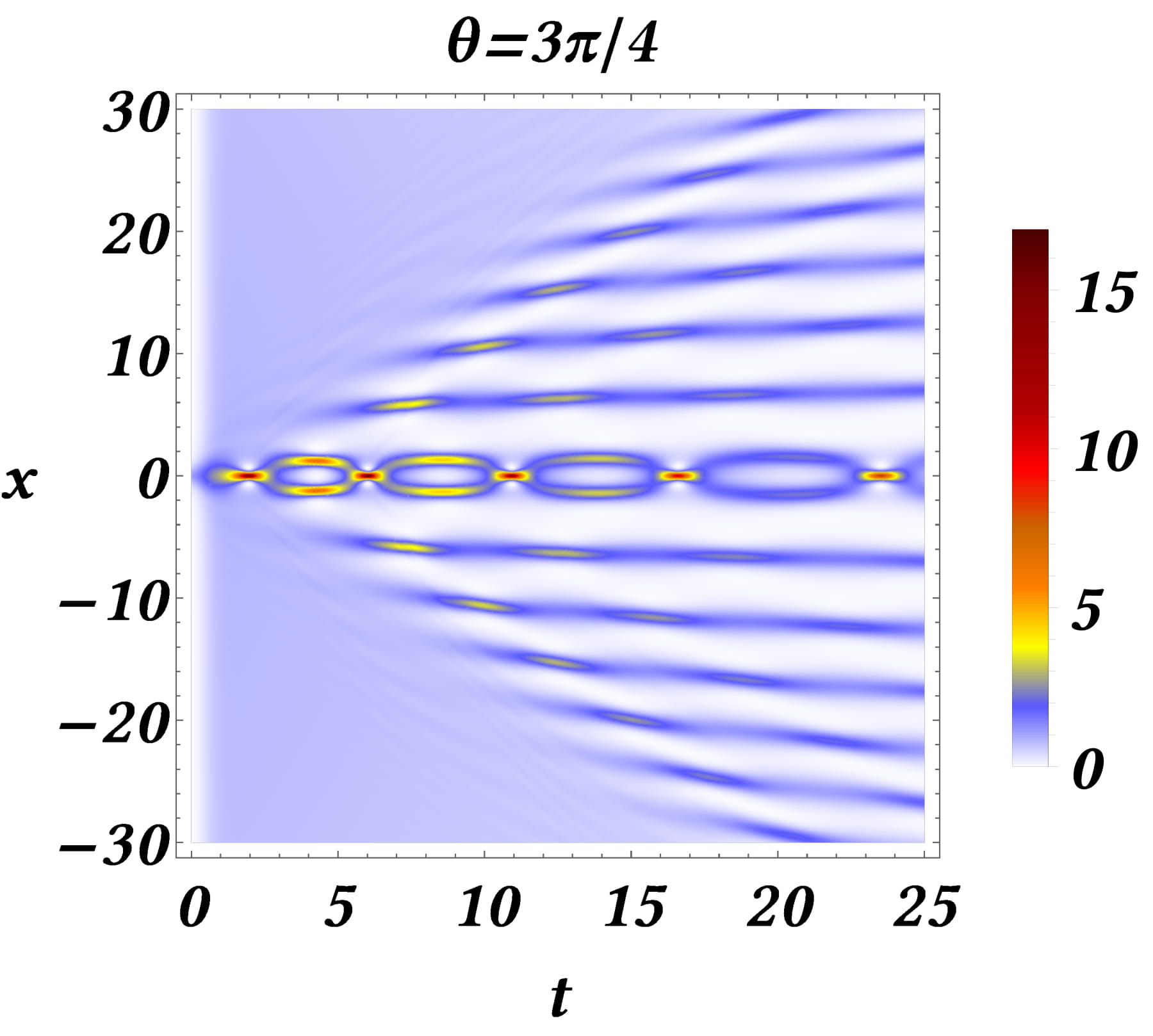}\\[10pt]
		\includegraphics[scale=0.12]{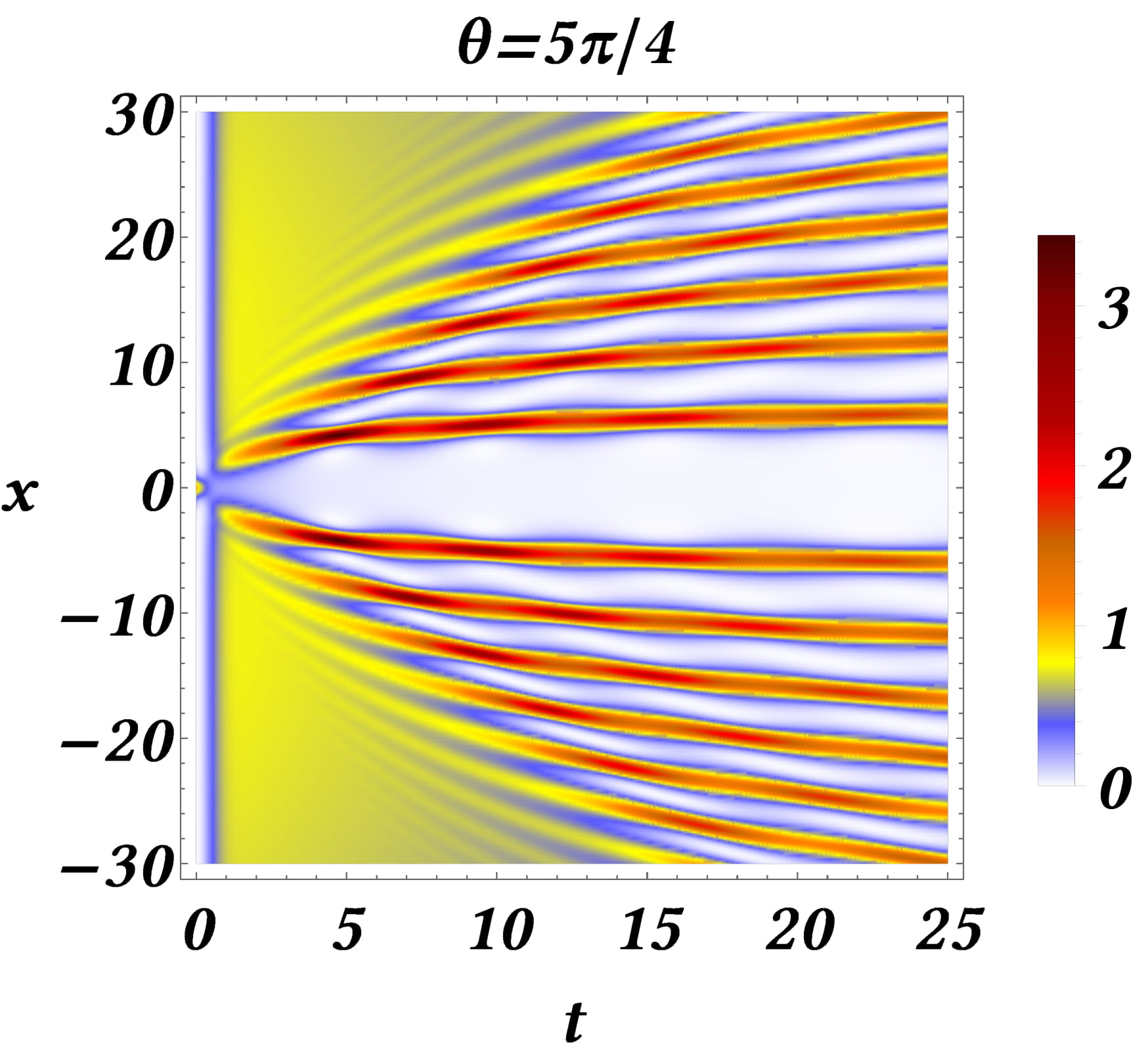}&\hspace{0.2cm}\includegraphics[scale=0.12]{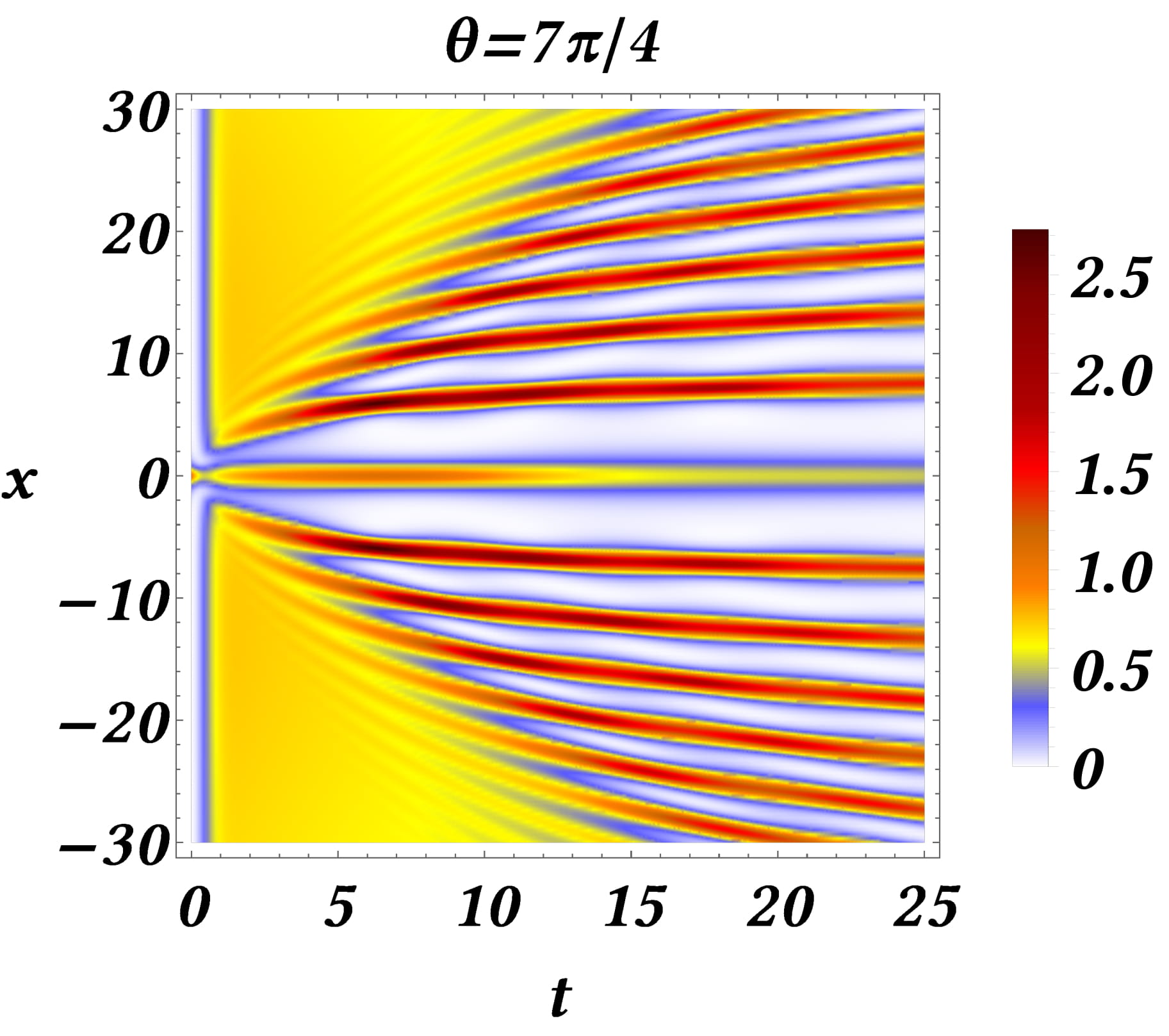}
	\end{tabular}
	\caption{(Color Online) The  effect of phase parameter $\theta$ of the driver \eqref{wlf} in the structure of the emerging spatiotemporal patterns. Top left panel: $\theta=\pi/4$. Top right panel: $\theta=3\pi/4$. Bottom left panel: $\theta=5\pi/4$. Bottom right panel: $\theta=7\pi/4$. Parameters:: $\gamma=0.01$, $L=250$, $\Gamma=1$, $\sigma_x=100$ and $\sigma_t=0.5$.}
	\label{figure9NN}
\end{figure}
We remark that if the initial condition had the 
form of an exponentially localized pulse on top of a finite background
(see also
Refs.~\cite{All2,Yang1,Yang2}) then the numerical solution would follow
more closely the 
dynamics in Ref.~\cite{BM}, albeit with some of the gain/loss-induced differences discussed above.


\paragraph{Dependence of the dynamics on the phase $\theta$ of the driver.}
One of the most interesting features of the dynamics of \eqref{eq1} to highlight, is its dependence on the phase parameter $\theta$ of  the forcing \eqref{wlf}. We remark that in contrast with the integrable limit $\Gamma=\gamma=0$, or the damped but unforced $\Gamma=0,\;\gamma>0$ limit, the damped and forced NLS \eqref{eq1} is not invariant under the gauge change of variables $u\rightarrow v\exp(\mathrm{i}\theta)$. Figure \ref{figure9NN} depicts the distinct spatiotemporal patterns produced when the phase $\theta$ of the forcing is changing counter-clockwise along the diagonals.
Here, we assume variation of a constant phase angle along the complex plane
without a spatial variation thereof.
The initial condition is the quadratically decaying \eqref{eq4} and
the rest of parameters are fixed as in the study of
Fig. \ref{figure1}; similar patterns are produced by the exponentially
decaying \eqref{eq4a}.  The upper left panel repeats (for completeness
of the presentation and for a shorter time span) the pattern of
Fig. \ref{figure8NN}  associated with the emergence of PRW as first
event, when the phase of the forcing is $\theta=\pi/4$. When
$\theta=3\pi/4$ the pattern changes drastically, as shown in the upper
right panel: we observe in the center, the emergence of a breathing
mode
reminiscent of the core of a Kuznetsov-Ma (KM) breather state. The bottom row shows the patterns when $\theta=5\pi/4$ and $\theta=7 \pi/4$.
 The dynamics in these cases are not producing extreme-wave events but
 are reminiscent of the waveforms emerging as a result of
 the modulational instability in its nonlinear stage as analysed in Ref.~\cite{BM}. Yet, we observe a remarkable difference between these two cases, in the evolution of the center: while in both cases the division of the spatial domain in the left and right  $x$-periodic fields is similar, when $\theta=5\pi/4$ a density hump is absent in the center, but is present with an almost stationary evolution when $\theta=7\pi/4$. 
We remark that a  complex mixture of the above patterns can be
produced when varying $\theta$ between the above prototypical
scenarios (not shown herein). 

Therefore, varying the phase angle $\theta$ of the
   driver, we may reconstruct  the prototypical spatiotemporal
   patterns associated to the emergence of extreme-wave events
   (including PS rogue waves and semi-classical type dynamics, among
   others)  and the patterns associated with the nonlinear stage of the modulational instability. 
 The above study on the dependence of the driver's phase could be of
 relevance in the context of weakly nonlinear waves in the presence of
 wind forcing \cite{onorato1,brunetti, DHZ19}.
Particularly in~\cite{DHZ19}, phase-shift effects could be
potentially induced due to the stochastic nature of the driving. While
the latter setup is not identical to the one considered herein, its
exposition of the stochastic nature of the wind forcing (and
the associated prefactor) lends relevance to the consideration of a phase variation herein.



\begin{figure}[tbp]
\hspace{-0.5cm}
\begin{center}
\includegraphics[scale=0.093]{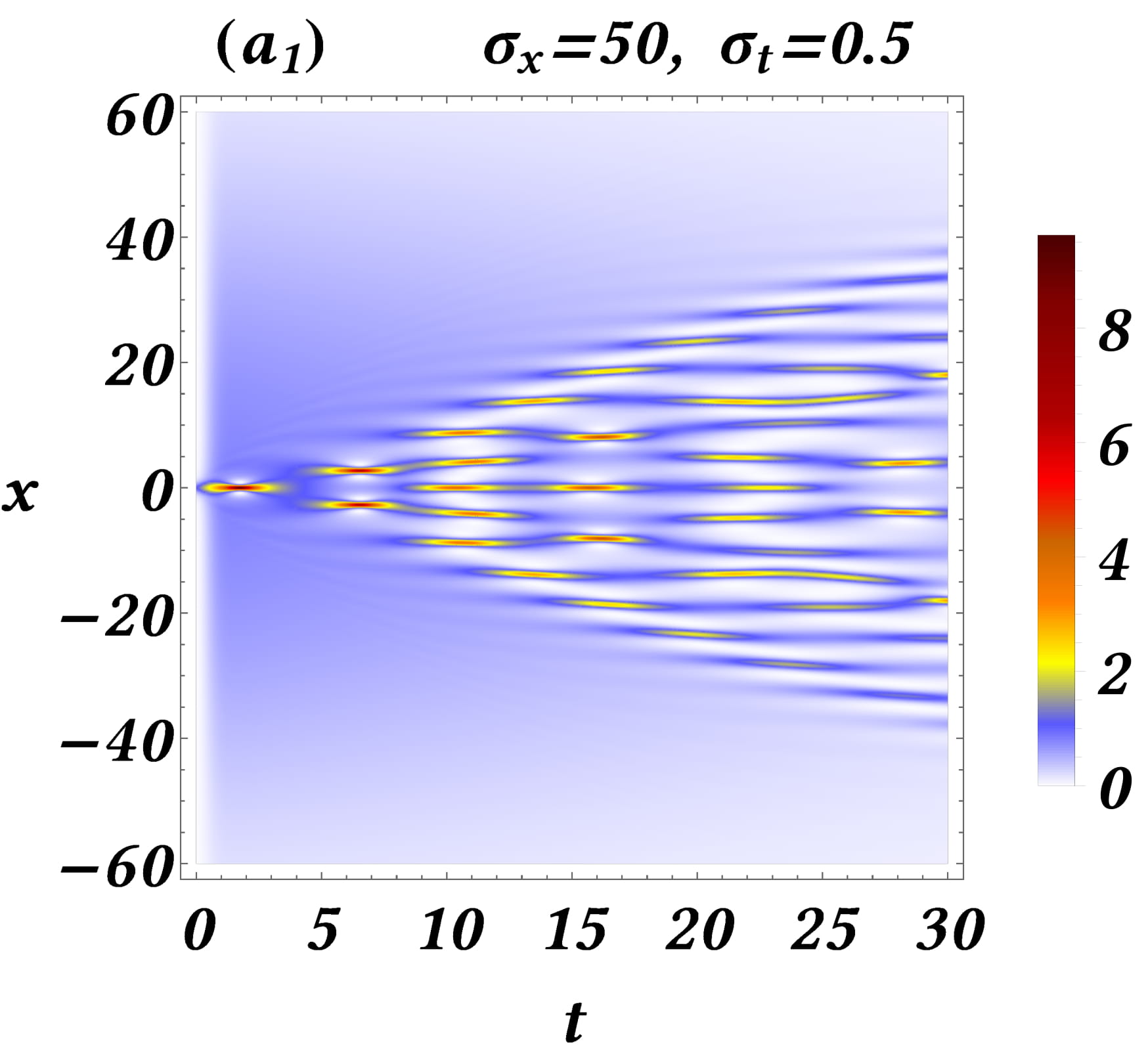}
\includegraphics[scale=0.093]{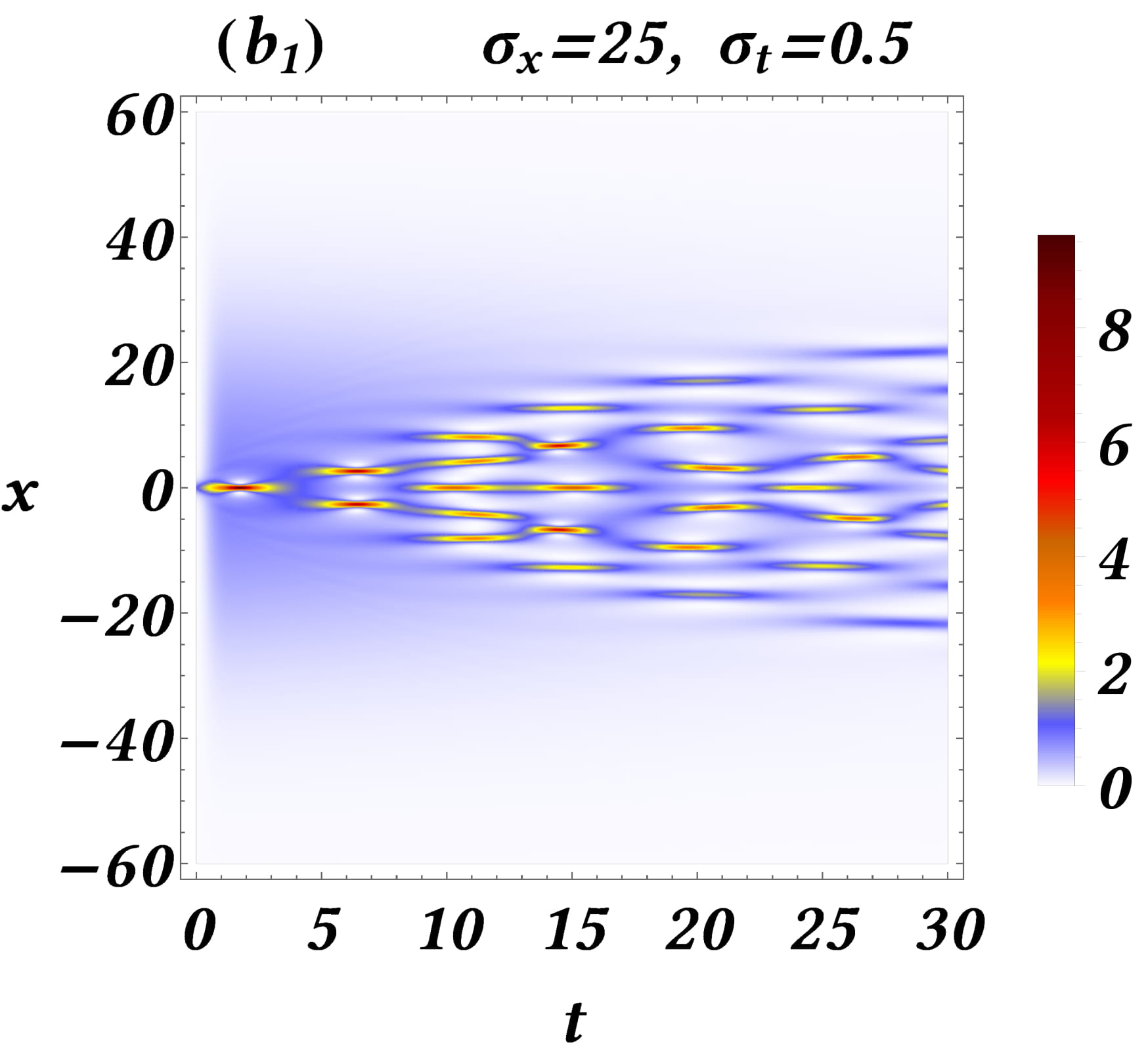}
\includegraphics[scale=0.093]{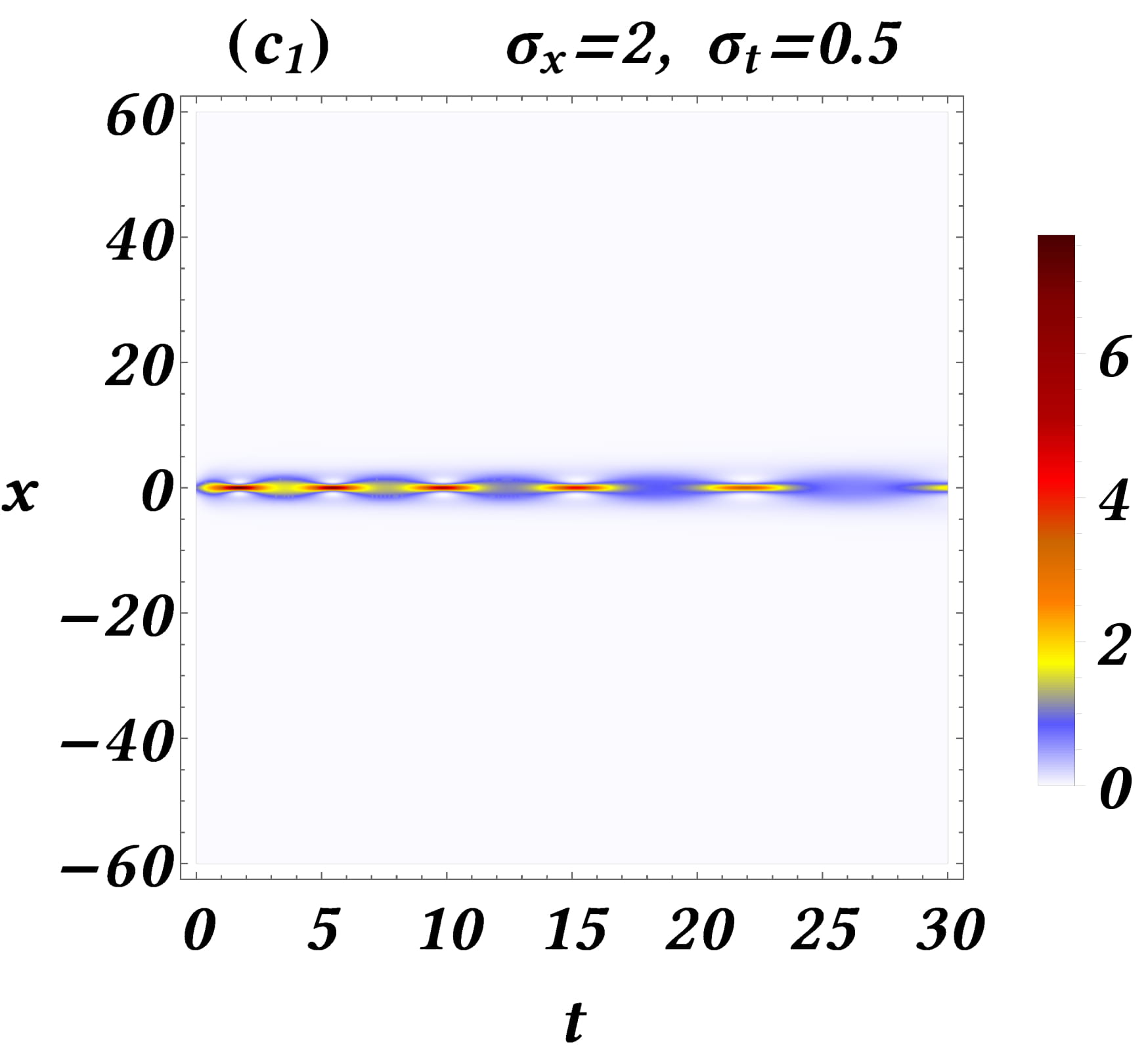}\\
\hspace{-0.8cm}\includegraphics[scale=0.093]{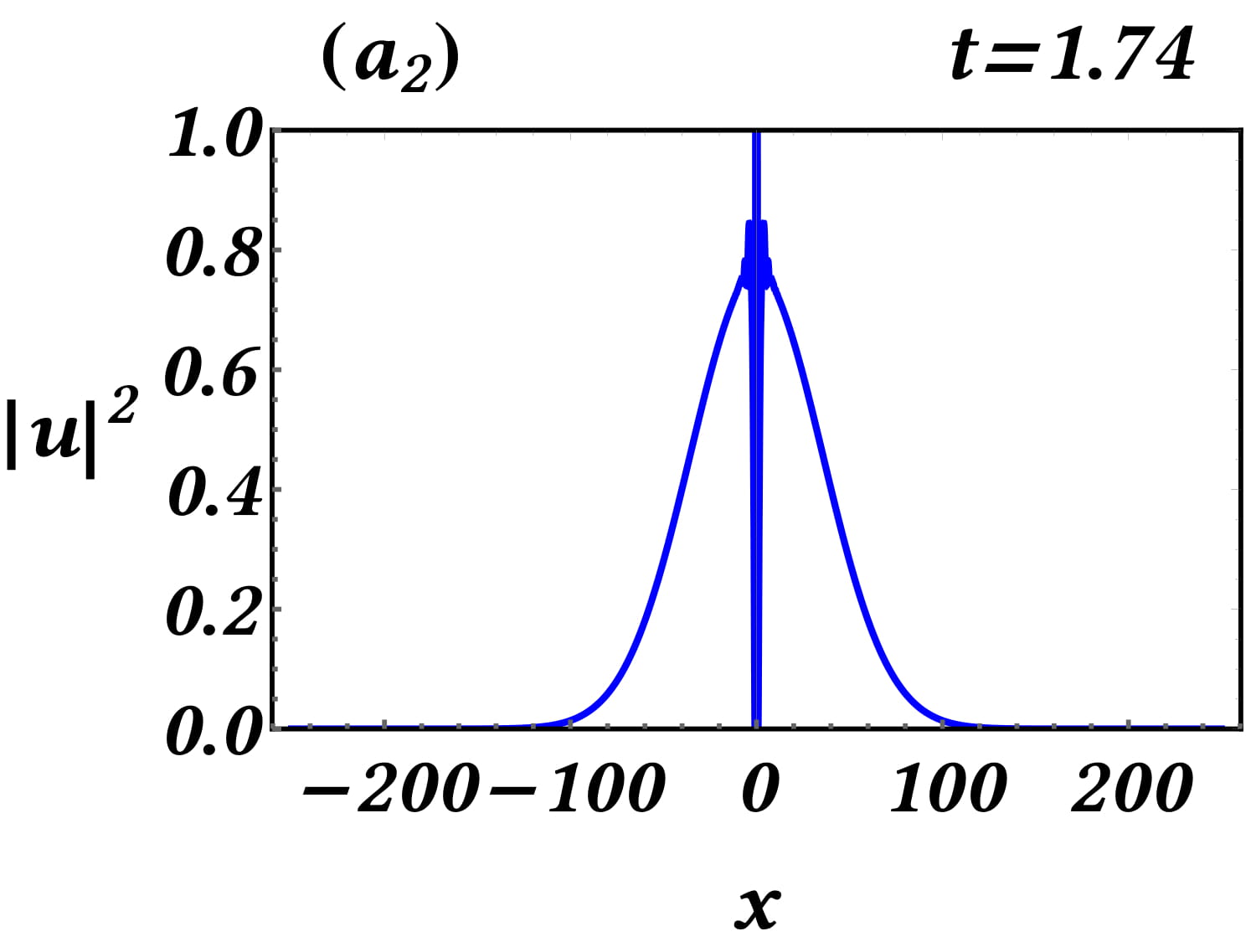}\hspace{0.8cm}
\includegraphics[scale=0.093]{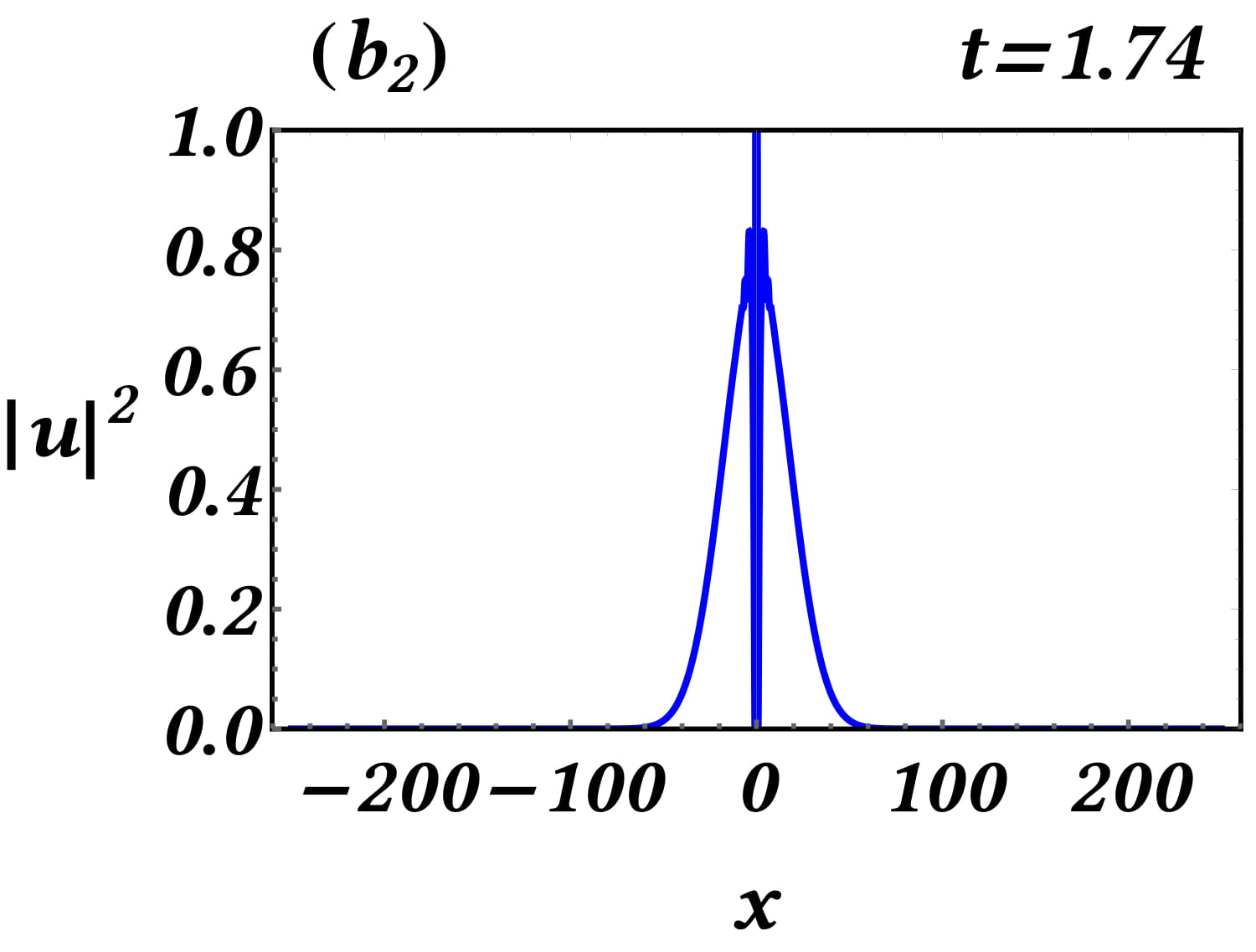}\hspace{0.8cm}
\includegraphics[scale=0.093]{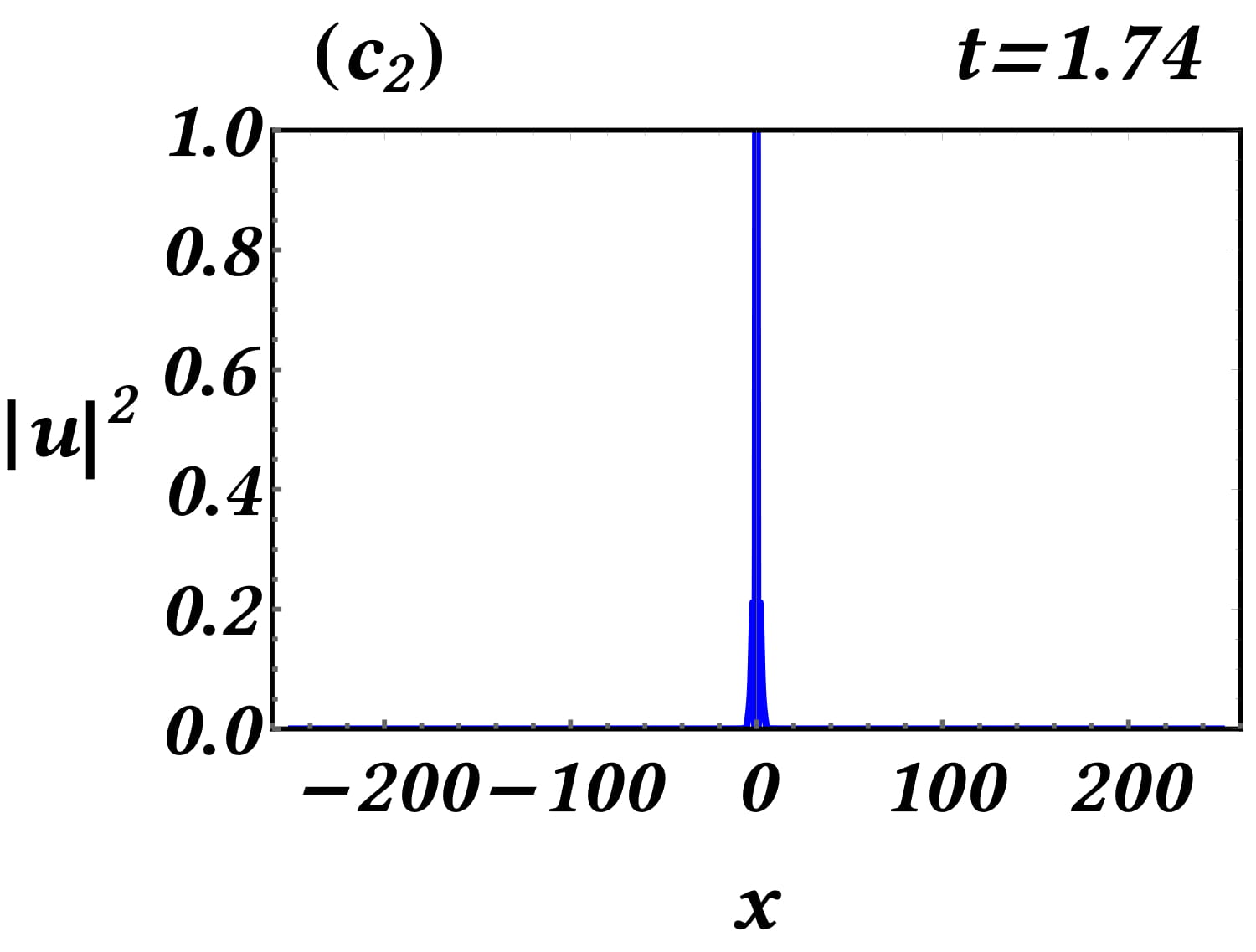}
\end{center}
	\caption{(Color Online) Top row:  A sequence of contour plots of the spatiotemporal evolution of the density for fixed $\sigma_t=0.5$ and decreasing values of $\sigma_x=50$, $25$, $2$. Bottom row: Snapshots of the evolution of the density at $t=1.74$ associated to each contour plot of the top row. Parameter values are $\gamma=0.01$ and $\Gamma=1$, $L=250$.}
	\label{figure10NN}
\end{figure}

\paragraph{Dependence of the dynamics on $\sigma_x$ and $\sigma_t$} 
Another important feature of the system is that the driver's spatial and temporal 
scales $\sigma_x$ and $\sigma_t$ affect the space-time localization of the 
numerical solution. This is verified in  Fig.~\ref{figure10NN}. There, the top row panels ($a_1$), ($b_1$), ($c_1$) show contour plots of the 
spatiotemporal evolution of the density, for fixed  $\sigma_t=0.5$ and 
$\sigma_x=50, 25, 2$ (left to right), respectively. Each of the bottom row panels  ($a_2$)-($b_2$)-($c_2$)  depicts 
snapshots of the density, at $t=1.74$, corresponding to the contour plots of the top row. Evidently, $\sigma_x$ controls the width of the decaying support of the emergent
wave train. Particularly, we observe that, for $\sigma_x=2$, the width of the support is 
considerably reduced and the solution is strongly localized around the center, exhibiting oscillations in time. 

\begin{figure}[tbp]
	\begin{center}
	\includegraphics[scale=0.12]{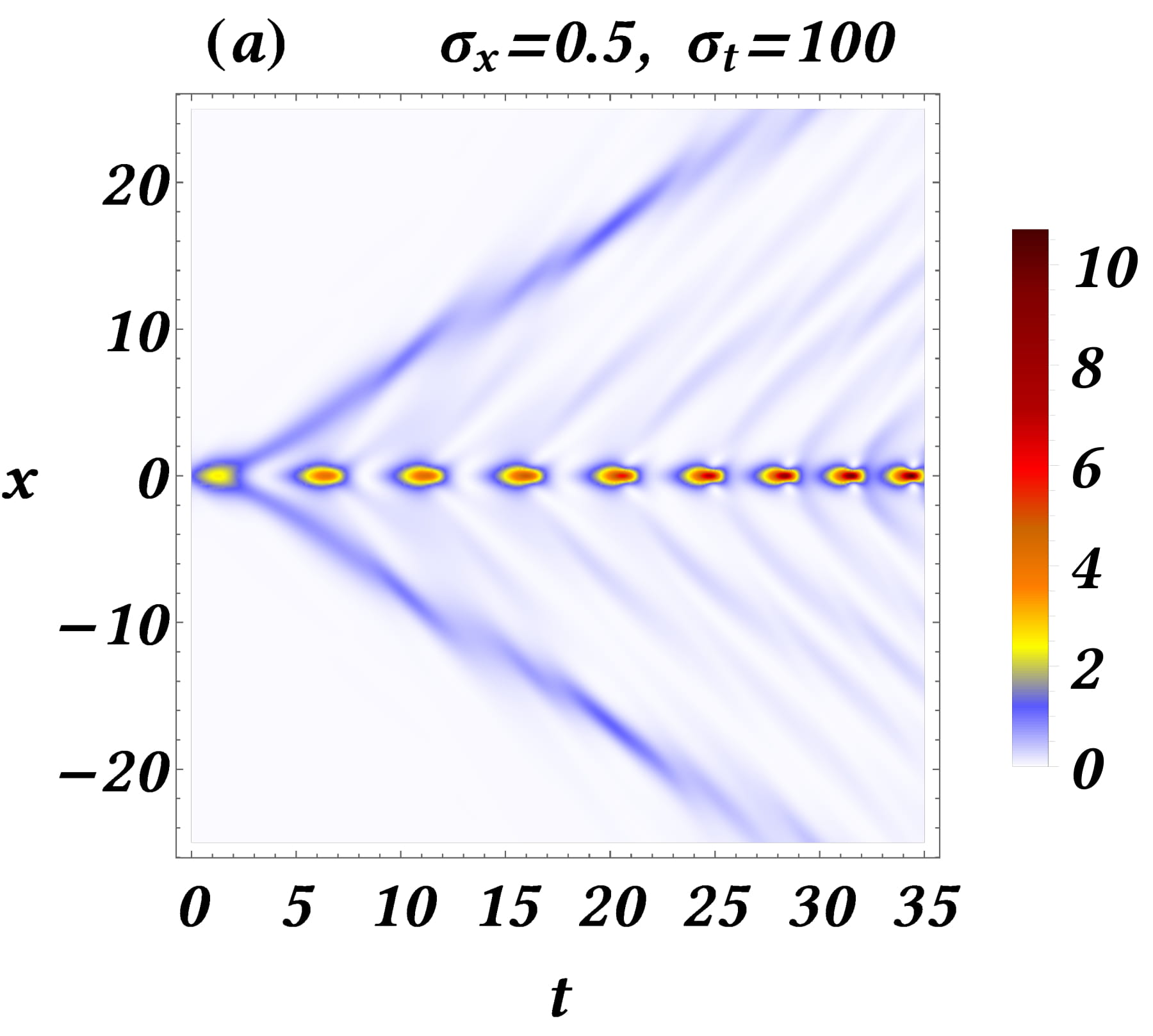}
	\includegraphics[scale=0.12]{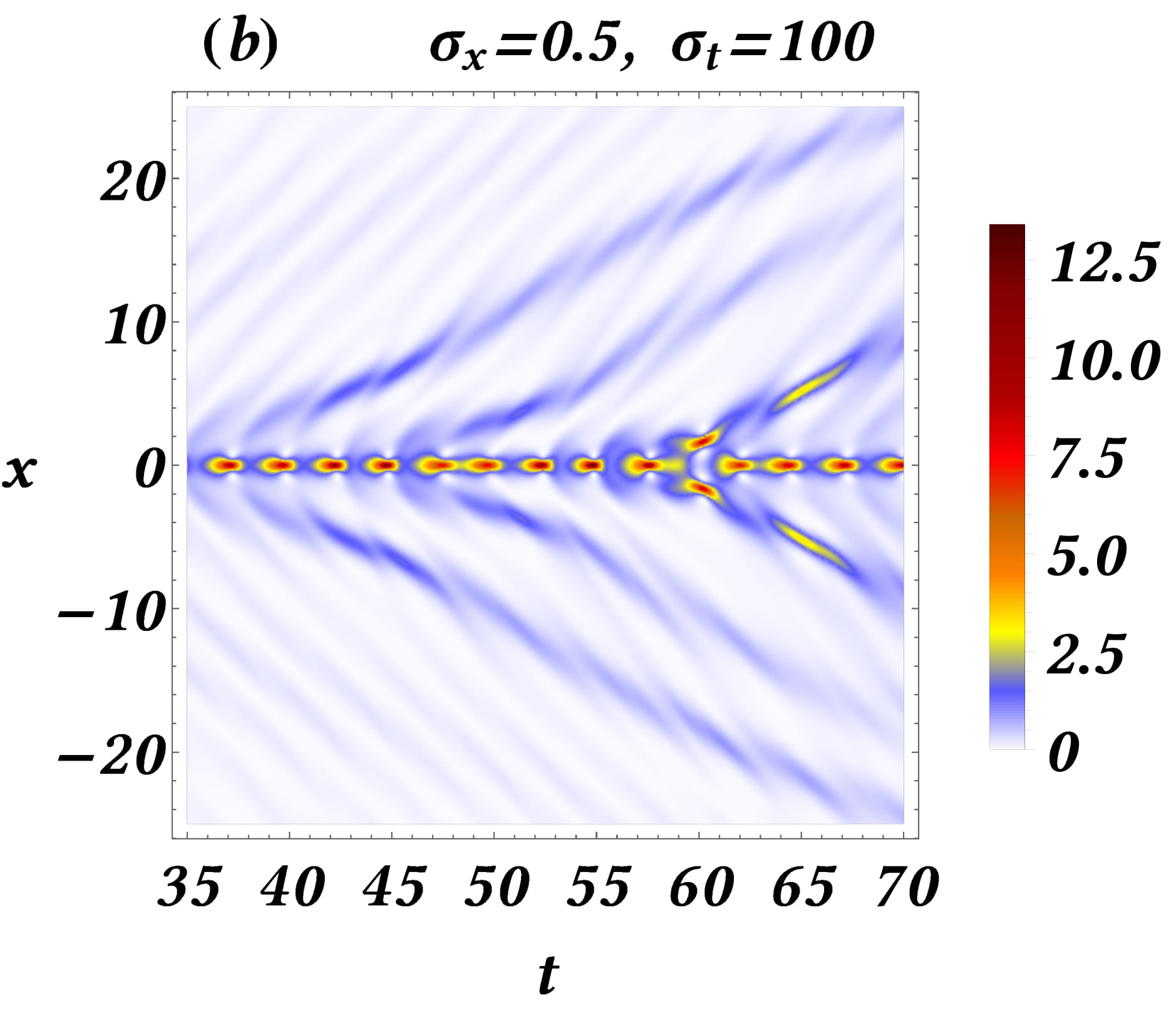}
	\end{center}
	\caption{(Color Online) Panel (a): Contour plot of the spatiotemporal evolution of the density of the initial condition (\ref{eq4}), for $\gamma=0.01$, $\Gamma=1$, and for spreads of the forcing $\sigma_x=0.5$, $\sigma_t=100$, for $t\in [0,35]$.  Panel (b): Continuation of the evolution depicted in panel (a), for $t\in [35,70]$. }
	\label{figure11NN}
\end{figure}

We also explore, in Fig.~\ref{figure11NN}, the impact of increasing the value of $\sigma_t$ to $\sigma_t=100$ (for sufficiently small $\sigma_x=0.5$).
Panel (a) shows the spatiotemporal evolution of the density of the numerical solution, for $t\in [0,35]$. We can observe a dominating oscillation around the center and 
small-amplitude linear waves emanating from the main pattern. On  one hand, 
the large acting time of the forcing enhances time-periodic oscillations; 
on the other hand, the small value of $\sigma_x$ reduces considerably the effect of the formation of the decaying support and the localization width of the  
solution. 
Panel (b) depicts the continuation of the evolution shown in panel (a) for $t\in [35, 70]$; it is observed that the breathing oscillations feature a significant lifetime 
prior to their eventual decay.

\paragraph{Dynamics in the case of purely spatial forcing} 

Having identified the role of the temporal localization of the driver above, we proceed 
by examining the case of a driver showing solely spatial dependence, namely: 
\begin{equation}
\label{Gaussx}
\begin{array}{c}
f(x)=g(x)\exp(i\theta),\;\;
\hbox{where}\;\;\displaystyle g(x)=\sqrt{2}\Gamma\exp\left(-\frac{x^2}{2\sigma_x^2}\right).
\end{array}
\end{equation}

\begin{figure}[tbp]
	\begin{center}
\includegraphics[scale=0.11]{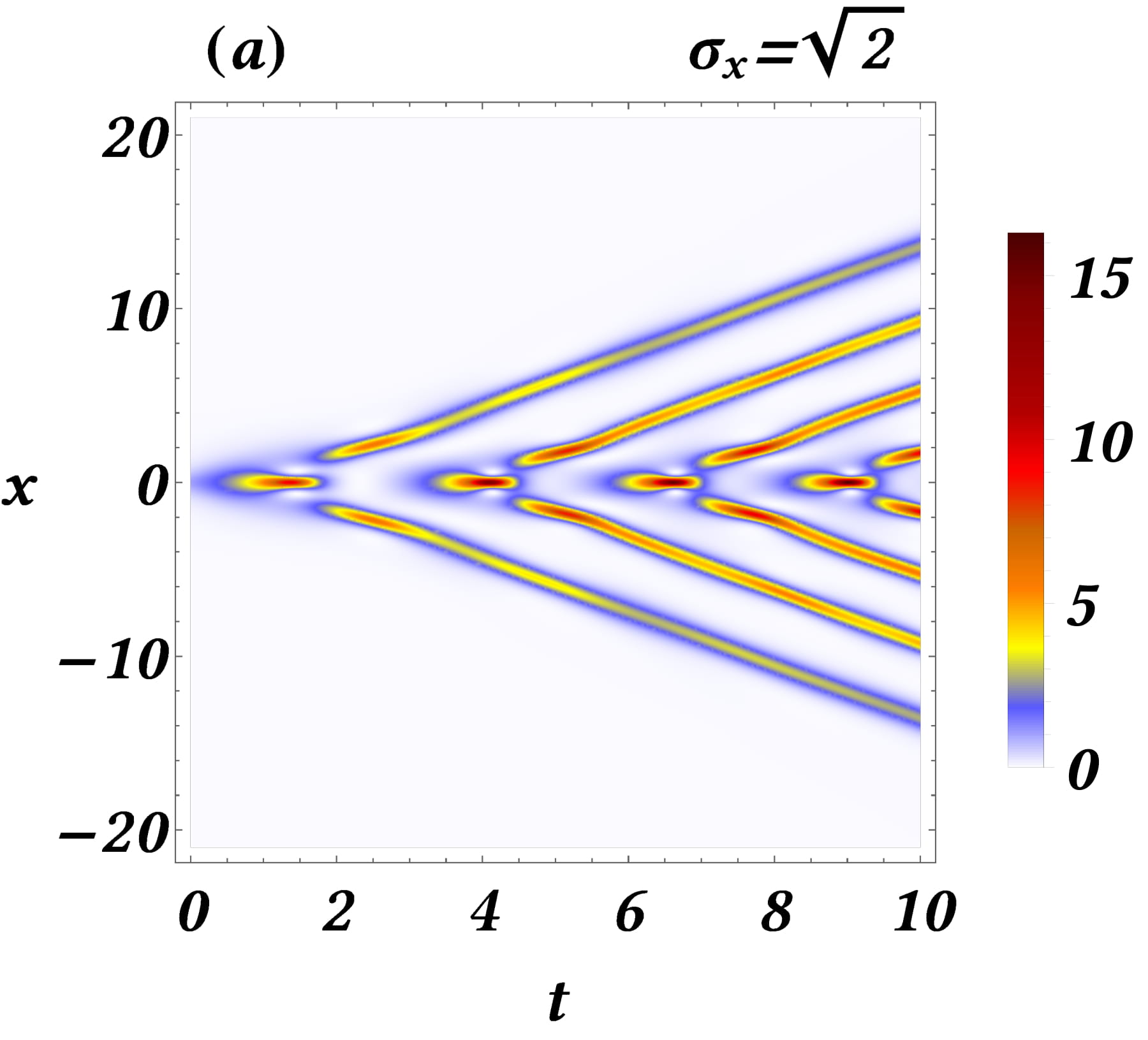}
\includegraphics[scale=0.11]{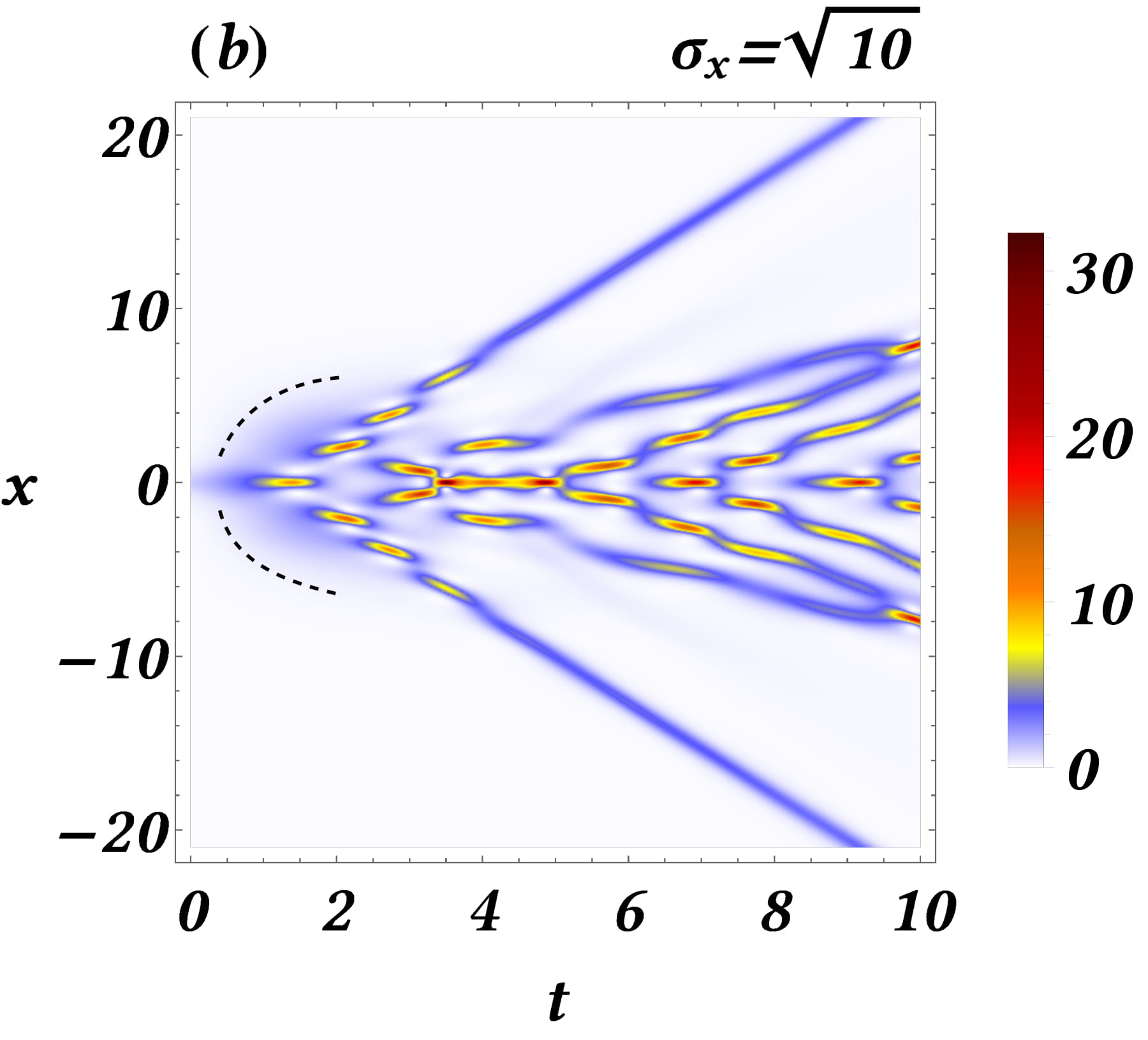}\\
\includegraphics[scale=0.11]{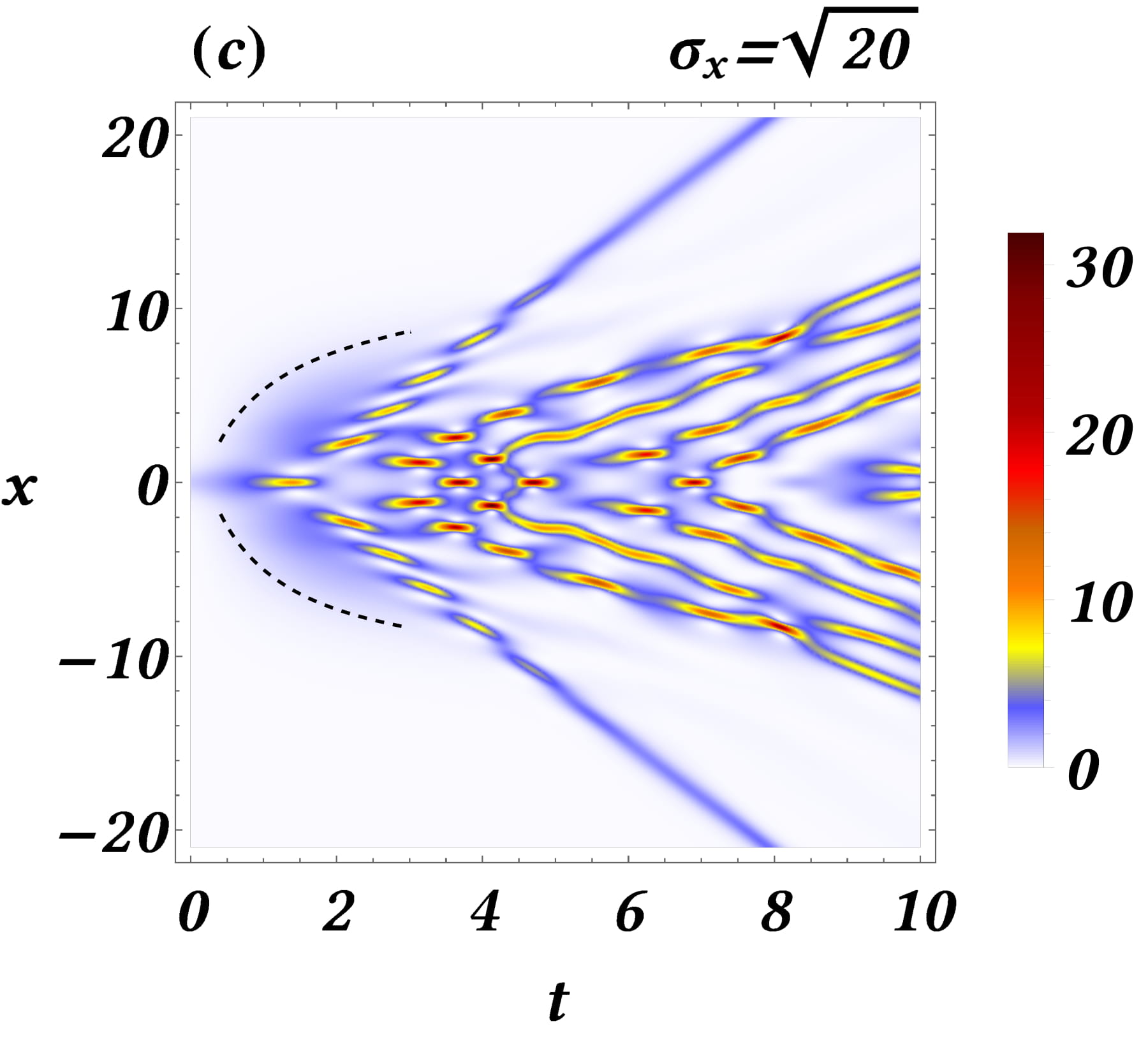}
\includegraphics[scale=0.11]{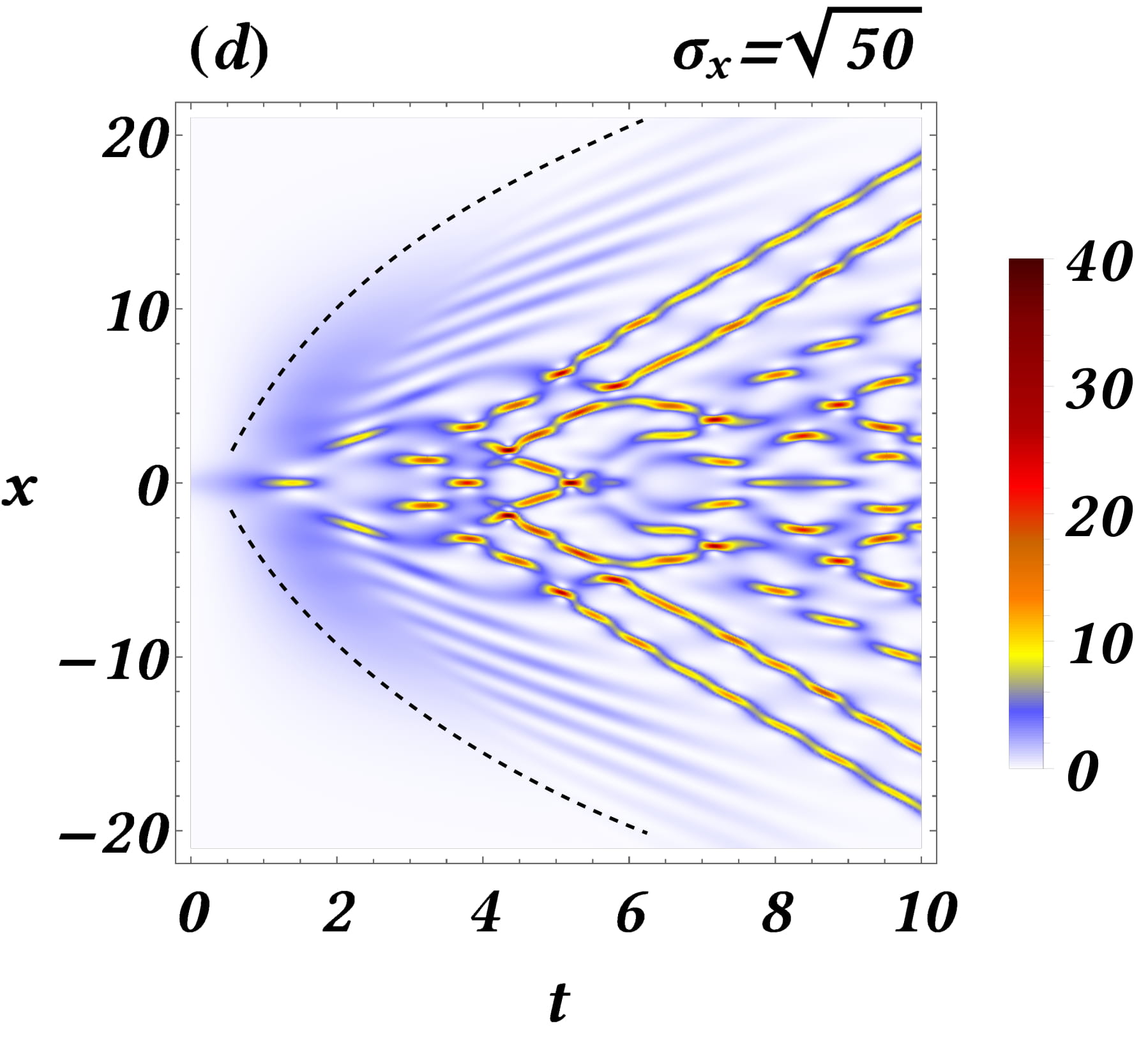}\\
	\hspace{-0.8cm}\includegraphics[scale=0.09]{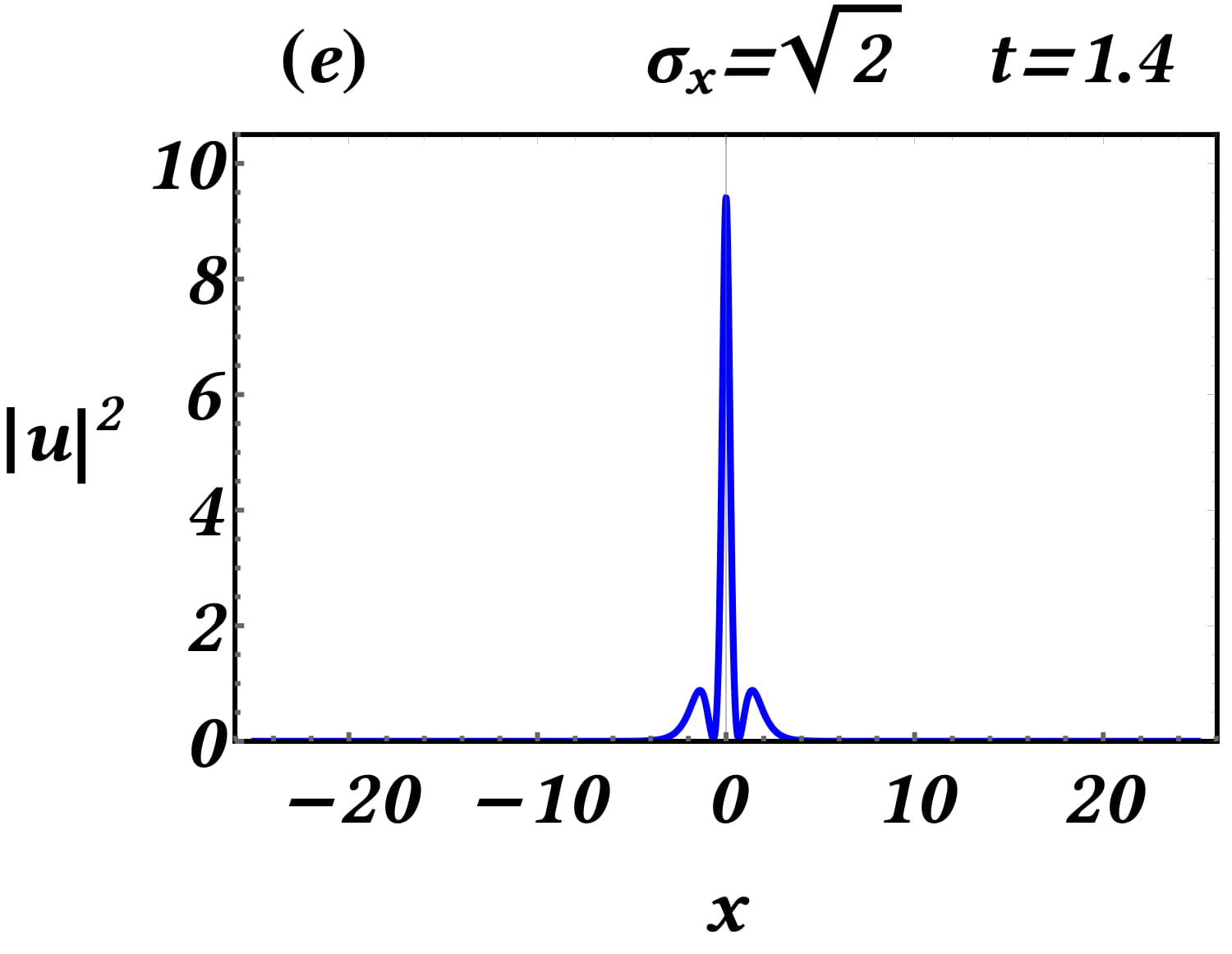}\hspace{0.8cm}
\includegraphics[scale=0.09]{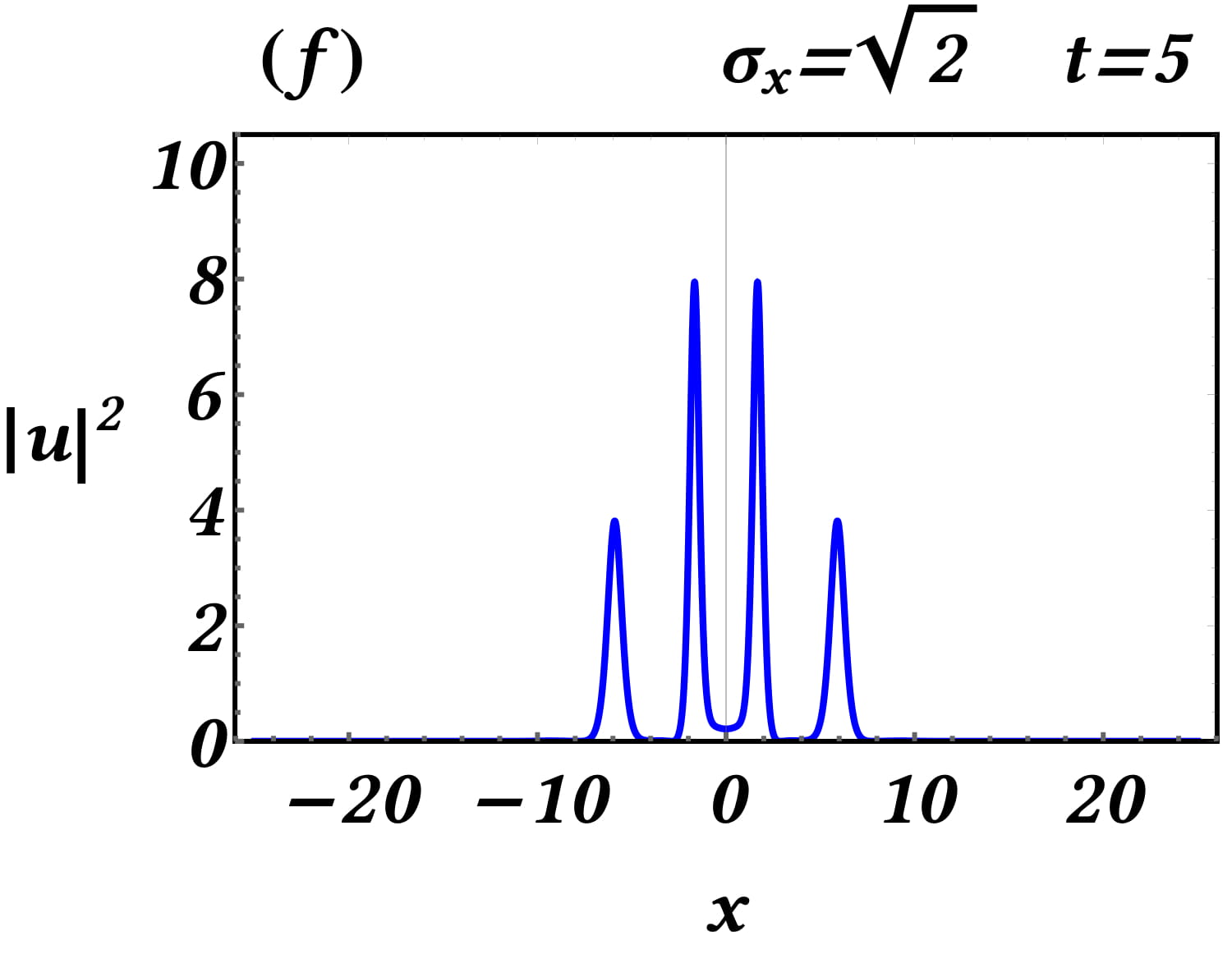}\hspace{0.8cm}
\includegraphics[scale=0.09]{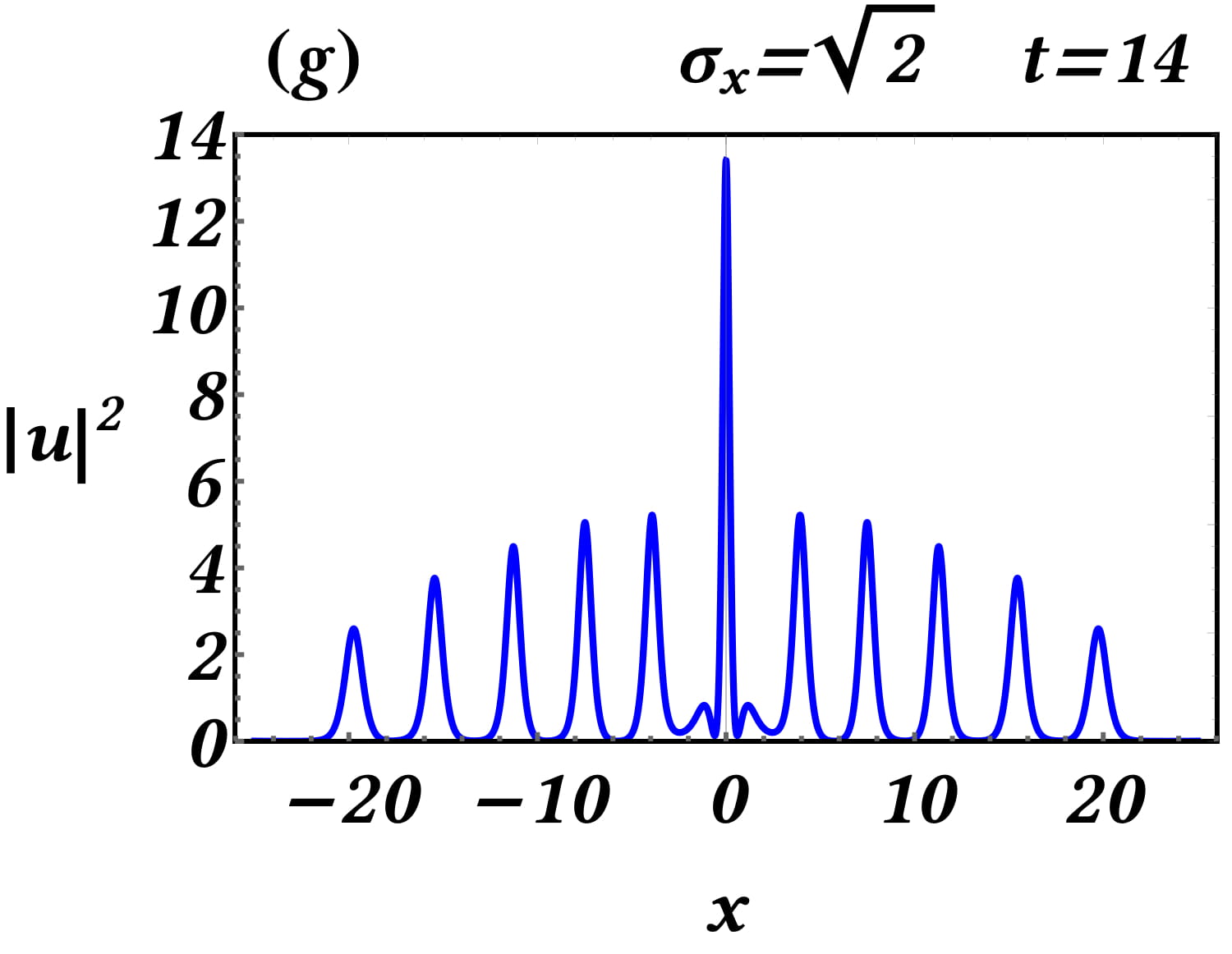}
\end{center}
	\caption{(Color Online) Top and middle rows: A sequence of contour plots of the spatiotemporal evolution of the density for the initial condition \eqref{eq4}, in the presence of the spatially-dependent forcing \eqref{Gaussx}. Panel (a):  $\sigma_x=\sqrt{2}$. Panel (b): $\sigma_x=\sqrt{10}$. Panel (c): $\sigma_x=\sqrt{20}$. Panel (d): $\sigma_x=\sqrt{50}$. Bottom row [panels (e)-(g)]: snapshots of the evolution of the density corresponding to the spatiotemporal dynamics shown in panel (a), for $\sigma_x=\sqrt{2}$.}
	\label{figure12NN}
\end{figure}

The results in this case portray similarities, but also significant differences, if compared to the  case of spatiotemporal driving. We will examine the dynamics of the system for increasing values of $\sigma_x$, starting from the  value of $\sigma_x=\sqrt{2}$. The contour plot of the spatiotemporal evolution of the density in this case,  is shown in panel (a), and the corresponding profile snapshots are depicted in panels (e), (f), (g)
of Fig.~\ref{figure12NN}. 
In the contour plot, 
we observe the emergence of a triangular spatiotemporal region, 
occupied by a traveling wavetrain which appears as a set of straight lines each one representing a solitonic-like structure (see snapshots). This setting 
is reminiscent of the defect scenario classified as flip-flop,
 emitting wavetrains alternately to the left and right -- see Ref.~\cite{sandstede}.

\begin{figure}[tbp]
	\begin{center}
\includegraphics[scale=0.12]{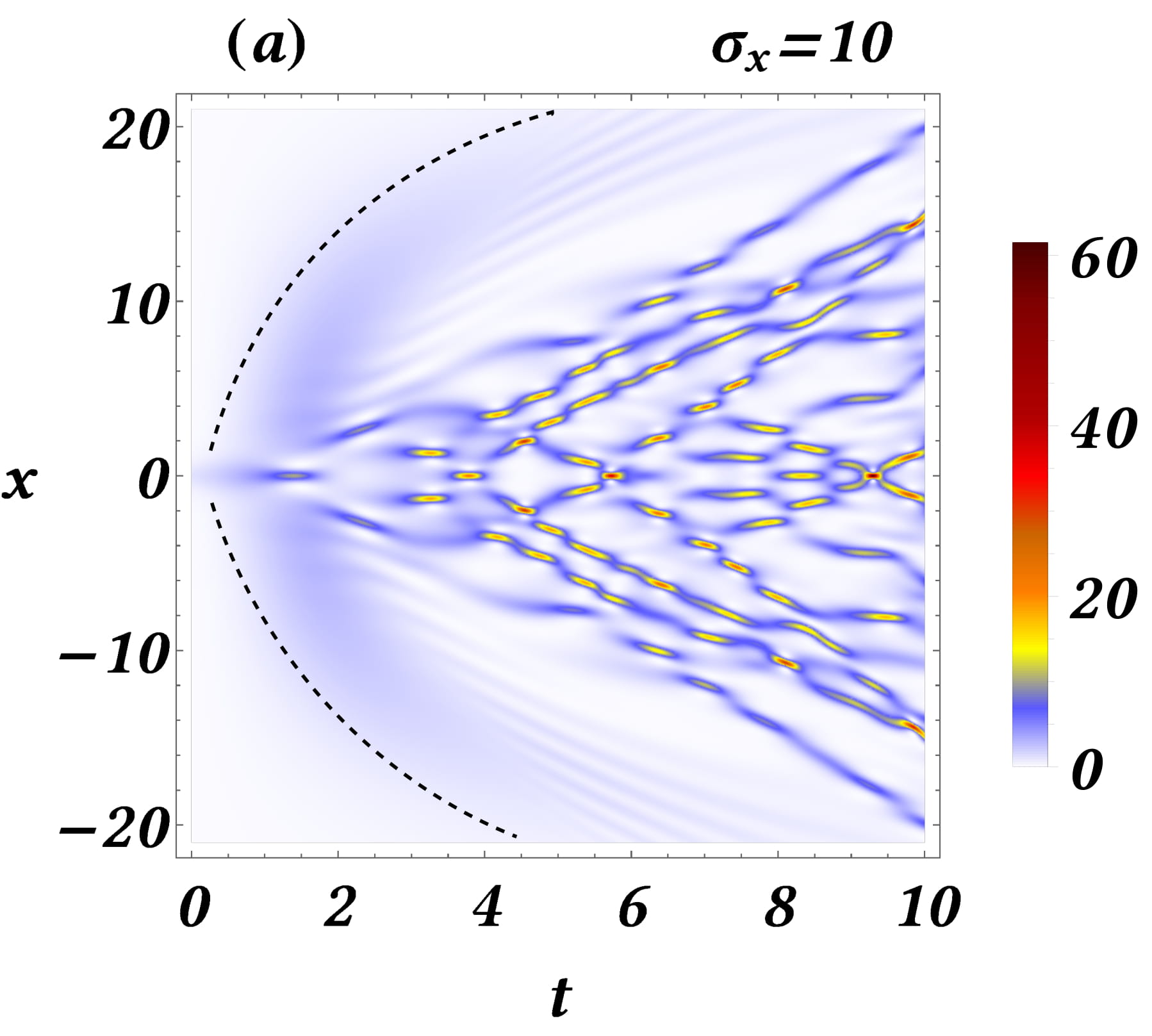}		
\includegraphics[scale=0.12]{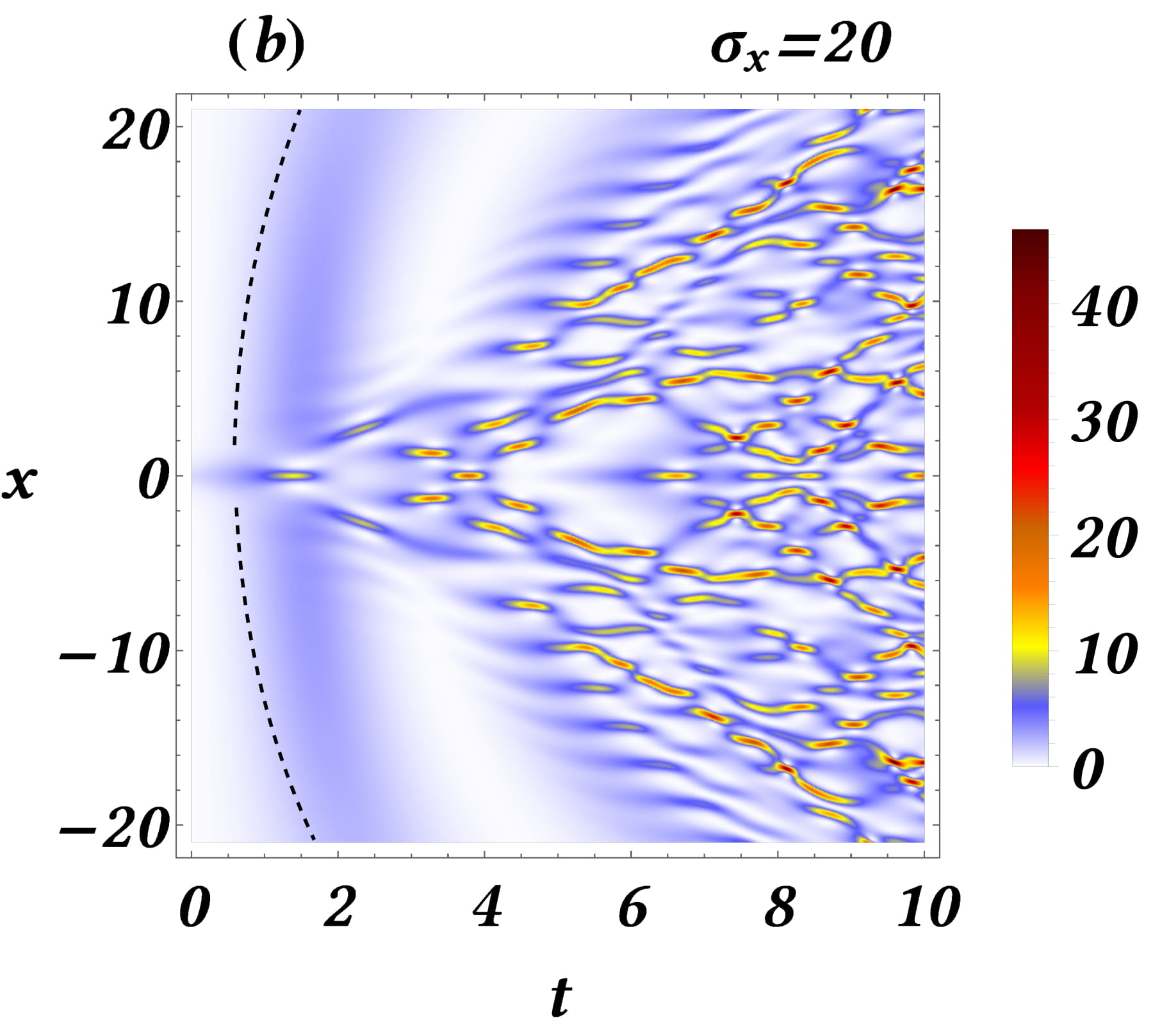}\\
		\hspace{-0.7cm}\includegraphics[scale=0.095]{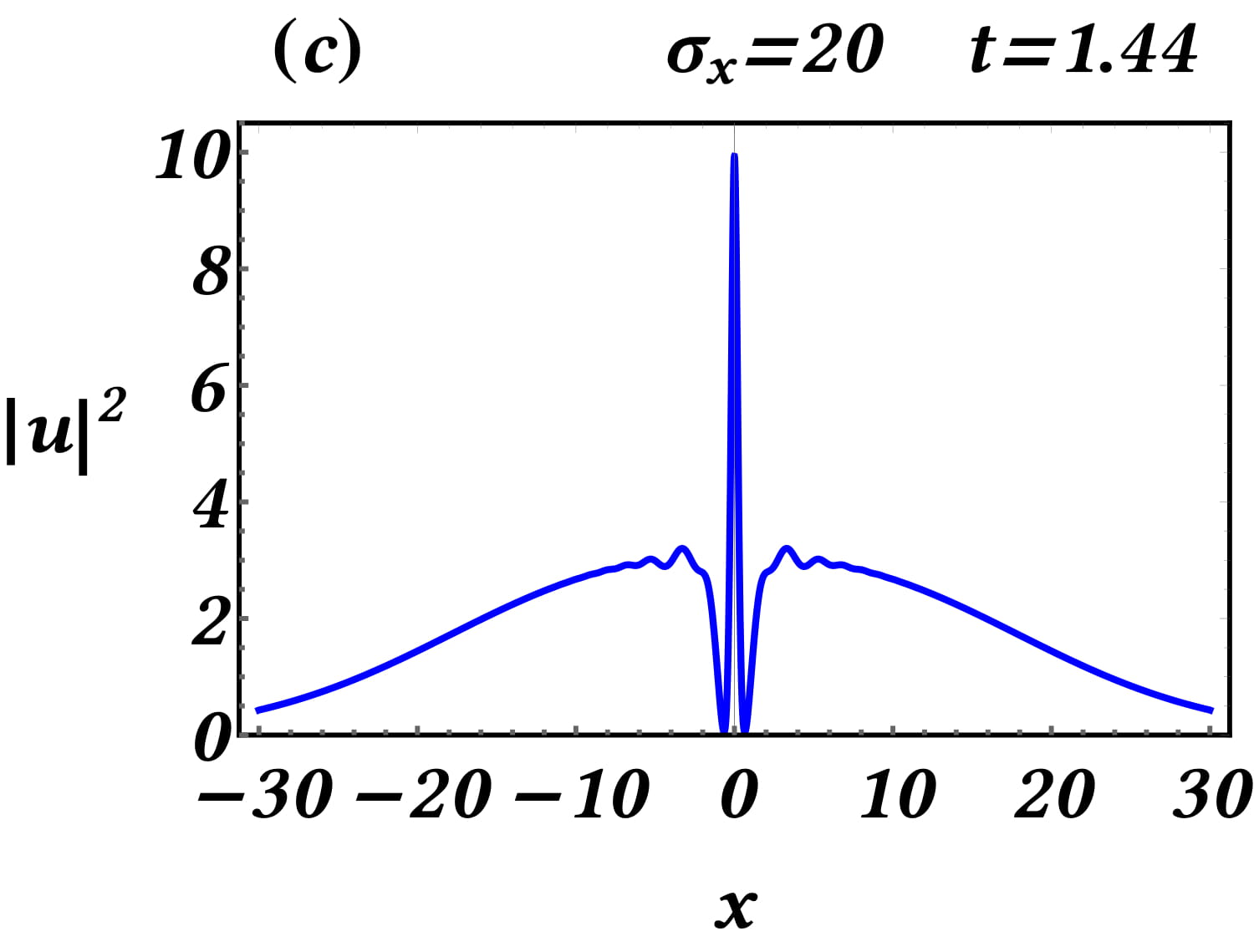}\hspace{0.8cm}
		\includegraphics[scale=0.095]{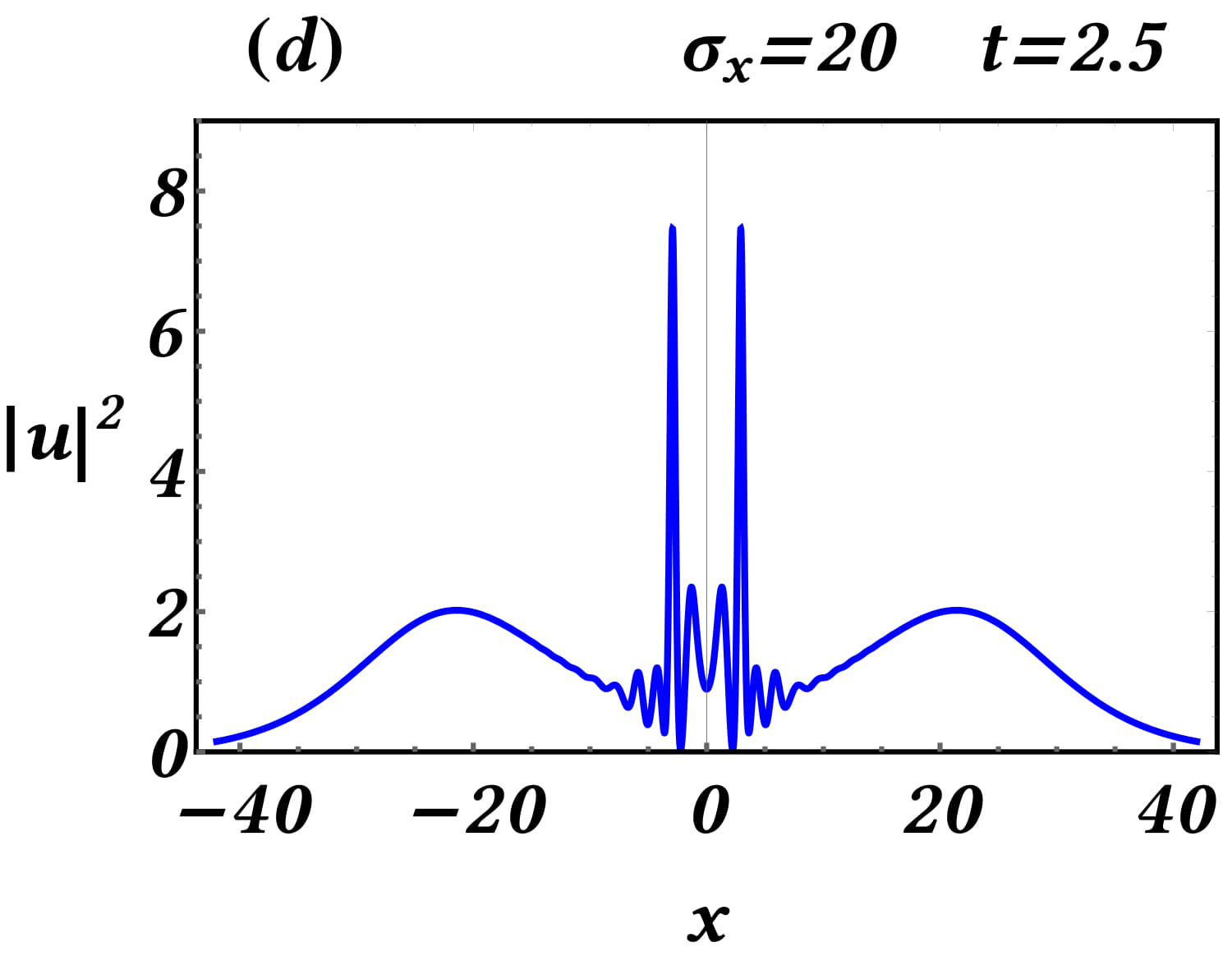}\hspace{0.8cm}
		\includegraphics[scale=0.095]{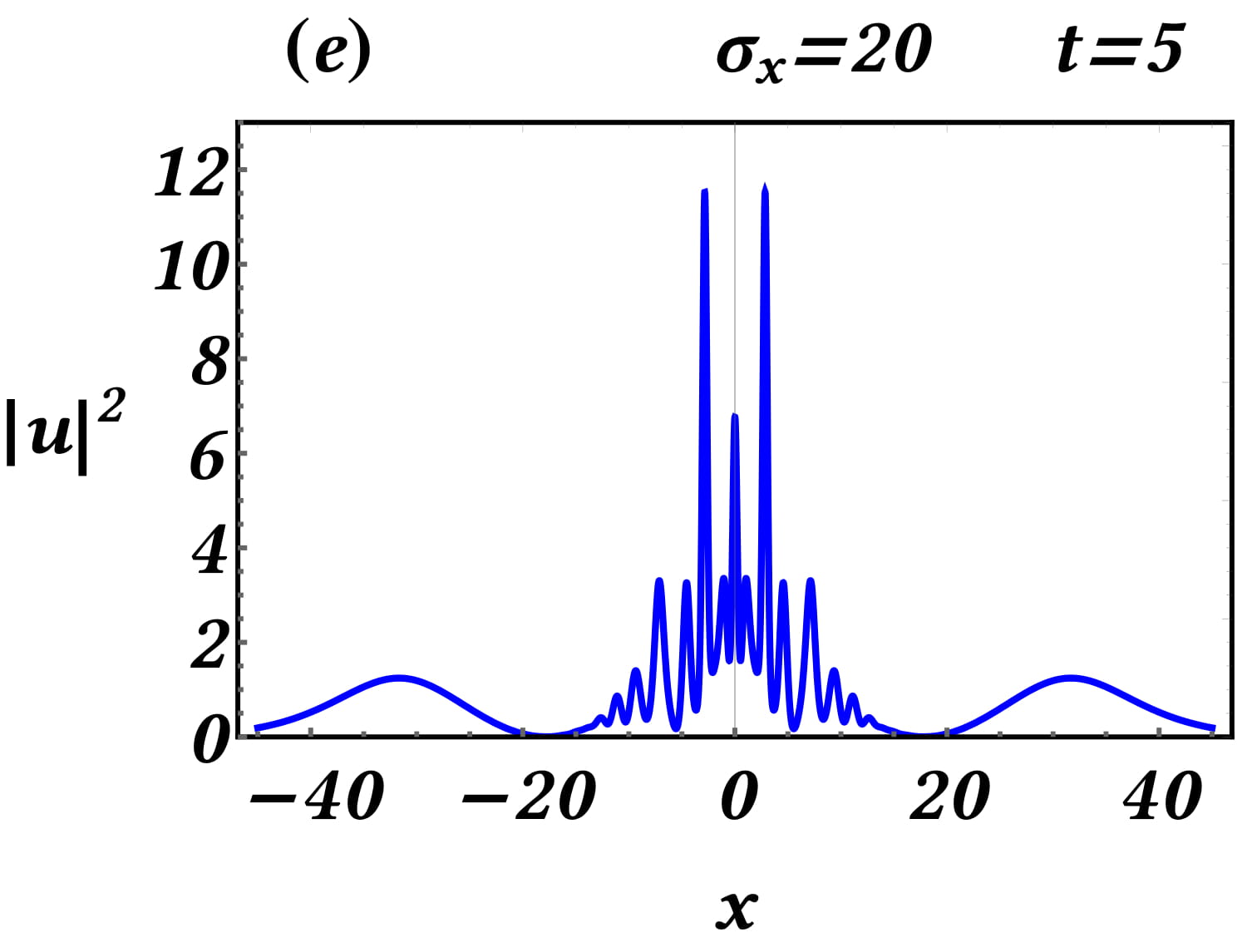}
	\end{center}
	\caption{(Color Online) Top row: Panels (a) and (b) depict
          contour plots of the spatiotemporal evolution of the
          density, for  $\sigma_x=10$ and $\sigma_x=20$, respectively,
          in the presence of the spatially-dependent forcing
          \eqref{Gaussx} for the   initial condition of
          Eq.~\eqref{eq4}.
          Bottom  row [panels (c)-(e)]: snapshots of the evolution of the density corresponding to the spatiotemporal dynamics shown in panel (b), for $\sigma_x=20$.}
	\label{figure13NN}
\end{figure}

The situation, however, drastically changes for larger values of $\sigma_x$. 
First of all, we observe how the straight lines of the interior of the
triangular domain of panel (a) are gradually replaced by a region of
large amplitude oscillations shown in panels (b) and (c). Eventually,
the whole triangular region is consumed by these oscillations, as
shown in  panel (d).  On the other hand, the slope of the exterior
straight lines of the triangular region increases as it can be seen in
panels (a) - (c), until they disappear in panel (d). Moreover, it can
be observed that the support of these solutions exhibits interesting
dynamics itself (the region where it becomes significant is marked by
dotted curves): First, it is progressively expanded within the spatial
domain, for increasing $\sigma_x$ as  is illustrated in panels (b) and
(c), depicting the dynamics for $\sigma_x=\sqrt{10}$ and
$\sigma_x=\sqrt{20}$, respectively. This phenomenon can be connected
to the corresponding one in the case of spatiotemporal driving of
Fig.~\ref{figure10NN} where also the support is expanding for
increasing values of $\sigma_x$. The second effect related to the
support, was already observed in panels (b) and (c) of
Fig.~\ref{figure12NN}, but it is enhanced for even larger values of
$\sigma_x$; the support is decomposed into two symmetric outward
moving parts (replacing the straight-line paths of the travelling
solitons observed in panels (b) and (c)), and a middle part.  This
effect can be more easily detected in the contour plot (d) of Figure
\ref{figure12NN} (for $\sigma_x=\sqrt{50}$), and the contour plots (a)
and (b) of Figure \ref{figure13NN} (for $\sigma_x=10$ and
$\sigma_x=20$, respectively). This dynamics of the support is
highlighted in the snapshots shown in the panels (c)-(e) of Figure
\ref{figure13NN}.  Furthermore, the middle part, sustaining
large-amplitude spatiotemporal oscillations, expands in its own right
forming chaotic patterns reminiscent of the ones studied in
\cite{NB86, nobe1},
as  can be seen in panels (d) and (e) of Figure \ref{figure13NN}. 
Thus, it is evident that the magnitude of $\sigma_x$ drastically affects
the nature of the resulting patterns in the case of purely spatial driving.

\paragraph{Comment on the dynamics of the integrable limit $\gamma=0$, $f=0$.}

We complete this Section by briefly commenting on the dynamics 
of the integrable limit, $\gamma=0$ and $f=0$, with the initial condition  
\begin{equation}
u_0(x)=\frac{1}{1+\frac{x^2}{\sigma^2}}.
\label{eq9b}
\end{equation}
In such a situation, we have found (results not shown here) that the dynamics 
is reminiscent of that observed in Ref.~\cite{BS}. In particular, for $\sigma=1$ 
(in this case, (\ref{eq9b}) coincides with \eqref{eq4}), the solution ``locks'' to a 
soliton of amplitude $\approx 0.73$. As $\sigma$ is increased, we 
progressively see a transition from oscillatory spatiotemporal patterns 
to a ``Christmas tree'' dynamical behavior, resembling the one at the early stages of Fig.~\ref{figure8NN}. 

The above results are, in turn, reminiscent of the initial data width variation reported  
in Ref.~\cite{BS} going from the solitonic limit of small width to 
the {\it semiclassical limit} of large width (see the connection in~\cite{BS} 
with the rigorous work of~\cite{BM1}). 
In addition, we should underline a significant difference between these emerging patterns and the case examples of our previous numerical experiments
involving the external drive. 
In the case where the drive is present, the width of the emerging support 
in which the extreme event lies, is considerably wider, 
connected to the width of the driving. On the other hand, in the integrable
case, due to the absence of such forcing, there is no emergent support and the background, 
on top of  which the extreme events are hosted, is only dictated by the initial condition.


\section{Discussion and Conclusions}
\label{conclusions}

In this work, direct numerical simulations revealed the excitation of extreme wave events for the linearly damped NLS equation supplemented with vanishing boundary conditions, in the presence of a spatiotemporally localized forcing. The driver assumed the form of a Gaussian function, and the dynamics emerged as a
result of  quadratically and  exponentially  decaying initial conditions. 
We found that when the spatial localization scale
of the driving is considerably larger than the scale of its 
temporal localization, the excited Peregrine rogue waveforms appear on top of 
a decaying support, and the transient dynamics, prior to decay, resembles
the one of the semi-classical NLS 
\cite{BM1,BM2}; nevertheless, the extreme events in our case do not appear 
as a lattice of Peregrine waveforms arising at points corresponding to the poles of the tritronqu{\'e}e solution of the 
Painlev{\'e}-I equation. 
The role of the driving/forcing strengths and of the localization
widths of the initial conditions in suppressing/enhancing and
modifying the dynamical features observed were
also elucidated. Another remarkable feature revealed
concerned the critical role of the phase-parameter of the forcing. Varying this parameter, one may produce a wealth of spatiotemporal patterns, including the fundamental ones reminiscent of the dynamics in the integrable limit. These involve the emergence of Peregrine rogue waves (as discussed above), Kuznetzov-Ma like breathers, even the division of the spatial domain in spatially periodic structures reminiscent of those analyzed at the nonlinear stage of modulation instability. Finally, the integrable limit for
the same type of initial data was explored for completeness; in this case,
the width of the initial data plays a crucial role in determining
the eventual fate of the algebraically localized initial
condition.

Via comparison of the dynamics produced by varying the space and time localization width of the driver, we are led to expect  PRW-type structures to form
in the presence of effectively strong spatially extended external forces, acting for short times, while transient, large-amplitude, almost time-periodic
breather-like structures emerge, when the acting time of the external driving is comparable to its spatial localization width. Particularly, our findings seem to further support one of our concluding remarks in  \cite{All1}, that suitable additional effects may contribute for the robustness of such waveforms, making them more easily observable.  Also, connecting these findings to the linearly damped and linearly forced NLS counterpart, \cite{Kharif1, Kharif2, EPeli}, we note that in the
work of~\cite{EPeli}, an effectively strong linear driving which is acting for  short times facilitates the occurrence of extreme wave conditions.

The results herein constitute a starting point for further studies,
such as analytically estimating the decay rates of the numerical solutions, examining the effect of different external drivers and initial conditions possessing different types of decaying rates.
In particular, exploring the ``separatrix'' between
the decay rates that favor the scenario of~\cite{BM1}, vs. those that
lead to the dynamics described in~\cite{BM} would be especially important
to clarify. Of exceptional interest could be the implementation of the deterministic and statistical diagnostics introduced in \cite{Tikan}, in the context of damped and forced NLS models. As an additional direction, the consideration of discrete
counterparts
to the presented phenomenology
in the form of the  discrete linearly damped and forced NLS
equation~\cite{KevreDNLS} will be of interest in their own right. Finally, some other interesting future directions would be to examine similar behaviors in systems of coupled  NLSEs \cite{BorPT,BorCD,BorMAN} or in systems where nonlinearity and dispersion management is applied \cite{BorDPJ}.
Relevant works are in progress, and will be reported in future publications.

{\bf Acknowledgments.} The authors acknowledge that this work was 
made possible by NPRP Grant No. $8-764-160$  and NPRP Grant No. $9-329-1-067$ from Qatar
National Research Fund (a member of Qatar Foundation). 
The findings achieved herein are solely the responsibility
of the authors.



\begin{thebibliography}{99}
\providecommand{\natexlab}[1]{#1}
\expandafter\ifx\csname urlstyle\endcsname\relax
\providecommand{\doi}[1]{doi:\discretionary{}{}{}#1}\else
\providecommand{\doi}{doi:\discretionary{}{}{}\begingroup
	\urlstyle{rm}\Url}\fi

\bibitem{Ablo} M.\,J. Ablowitz and H. Segur, {\it Solitons and Inverse Scattering Transform}
(SIAM, 1981).
%
\bibitem{ablo2} M.\,J. Ablowitz, {\it Nonlinear dispersive waves: Asymptotic analysis and solitons}
(Cambridge University Press, 2011).
%
\bibitem{NB86} K. Nozaki and N. Bekki, {\em Low dimensional chaos in a driven damped  nonlinear Schr{\"o}dinger equation}, Phys. D \textbf{21}, 381--393 (1986).

\bibitem{CLM} D. Cai, D.\,W.  McLaughlin and K.\,T.\,R. McLaughlin, {\em The nonlinear Schr{\"o}dinger equation as
	both a PDE and a dynamical system}, Handbook of dynamical systems, vol. \textbf{2}, 599--675. North-
Holland, Amsterdam, 2002.
%
\bibitem{nobe1} K. Nozaki and N. Bekki, {\em Chaos in a perturbed nonlinear Schr{\"o}dinger equation}, Phys. Rev. Lett. {\bf 50}, 1226--1229 (1983).
%
\bibitem{nobe2} K. Nozaki and N. Bekki, {\em Solitons as attractors of a forced dissipative nonlinear Schr{\"o}dinger equation},
Phys. Lett. A {\bf 102}, 383--386 (1984).
%
\bibitem{Li} Y. Li and D. W. McLaughlin, {\em Morse and Melnikov Functions for NLS PDE's}, Comm. Math. Phys. {\bf 162}, 175--214 (1994).
%
\bibitem{Wig1} G. Haller and S. Wiggins, {\em Multi-pulse jumping orbits and homoclinic trees in a modal truncation of the damped-forced nonlinear Schr{\"o}dinger equation},  Physica D {\bf 85}, 311--347 (1995).
%
\bibitem{kai}  D. Cai, D. W. McLaughlin, and J. Shatah, {\em Spatiotemporal chaos and effective stochastic dynamics for a near-integrable nonlinear system},
Phys. Lett. A {\bf 253}, 280--286 (1999).
%
\bibitem{eli} E. Shlizerman and V. Rom-Kedar, {\em Parabolic Resonance: A Route to Hamiltonian Spatiotemporal Chaos},
Phys. Rev. Lett {\bf 102}, 033901, 1-4 (2009).
%
\bibitem{Ghid88} J.\,M. Ghidaglia, {\em Finite dimensional behavior for the weakly damped driven Schr{\"o}dinger equations}, Ann. Inst. Henri Poincar\'{e} \textbf{5}, 365--405 (1988). 
%
\bibitem{XW95} X. Wang, {\em An energy equation for the weakly damped driven nonlinear Schr{\"o}dinger  equations and its application to their attractors},   Phys. D \textbf{88}, 167--175 (1995).
%
\bibitem{Goubet1} O. Goubet, {\em Regularity of the attractor for the weakly damped nonlinear Schr{\"o}dinger  equations}, Applicable Anal. \textbf{60}, 99--119 (1996).
%
\bibitem{Goubet2} O. Goubet, {\em Regularity of the Attractor
for Schr{\"o}dinger Equation}, Appl. Math. Lett. \textbf{10}, 57--59 (1997).
%
\bibitem{Goubet3} O. Goubet, {\em Regularity of the attractor for a weakly damped nonlinear Schr{\"o}dinger equation in $\mathbb{R}^2$}, Adv. Differential Equations \textbf{3}, 337--360 (1998). 
%
\bibitem{Goubet2a} O. Goubet, {\em Global attractor for weakly damped nonlinear  Schr{\"o}dinger equations in $L^2(\mathbb{R})$},  Nonlinear Anal. \textbf{71}, 317--320 (2009).
%
\bibitem{Lauren95} P. Lauren\c{c}ot, {\em Long-time behaviour for weakly damped driven nonlinear Schr{\"o}dinger equations in $\mathbb{R}^N$, $N\leq 3$}, Nonlinear Differential Equations and Applications NoDEA \textbf{2}, 357--369 (1995).
%
\bibitem{All2} Z.\,A. Anastassi, G. Fotopoulos, D.\,J. Frantzeskakis, T.\,P. Horikis, N.\,I. Karachalios, P.\,G. Kevrekidis, I.\,G. Stratis and K. Vetas, {\em Spatiotemporal algebraically localized waveforms for a nonlinear
	Schr{\"o}dinger model with gain and loss}, Phys. D \textbf{355}, 24--33 (2017).
%
\bibitem{ZNA2019} N. I. Karachalios, P. Kyriazopoulos and K. Vetas, {\em Excitation of Peregrine-type waveforms from vanishing initial conditions in the presence of  periodic forcing}, Z. Naturforsch. \textbf{A 75}, 371--382 (2019).
%
\bibitem{k2a} E. Pelinovsky and C. Kharif (eds.), {\it Extreme Ocean Waves} (Springer, New York, 2008).
%
\bibitem{k2b} C. Kharif, E. Pelinovsky, and A. Slunyaev, {\it Rogue Waves in the Ocean} (Springer, New York, 2009).
%
\bibitem{k2c} A.\,R. Osborne, {\it Nonlinear Ocean Waves and the Inverse Scattering Transform} (Academic Press, Amsterdam, 2010).
%
\bibitem{k2d} M. Onorato, S. Residori and F. Baronio, {\it Rogue and Shock Waves in Nonlinear Dispersive Media} (Springer-Verlag, Heidelberg, 2016).

\bibitem{H_Peregrine} D.\,H. Peregrine, {\em Water waves, nonlinear Schr{\"o}dinger equations and their solutions},
J. Austral. Math. Soc. B \textbf{25}, 16--43 (1983).
%
\bibitem{kuz} E.\,A. Kuznetsov, {\em Solitons in a parametrically unstable plasma}, Sov. Phys.-Dokl. {\bf 22}, 507--508 (1977).
%
\bibitem{ma} Y.\,C. Ma, {\em The Perturbed Plane‐Wave Solutions of the Cubic Schr{\"o}dinger equation}, Stud. Appl. Math. {\bf 60}, 43--58 (1979).
%
\bibitem{akh} N.\,N. Akhmediev, V.\,M. Eleonskii, and N.\,E. Kulagin, {\em Exact first order solutions of the nonlinear Schr{\"o}dinger equation},
Theor. Math. Phys. {\bf 72}, 809--818 (1987).
%
\bibitem{dt} K.\,B. Dysthe and K. Trulsen, {\em Note on breather type solutions of the NLS as models for freak-waves}, Phys. Scr. {\bf T82}, 48--52 (1999).
%
\bibitem{hydro} A. Chabchoub, N. P. Hoffmann and N. Akhmediev, {\em Rogue Wave Observation in a Water Wave Tank}, Phys. Rev. Lett. {\bf 106}, 204502, 1--4 (2011).
%
\bibitem{opt2} B. Kibler, J. Fatome, C. Finot, G. Millot, F. Dias, G. Genty, N. Akhmediev and J. M. Dudley, {\em The Peregrine soliton in nonlinear fibre optics},
Nature Phys. {\bf 6}, 790--795 (2010).
%
%
\bibitem{laser} C. Lecaplain, Ph. Grelu, J. M. Soto-Crespo, and N. Akhmediev, {\em Dissipative Rogue Waves Generated by Chaotic Pulse Bunching in a Mode-Locked Laser},
Phys. Rev. Lett. {\bf 108}, 233901, 1--4 (2012).
%
\bibitem{He} A.\,N. Ganshin, V.\,B. Efimov, G.\,V. Kolmakov, L.\,P. Mezhov-Deglin and P.\,V.\,E. McClintock, {\em Observation of an Inverse Energy Cascade in Developed Acoustic Turbulence in Superfluid Helium}, Phys. Rev. Lett. {\bf 101}, 065303, 1--4 (2008).
%
\bibitem{plasma} H. Bailung, S.\,K. Sharma, and Y. Nakamura, {\em Observation of Peregrine Solitons in a Multicomponent Plasma with Negative Ions}, Phys. Rev. Lett. {\bf 107}, 255005, 1--4 (2011).
%
%
\bibitem{devine} A. Ankiewicz, N. Devine, N. Akhmediev, {\em Are rogue waves robust against perturbations?},
Phys. Lett. A {\bf 373}, 3997--4000 (2009).
\bibitem{calinibook} A. Calini and C. M. Schober,
{\em Rogue Waves in Higher Order Nonlinear Schrödinger Models}, pp.~31--51 in Ref.~\cite{k2a}.
%
\bibitem{NRbor5a} Y.  Wang,  L. Song,  L. LI  and B. A. Malomed, {\em High-power pulse trains excited by modulated continuous waves}, J. Opt. Soc. Am. B \textbf{32}, 2257--2263 (2015).
%
%
\bibitem{NR5Wang} L. H. Wang, K. Porsezian, and J. S. He, {\em Breather and rogue wave solutions of a generalized nonlinear Schr{\"o}dinger equation},  Phys. Rev. E {\bf 87}, 053202, 1--10 (2013).

%
\bibitem{NR4Anki} A. Ankiewicz, Y. Wang, S. Wabnitz, and N. Akhmediev, {\em Extended nonlinear Schr{\"o}dinger equation with higher-order odd and even terms
	and its rogue wave solutions},
Phys. Rev. E {\bf 89}, 012907, 1--9 (2014).
%
\bibitem{NRbor6} Y. Yang, Z. Yan and B. A. Malomed, {\em Rogue waves, rational solitons, and modulational instability in an integrable fifth-order nonlinear  Schr{\"o}dinger equation}, Chaos \textbf{25}, 103112, 1--9 (2015).
%
\bibitem{BorPT} Y. V. Bludov, R. Driben, V. V. Konotop and B. A. Malomed, {\em Instabilities, solitons and rogue waves in
	$\mathcal{PT}$ -coupled nonlinear waveguides}, J. Opt. \textbf{15}  064010, 1--7 (2013).
%
\bibitem{BorCD} H. N. Chan, B. A. Malomed, K. W. Chow and E. Ding, {\em Rogue waves for a system of coupled derivative nonlinear Schr{\"o}dinger equations}, Phys. Rev. E \textbf{93}, 012217, 1--10 (2016).
%
\bibitem{BorMAN} W. P. Zhong,  M. Beli\'{c} and B. A. Malomed, {\em Rogue waves in a two-component Manakov system with variable coefficients
	and an external potential}, Phys. Rev. E  \textbf{92}, 053201, 1--5 (2015).
%
\bibitem{BorDPJ} J. Cuevas-Maraver, B. A. Malomed, P.G.Kevrekidis and D.J.Frantzeskakis, {\em Stabilization of the Peregrine soliton and Kuznetsov--Ma breathers by means of nonlinearity and dispersion management}, Phys. Lett.  A \textbf{382}, 968-972 (2018).
%
%
%
%
\bibitem{Kharif1} C. Kharif and J. Touboul, {\em Under which conditions the Benjamin-Feir instability may spawn an extreme wave event: A
	fully nonlinear approach}, Eur. Phys. J. Special Topics \textbf{185}, 159–-168 (2010).
%
\bibitem{Kharif2} C. Kharif, R. A. Kraenkel, M. A. Manna
and R. Thomas, {\em The modulational instability in deep water under the action of wind and dissipation}, J. Fluid Mech. \textbf{664}, 138–149 (2010).
%
\bibitem{BM1} M. Bertola and A. Tovbis, {\em Universality for the focusing Nonlinear Schr{\"o}dinger
	Equation at the gradient catastrophe point:
	Rational breathers and poles
	of the Tritronqu\'{e}e solution to Painlev\'{e}},  Comm. Pure Appl. Math. \textbf{66}, 678--752 (2009).
%
\bibitem{BM2} R.\,H. J. Grimshaw and A. Tovbis, {\em Rogue Waves: analytical predictions},  Proc. R. Soc. A \textbf{469}, 20130094 (2013). 
  %
\bibitem{Rev1_b} A. Tikan, C. Billet, G. El, A. Tovbis, M. Bertola,  T. Sylvestre,
F. Gustave, S. Randoux, G. Genty, P. Suret, and J. M. Dudley, {\em Universality of the Peregrine Soliton in the Focusing Dynamics
	of the Cubic Nonlinear Schr\"{o}dinger Equation}, Phys. Rev. Lett. \textbf{119}, 033901, 1--6,  (2017).
%
\bibitem{BS} E.\,G. Charalampidis, J. Cuevas-Maraver, D.\,J. Frantzeskakis, and P.\,G. Kevrekidis, {\em Rogue Waves in Ultracold Bosonic Seas}, Rom. Rep. Phys. \textbf{70} 504, 1--26 (2018).
  %
\bibitem{Yang1} G. Yang, L. Li and S. Jia, {\em Peregrine rogue waves induced by the interaction between a continuous wave and a soliton}, Phys. Rev. E \textbf{85}, 046608 (2012).
%
\bibitem{Yang2} G. Yang, Y. Wang,  Z. Qin,  B.\,A. Malomed,  D. Mihalache and L. Li, {\em Breatherlike solitons extracted from the Peregrine rogue wave} Phys.  Rev. E \textbf{90}, 062909 (2014).
\bibitem{BM} G. Biondini and D. Mantzavinos, {\em Universal nature of the nonlinear stage of modulational instability}, Phys. Rev. Lett. \textbf{116}, 043902 (2016).
%

 %
\bibitem{sandstede} B. Sandstede, A. Scheel, {\em Defects in oscillatory media: toward a classification}, SIAM J. Appl. Dyn. Syst. {\bf 3}, 1--68 (2004).
  %
\bibitem{All1} J. Cuevas Maraver, P. G. Kevrekidis, D. J. Frantzeskakis, N. I. Karachalios, M. Haragus and G. James, {\em Floquet analysis of Kuznetsov-Ma breathers: A path towards spectral stability of rogue waves}, Phys. Rev. E \textbf{96}, 012202, 1--8 (2017).
%
\bibitem{EPeli} A. Slunyaev, A. Sergeeva and E. Pelinovsky, {\em Wave amplification in the framework of forced nonlinear Schr{\"o}dinger 
	equation: The rogue wave context}, Phys. D \textbf{303}, 18--27 (2015).
%
\bibitem{Tikan} A. Tikan, {\em Effect of local Peregrine soliton emergence on statistics of random waves in the 1-D focusing Nonlinear Schr{\"o}dinger equation}, \url{https://arxiv.org/abs/1905.11938}.

\bibitem{KevreDNLS} P.\,G. Kevrekidis, {\em The Discrete Nonlinear Schr\"odinger Equation: Mathematical Analysis, Numerical Computations and Physical Perspectives} (Springer, 2009).
%
\bibitem{onorato1} M. Onorato and  D. Proment, {\em Approximate rogue wave solutions of the forced and damped nonlinear Schrödinger equation for water waves},  Phys. Lett. A \textbf{376}, 3057
(2012).
%
\bibitem{brunetti} M. Brunetti, N. Marchiando, N. Berti, J. Kasparian, {\em Nonlinear fast growth of water waves under wind forcing}, Phys. Lett. A \textbf{378}, 1025 (2014).
%
\bibitem{DHZ19} L. Dostal, M. Hollm and E. Kreuzer, {\em Study on the behavior of weakly nonlinear water waves in the presence of Random Wind Forcing}, \url{https://arxiv.org/abs/1909.11761}.


\end{thebibliography}
\end{document}